# Does the brain behave like a (complex) network? I. *Dynamics*

D. Papo[1,2], J.M. Buldú[3]

[1] *Department of Neuroscience and Rehabilitation, Section of Physiology, University of Ferrara, Ferrara (Italy)*
[2] *Center for Translational Neurophysiology, Fondazione Istituto Italiano di Tecnologia, Ferrara (Italy)*
[3] *Complex Systems Group & G.I.S.C., Universidad Rey Juan Carlos, Madrid (Spain)*
\* Email address: david.papo@unife.it

| ARTICLE INFO | ABSTRACT |
|---|---|
| *Keywords*:<br>Brain dynamics<br>Brain topology<br>Criticality<br>Functional networks<br>Spatial networks<br>Adaptive networks<br>Structure-dynamics relationship<br>Disorder<br>Emergence<br>Topological phase transitions<br>Intrinsic property<br>Network controllability | Graph theory is now becoming a standard tool in system-level neuroscience. However, endowing observed brain anatomy and dynamics with a complex network structure does not entail that the brain actually works as a network. Asking whether the brain behaves as a network means asking whether network properties count. From the viewpoint of neurophysiology and, possibly, of brain physics, the most substantial issues a network structure may be instrumental in addressing relate to the influence of network properties on brain dynamics and to whether these properties ultimately explain some aspects of brain function. Here, we address the dynamical implications of complex network, examining which aspects and scales of brain activity may be understood to genuinely behave as a network. To do so, we first define the meaning of *networkness*, and analyse some of its implications. We then examine ways in which brain anatomy and dynamics can be endowed with a network structure and discuss possible ways in which network structure may be shown to represent a genuine organisational principle of brain activity, rather than just a convenient description of its anatomy and dynamics. |

## 1. Introduction

Representing brain anatomy, dynamics and function as a network is now becoming standard in neuroscience. A network is a deceptively simple structure, in essence, a collection of units and relations between them, but with a plethora of characteristic properties which can be extracted from the relation pattern. Endowing a system with a network structure does not necessarily entail that this system actually works as a network and that the properties of the associated network are those of the underlying system. Rather than affording descriptions highlighting essential aspects of whatever brain aspect one wishes to characterise, a network representation may simply allow operations on data, such as comparing conditions, evaluating differences, distances, and related concepts, or even classifying populations. Nonetheless, the set of operations permitted on the reconstructed network structure may differ from the set of operations actually carried out by the system, particularly at system-level scales. Therefore, a fundamental question in network neuroscience is whether a network structure reflects genuine aspects of brain organisation and phenomenology, or else is an epiphenomenon of coordinated dynamical activity in the same way as spatio-temporal electrical field fluctuations are sometimes thought of, and therefore merely a convenient set of tools.

It is therefore essential to evaluate the extent to which a network representation, particularly at system-level scales, can reveal fundamental aspects of brain dynamics, whether it can produce specific brain phenomenology, and ultimately whether it genuinely documents the way the brain carries out the functions it is supposed to fulfil, the extent to which they show robustness with respect to the way they are equipped with a network structure or the way in which the structure is parsed when analysing it (Atmanspacher and beim Graben, 2007; Haimovici and Marsili, 2015). Addressing these issues involves understanding the meaning, conditions, and implications of networkness, but also understanding what may count as a convincing explanation. It also involves answering the following related questions: are network properties telling us something more than pure connectivity? Is there a scale at which we can consider that the collection of nodes and their connections starts being a network? What aspects of brain activity genuinely behave as a network? What aspects of known brain phenomenology can be ascribed to its network structure? Are there genuine topology-driven neural phenomena? Is there some network-based equivalent of *dys*connection syndromes? In fine, how much of brain physics network structure can (correctly) represent? Can network theory help highlighting as yet unknown phenomenology? What are the minimal ingredients a network structure should incorporate to reflect/highlight properties of brain physics?

In the remainder, we first discuss the meaning of networkness and the general conditions under which a complex system has genuine network structure. We then review ways in which neural systems can meaningfully be equipped with a network structure in different domains and at various spatial and temporal scales and analyse some implications for neural systems' dynamics of equipping a neural system with a network structure. Throughout, the discussion focuses on bare dynamics, as opposed to genuine functional brain activity (Papo, 2019; Korhonen et al., 2021), which will instead be formally defined, differentiated from the former and examined in a companion paper (Papo and Buldú, in preparation). Here, we treat dynamics in a way that is divorced from considerations of neural systems' ability to perform a given task and, to a large extent, from its neurophysiological properties, a stance often implicitly adopted in network neuroscience studies, and ultimately the brain as a spatially-discrete, continuous or discrete time dynamical system. (See also *§4. Concluding remarks*). most basic network structure, i.e. simple networks, possibly time varying, but evaluated at zero-lag

## 2. Networkness: meaning and conditions

With 1 mm$^3$ of cerebral cortex estimated to contain ~$10^5$–$10^6$ neurons and ~$10^{10}$ synapses, and several kilometres of wires connecting neurons (Braitenberg and Schütz, 2013), the underlying neural tissue can be thought of as approximately continuous (Robinson, 2013). As a result, brain activity can be modelled as the output of an underlying dynamical system and treated as a field $\mathcal{F} = (\vec{s}, t)$, where $\vec{s}$ is the position in the anatomical space and $t \in \mathbb{R}$ is the physical time, at scales ranging from observational to developmental and evolutionary. $\mathcal{F}$ can be a scalar, a vector or a tensor field, but can also have further structure, including symmetries. It can also be associated with the system's frequency domain, or phase space.

On the other hand, if one represents neural tissue as a set of cables connecting separate entities (neurons, modules, etc.) and brain activity as some dynamics occurring within this physical system, it is straightforward to endow both brain anatomy and various aspects of brain activity with a graph or network representation (Bullmore and Sporns, 2009; Fornito et al., 2016). As in the continuous representation, a network structure need not be limited to this real space representation and may be associated with other aspects of brain anatomy, dynamics, and function. In its simplest form, a graph is a structure $N = (V, E)$, where $V$ is a finite set of *nodes* or *vertices* and $E \subseteq V \otimes V$ a set of pairs of $V$ called *links* or *edges*. The edges can carry a weight, parametrising the strength of an interaction, and a direction, where edges are ordered (see *Table 1*). Moreover, networks can be static or time-varying (Holme and Saramäki, 2012). More subtly, edges may be evaluated at zero-lag or may contain time delays (Cabral et al., 2011; Deco et al., 2008; Petkoski et al. 2023). Most of the time, we will resort to arguably the most basic class of graphs, called *simple* graphs, i.e. graphs comprising no loops and multiple edges between pairs of nodes. The phenomenological implications of graph theoretical structure relaxing the constraints of simple networks' assumptions will be discussed elsewhere (Papo and Buldú, in preparation).

The sheer number of components calls for the use of statistical approaches, including coarse-graining (Chow and Karimipanah, 2020). Crucially, in this approach, which is predicated upon statistical mechanics (Albert and Barabási, 2002; Dorogovtsev et al., 2008) a combination of mechanics and laws of large numbers, the identity of nodes and links loses importance, at least *prima facie*, as the network's properties are statistical in nature.

A network structure comes with properties of different types (see *Table 2*). In addition to its purely *combinatorial* properties (Bollobás, 1986), a network structure is characterised by *topological* and *geometric* properties at all scales (Boccaletti et al., 2006), as well as measurable symmetries (Garlaschelli et al., 2010; Pecora et al., 2014; Dobson et al., 2022). Combinatorics relates to the ways discrete objects can be counted, arranged, and constructed (Erickson, 2014). It may play a central role in covering and packing problems, and therefore in neural codes and in functional robustness. Topology is concerned with properties that are invariant under continuous deformation without tearing or cutting, e.g. bending, twisting, and stretching, but does not consider other properties such as size or orientation. Topological spaces constitute the broadest context in which the notion of a continuous function makes sense. In particular, topology provides a fundamental way to define *nearness* and, hence, properties such as *continuity, connectedness,* or *compactness*. This naturally provides a flexible notion of distance, a useful property given the complex spatial structure of brain anatomy, and functional quasi-invariance in the face of its variability across subjects. A topological structure allows comparing systems of different metric sizes and differing local properties and, to some extent, converting local into global properties (Ghrist, 2014; Simas et al., 2015), providing robust features to characterise brain activity. At experimental scales, brain anatomy can be identified with a static *topological space*[1] $\mathfrak{I} = (N, \mathfrak{R}_N)$, where $\mathfrak{R}_N$ is a relation on *N*'s elements. This structure may have additional superstructure, e.g. a group G, acting on *N*. Brain dynamics can then be treated as a *topological dynamical system*, i.e. a triple $(\mathfrak{I}, G, T_G)$, where $T_G$ is a continuous function $T_G: G \times \mathfrak{I} \to \mathfrak{I}$. (See *§3.1.1 Structure-dynamics: the role of time scales* and *§3.5 Topological phase transitions*). The properties of interest in geometry are those that are invariant under rigid transformation, e.g. translation, rotation, or reflection. These properties are invariant under *isometries*, i.e. distance-preserving transformations. Importantly, while most of them, such as lengths, angles, volumes are reference- or embedding-dependent some, e.g. Gaussian curvature, are intrinsic[2], and therefore constitute an invariant property of the metric on the considered space.

Equipping a system with a network structure may come with some noticeable advantages. First, it constitutes a natural, versatile, and inherently multiscale characterisation of multi-body systems' structure. Second, a network structure allows understanding network features as statistical properties and individual systems as instances of some *network ensemble* with given properties (Park and Newman, 2004; Bianconi, 2007, 2009). (See *Table 2*) Thus, network descriptions may have a certain degree of universality, and the sort of robustness that this property confers. Furthermore, a network structure can be perturbed and, to some extent, controlled (Liu and Barabási, 2016) or steered to desired configurations (Gutiérrez et al., 2012, 2020). (See *Table 2*) Altogether, a network structure can help in addressing fundamental neuroscientific issues such as functional localisation and parcellation and affords the ability to predict but also retrodict (by considering observed structure as the result of some process) the system, to quantify its evolvability and to act upon it.

| Table 1. Basic network terminology: constituents and connectivity | |
|---|---|
| Brain network | Any representation of the brain anatomy or activity with a predefined (global or partial) parcellation, where each brain region corresponds to a node and connections between nodes are quantified according to a certain anatomical or dynamical criterion. |
| Brain node | A *node i* represents any brain partition that may have an anatomical or functional meaning, from a single neuron or a voxel, to large brain regions or regions of interest (ROI). |
| Brain link | A *link l{i, j}* quantifies any kind of interaction between two (or more) brain nodes *i* and *j*. Interactions can be physical (from synapse to large tracts) or dynamical (e.g., coordinated dynamics between two nodes), evolve with time and co-exist at different temporal and spatial scales. |
| Brain connectivity | Brain connectivity is in general defined in the anatomical space and can refer to the presence of anatomical fibres, or statistical dependencies between the activity of different nodes or causal interactions ("effective connectivity") between units identified with nodes within the nervous system. |
| Sparsity | A network with *N* vertices and *K* edges is said to be sparse if *K* is much smaller than the possible maximum number of links a network with *N* nodes in principle can have, and the node degree is finite for $N \to \infty$. |

| Table 2. Network structure | |
|---|---|
| Adjacency matrix | The *adjacency matrix* $A_{ij}$ of a network accounts for the connections between any pair of nodes *i* and *j*. Most network properties can be extracted from the adjacency matrix. |
| Degree and degree distribution | The *degree* of a node refers to the number of connections that the node has with other nodes in the network. The *degree distribution* of a network is the probability distribution of nodes having a specific degree value. In a *random network*, the degree distribution decays exponentially, whereas *scale-free networks* have a power-law degree distribution, where few nodes possess very high degrees while most nodes have low degrees. The degree distribution constrains several network properties, e.g. information propagation or robustness to node failure. |
| Laplacian matrix | The *Laplacian matrix* $\mathcal{L}_{ij}$ is a transformation of $A_{ij}$, such as $\mathcal{L}_{ij} = K - A_{ij}$, where *K* is a diagonal matrix and the elements $K_{ii}$ accounts for the degree of node *i*, so that $\mathcal{L}_{ij}$ is a zero-row sum matrix. The *Laplacian matrix* allows extracting both topological and dynamical information contained in the graph spectrum, e.g. the number of connected components or cycles and the diffusion or vibration modes taking place on the graph. |
| Random, small-world, scale-free and fractal networks | A *random network* is a type of network where the connections between nodes are established in a random or probabilistic manner. There are different ways to create random networks, but most extended is the Erdős-Rényi (ER) model, where links between nods are selected randomly according to a probability *p* (Erdős and Rényi, 1959). A *small-world network* is a network with (i) high *clustering coefficient* and (ii) low *shortest path length*. The clustering coefficient quantifies the local connection density within a network and can be assessed through either the proportion of closed triplets or the likelihood that two neighbors of a specified node exhibit a direct connection. The shortest path length denotes the average of the shortest distance (i.e., the path with the lowest number of traversed links) between any pair of nodes. In small-world networks, the shortest path length experiences logarithmic growth with an increasing number of nodes. Small-world properties are thought to play a crucial role in network information processing capabilities. *Scale-free networks* are characterised by degree distribution that follows a power-law, which means that a few nodes have a very high number of connections (hubs), while most nodes have relatively few connections (Barabási and Albert, 1999). This particular degree distribution has been reported in a variety of real networks with an exponent between $2 < \gamma < 3$. |
| Network modularity | The presence of communities or clusters is a common phenomenon in various types of complex networks characterised by groups of nodes densely interconnected among themselves, while having fewer connections to nodes outside their respective groups. *Network modularity* refers to a measure that quantifies the extent to which a network can be divided into modules and is used as a proxy for *community detection*. |
| Hierarchical network | A *hierarchical network* is a network structure where the nodes are organised into multiple levels or layers based on their connectivity and relationships. Nodes within the same level are more densely connected to each other, and connections between nodes in different levels tend to be fewer and more specific. |
| Network topology and geometry | Network *topology* refers to the way the links and nodes of a network are arranged to relate to each other. It can be thought of as a very loose geometry where what is required is a notion of nearness (and its preservation through maps) but not a metric. However, it is sometimes appropriate to consider networks as metric spaces, equipped with some distance. Network *geometry* can arise in various ways, each with its own specific metric (Boguñá et al., 2021). For instance, some networks may not have explicit Euclidean coordinates but can still exhibit certain forms of "hidden" or "latent" spatial organisation, where the relationships between nodes and edges into a space where distances and angles have meaningful interpretations. A network geometry can also be induced by dynamical processes unfolding on networks. |
| Network ensembles | A *network ensemble* refers to a collection or set of multiple network instances that share certain common characteristics or properties. These ensembles are used to study and analyse the statistical properties and behaviours of networks, or as a null |

---

[1] A topological space is a collection of objects, along with a collection of subsets referred to as open sets. Intuitively, a set *U* is open if, starting from any point in *U* and going in any direction, it is possible to move a little and still remain inside the set.

[2] A property is said to be *extrinsic* when it depends on the coordinate system in which it is embedded. *Intrinsic* properties of surfaces are properties that can be measured within the surface itself without any reference to a larger space.

| | |
|---|---|
| | model. Network ensembles are generated by creating multiple network instances, typically through random processes, by sampling from a distribution, or by applying different algorithms or parameters to generate the ensemble. |

| Table 3. Some dynamical concepts | |
|---|---|
| Dynamics *on/of* networks | A brain network is a dynamical system whose organisation changes at different spatial and temporal scales. At the same time, a set of variables may describe the dynamical state of each node. Therefore, we can distinguish between the dynamics of the whole network structure (dynamics *of* networks) and the dynamics of each of its nodes (dynamics *on* networks). |
| Synchronisation | *Synchronisation* designates the coordinated motion of two or more dynamical systems thanks to the existence of a certain interaction. Synchronisation is one of the fundamental phenomena supporting communication and information transfer in the brain. There exists a diversity of co-existing synchronisation scenarios (e.g., amplitude, phase, anti-phase or delay synchronisation) and a vast repertoire of metrics to detect and quantify the level of synchronisation. Topological and spectral properties of brain networks have been shown to play a key role in the synchronisation of systems of coupled oscillators. |
| Network observability | A system and, more particularly, a network, is said to be *observable* if it is possible to reconstruct the network's complete internal state from its recorded outputs. |
| Network control and targeting | The general goal of *network control* is to stabilise a trajectory of the networked dynamical system that would not be naturally attained in finite time (Liu and Barabási, 2016), while the one of *targeting* consists in attaining one of the system's trajectories achievable from its attractor basin, effectively forcing the system dynamics into forgetting its initial condition (Gutiérrez et al., 2012, 2020). This may involve perturbing node or link dynamics. For linear systems, controllability and observability are mathematically dual. |

## 2.1 Reducibility to network structure

Two fundamental questions relate to the *reducibility* of brain activity to a network representation. Can the brain be reduced to a network? What does such a reduction require? Does such a structure make brain dynamics and function *observable*, i.e. does the network structure retain sufficient information to adequately reconstruct the full system dynamics? These issues can somehow heuristically be divided into those that hinge on the method used to construct networks from experimental data and those that do not, although the same aspect may sometimes fall into both categories. Of course, the technical aspects of network reconstruction from experimental data have a profound impact on the observability of brain structure afforded by a network representation, and so have the variables and models chosen to describe it (Letellier and Aguirre, 2002; Aguirre et al., 2018). But a more fundamental question could also be addressed by supposing that the reconstruction from empirical data could be carried out optimally, i.e. in a way that reflects the true underlying structure of the system. Even in this theoretical case, to what extent and under what conditions a network representation allows recovering the states of a high-dimensional system such as the brain is still poorly understood.

### 2.1.1 Connectedness

A fundamental issue relates to the definition of the relation among the objects of the network structure. Network modelling in neuroscience is predicated upon connectivity. But is connectivity a necessary and sufficient condition for networkness?

The neuronal hardware has a distinctively cable-like structure at various spatial scales, a property which is naturally reflected by *connectivity*. However, at the algorithmic level, collective dynamics and function need not directly arise from a connected structure and may instead be a consequence of *collectivity* (Fraiman et al., 2009; Chialvo, 2010). The simultaneous activity of large enough neuronal populations can generate strong spatial extracellular voltage gradients, which neurons are sensitive to, due to the extracellular space conductivity, enhancing spike–field coherence and bias the preferred spiking phases (Buzsáki et al., 2012). This field-like activity, called *ephaptic coupling*, has been highlighted at various scales of neural activity (Weiss and Faber, 2010; Anastassiou et al., 2011; Anastassiou and Koch, 2015; Martinez-Banaclocha, 2018). While the fact that this effect can be mimicked by appropriate intracellular current injections renders its functional role unclear, the presence of ephaptic coupling implies that for field scales comparable to that of the node, the main effect would not be network-like, and below the node scale, neural architecture should be *exchangeable*, i.e. it should be symmetric under node label permutation. Moreover, collectivity may be used as a criterion to establish the scale at which the network structure emerges. Perhaps even more importantly, experimental results have shown that periodic activity can self-propagate by endogenous electric fields even when neuron-to-neuron physical connections are interrupted and that signal propagation due to cortical wave modes in highly folded areas may be uncorrelated with the fibre directions (Zhang et al., 2014; Qiu et al., 2015b; Chiang, et al., 2019; Shivacharan et al., 2019).

Connectedness is also not a sufficient condition. For instance, a small, connected network, i.e. a network with few nodes, would not in general constitute a genuine network in a statistical mechanical sense in which nodes and links are understood in a statistical sense. In such a system, connectivity would fulfil the functional role first ascribed to brain i.e. that of allowing neural signals to flow from one segregated functional module to another (Wernicke, 1874), a vision consistent with a box-and-arrow computer-like metaphor of the brain. In this framework, connectivity produces phenomenology by creating a link between well-identified computational units. In the statistical mechanics approach, a necessary condition would instead be that of *emergence* of collective behaviour from the interactions of a great number of components. In this approach, brain connectivity does more than simply connect different parts of a neural circuit allowing neural signals to flow across brain regions. Connectivity may in fact lie at the root of complex generic properties of brain activity (Osorio et al., 2010; Kozma and Puljic, 2015; Kozma and Freeman, 2016), acting as a control parameter of brain dynamics (Osorio et al., 2010; Sreenivasan et al., 2017).

### 2.1.2 Discretisability

A network structure can be seen as a discretisation of the continuous field $\mathcal{F}$, i.e. a map $\Phi \mapsto N$, from the space $\Phi$ of fields $\mathcal{F}$ to the space $N$ of networks $N$, where the former can be recovered from the latter in the infinite limit of $V$. This space contraction implies that $\Phi$ is a Hausdorff (separable) space. The relation $E$ is typically predicated upon *connectedness*, a property of topological spaces, whereby an entity cannot be represented as the sum of two separated parts. This property reflects both the cable-like structure appearing at all scales of brain anatomy, from the microscopic scales of local neuronal populations, to the macroscopic ones of giant fibres interconnecting distant brain regions, and the structure of dynamical connectivity between brain regions, which are believed to be required for the correct execution of a great number of cognitive tasks (Varela et al., 2001), and to be either reduced or increased in several neurological and psychiatric conditions (Friston, 1998; Stam, 2014).

If one identifies single cells (neurons, glias, etc.) as the microscopic scale of description, the brain as a whole can indeed be thought of as a discrete system. However, at mesoscopic scales, discretisation becomes more challenging and a continuous medium approximation can be adopted, wherein brain matter is thought of as approximately continuous, so that spatially organised neuronal ensembles of interacting populations can be described in terms of neural field equations under the assumption that space partitions are respectively much larger and much smaller than the microscopic and macroscopic ones (Lesne, 2007), corresponding for instance to the characteristic length of some descriptor of the dynamics or of function. Equipping the brain with a network structure requires mapping the underlying continuous field onto a discrete set of (usually, though not necessarily) point-wise nodes (Korhonen et al., 2021)[3]. Insofar as network vertices are by construction coarse-grained units, the dynamics of a discrete system effectively constitutes a kinetic equation, wherein some degrees of freedom are coalesced into an error term, irrespective of the scale at which the system is described. The conditions under which discretising the continuous field representation is non-singular and essentially without loss of information are poorly known. This is in part due to the fact that the discretisation steps are in general not intrinsic, for example they are not prescribed by the dynamics. The extent to which information is lost as a result of discretisation also depends in practice on the choice the scales of description, observation, variations, and correlations (Lesne, 2007). On the other hand, the network structure becomes genuinely relevant when continuous and discrete representations non-spuriously become inequivalent, the latter being better thought of as a convenient representation of the former in the opposite case. Such a difference would for instance result if the discrete representation constituted a singular limit of the continuous field, inducing a qualitative change in the system behaviour with irreversible loss of information.

Discretisation has several implications. First, much as embedding, it implies a dimensionality reduction and although this may capture functionally important phenomena at given scales, it may also affect the properties of the resulting system (Severino et al., 2016; Kafashan et al., 2018; Safari et al., 2017). (See *§3.5.2 The role of network dimension in brain dynamics*). From a topological viewpoint, the set of vertices should represent a *good cover* of the underlying (continuous) space, which in turn should be locally contractible[4]. Such a contraction is akin to the reduction of a mechanical system to its centre of mass, which is allowed by the system's symmetries. At a given scale, a node can be assumed to be *irreducible*[5]. On the other hand, a node can be thought of as resulting from a renormalisation process at scales not considered in the model. Ultimately, a network representation should constitute an effective field theory for the underlying physical theory, i.e. it should include the appropriate degrees of freedom to describe the system at a chosen length scale, and all the substructure and degrees of freedom at shorter distances should be negligible. However, evaluating the structure's relevance is not straightforward[6], while it is not completely clear under what conditions this renormalisation leads to functionally meaningful units, effective field theories typically work best when there is a large separation between the length scale of interest and the length scale of the underlying dynamics.

Discretising connections between a limited number of nodes implies that these constitute an average connectivity, approximating connections travelling indirectly through multiple polysynaptic paths that do not feature in the connection matrix between renormalised nodes (Galán, 2008; Robinson, 2012, 2013; Robinson et al., 2016). Finally, discretisation implies that the system possesses some *separation* property. This is typically used to support models of neuronal ensemble dynamics at various spatial scales as a set of coupled heterogeneous active systems, e.g. threshold oscillators (Ashwin et al., 2016; Sreenivasan et al., 2017). Defining oscillators may be intuitive at microscopic spatial scales but is far less straightforward at meso- and macroscopic scales. Note that this is both a reconstruction-dependent and independent issue. However, discretisation becomes a challenging task when dealing with spatially extended systems of largely unknown organisation and complex dynamics. In this case, collapsing the space to a network requires a further step logically preceding discretisation, i.e. parcellation. As a consequence, brain models wherein time-invariant spatially-embedded network nodes are endowed with some local dynamics (Deco et al., 2011; Cabral et al., 2011) are just as accurate as this parcellation, which itself involves prior assumptions on brain function (Korhonen et al., 2021).

### 2.1.3 Structure preservation

Ideally, the network structure should be *equivalent* (up to some transformation) to that of the underlying space (anatomical, dynamical, or functional). A faithful representation should then preserve the system's intrinsic properties and symmetries (Cross and Gilmore, 2010), and should therefore involve a structure-preserving discretisation. Two complementary questions should be addressed. First, what properties should be introduced or forbidden at microscopic scales for a network structure to allow a faithful representation of functional brain activity? Second, what properties of brain dynamics and function *can* networks account for or even highlight? Answering this question involves understanding what properties network microscopic scales should have to reproduce known anatomical patterns and dynamical phenomenology.

### 2.1.4 Intrinsicality

Our knowledge of brain network structure essentially hinges on two different approaches: network reconstruction methods from experimental data and computational models of brain activity. In some sense, both can be thought of as a particular *inverse problem* (Nguyen et al., 2017; Squartini et al., 2018), involving the reconstruction of connectivity kernels given a prescribed dynamics of the activity field (Coombes et al., 2014). Inverse problems are, by definition, ill-posed in the absence of boundary conditions. Specifying these conditions involves choices of varying degrees of arbitrariness. Both approaches comprise several aspects or steps that crucially depend on discretionary choices and assumptions (Korhonen et al., 2021), an essential problem that neither experimental nor theoretical neuroscience can obviate.

---

[3] In a sense, this is a problem akin to that of surface triangulation. Which manifolds have piecewise-linear triangulations is a complex topological question. Smooth compact surfaces can be triangulated (Jost, 1997), and topological manifolds of dimensions 2 and 3 are always triangulable by an essentially unique triangulation, each of these manifolds admitting a smooth structure, unique up to diffeomorphism. However, the anatomical and even more the dynamical and functional spaces can only approximatively be thought of as smooth and compact.
[4] A *cover* of a set $X$ is any family $\mathcal{U}$ of subsets $U_i$ with union $X$. $\mathcal{U}$ is a *good cover* of $X$ if every non-empty intersection of subsets $\bigcap_{i \in \sigma} U_i$ is *contractible*. A space $X$ is contractible if it is *homotopically equivalent* to a constant map, i.e. intuitively, if it can be continuously shrunk to a point of that space.
[5] A topological space is *irreducible* if it is not the union of two proper closed subsets, and an irreducible component is a maximal closed subspace that is irreducible for the induced topology. A similar definition can be given for geometric objects such as *algebraic sets* varieties, i.e. sets of solutions of systems of polynomial equations over the real or complex numbers. According to the Lasker–Noether theorem, any such object is the union of a finite number of uniquely defined algebraic sets, called its *irreducible components*.
[6] In the complex networks approach, *node* and *microscopic scale* are essentially synonymous. In a sense, defining a node and defining a module are comparable tasks, as the latter can be thought of as resulting from the coarse-graining of nodes at a lower scale.

One fundamental issue is then to evaluate the extent to which the properties of the reconstructed system are invariant with respect to a priori assumptions and discretionary choices. For instance, while topological features may show invariance with respect to some aspects of brain recording (Billings et al., 2021), different link metrics may induce different geometries (Amari and Nagaoka, 2007). The resulting space may for example be a Riemannian manifold[7] equipped with a metric on the tangent bundle (Lenglet et al., 2006; Krajsek et al., 2016). A Riemannian tangent space parameterisation allows preserving the geometry of connectivity, yielding intrinsic, i.e. reference independent properties. This has a clear meaning for the anatomical structure, as the tangent space parameterisation preserves the geometry of anatomy-embedded functional connectivity. Various studies suggest that such an assumption may be more appropriate than a Euclidean embedding of dynamical connectivity in terms of disease (Qiu et al., 2015a; Ng et al., 2016, 2017; Dadi et al., 2019; Pervaiz et al., 2020) and ageing prediction, subject identification (Venkatesh et al., 2020; Abbas et al., 2022), and of the design of brain connectivity interface (Congedo et al., 2017; Yger et al., 2017), and methodologically, in terms of harmonisation of multi-site data (Simeon et al., 2022).

However, its meaning is less clear for dynamical networks, as the dynamics may turn out to live in a rather complex space, and even less for the resulting functional space, which cannot in general be treated as a smooth manifold, even when the dynamical space is (Papo, 2017, 2019a). On the other hand, a network can in general be endowed with extrinsic geometrical properties. In fact, a network can always be embedded in a surface, provided it is of sufficiently high genus, i.e. it has a sufficient number of surface handles (Aste et al., 2005), and Nash's embedding theorem allows viewing a Riemannian manifold as a submanifold of a Euclidean space, providing extrinsic definitions of intrinsic properties. Finally, the relation used to define network links, for example considering directed as opposed to undirected links, or generalising the standard network structure to allow many-body interactions, affects several important network properties, including dimension and symmetries (Salnikov et al., 2018; Torres and Bianconi, 2020; Millán et al., 2020), and ultimately also the physics associated with the system's network structure. For example, symmetric connectivity readily accounts for equilibrium systems, whereas asymmetric coupling matrices reflect out-of-equilibrium systems with break-down of detailed balance.

## 2.2 Modelling the brain as a network

In its early stages, computational neuroscience essentially turned its attention to single-neuron dynamics. By the mid-twentieth century, it became an established idea that at least some information processing in the brain is performed at neural population as opposed to single-neuron level (Feldman and Cowan, 1975). Studies started looking into the dynamics of neuronal populations in order to find a theoretical framework for studying the collective behaviour of neuronal populations (Griffith, 1963a, 1965). The large-scale spatio-temporal dynamics of interacting neurons on a cortical surface is then represented as a continuum neural field (Coombes, 2010; Coombes et al., 2014). Neural field models describe mean-field neural dynamics at mesoscopic spatial scales (>0.5 mm), treating cortical activity as a superposition of traveling waves propagating through a physically continuous sheet of neural tissue. In neural field models, the large-scale dynamics of neuronal populations is modelled in terms of nonlinear integro-differential equations, whose kernel represents the connectivity's spatial layout (Bressloff, 2011, 2012, 2014; Coombes et al., 2012; Chow and Buice, 2015; Bressloff et al., 2016). Such models provide an important example of spatially-extended dynamical systems with nonlocal interactions. Neural fields can exhibit a rich phenomenology, including pulses, travelling fronts, or spiral waves (Ermentrout, 1998; Xu et al., 2023), and have been used to model wave propagation *in vivo* (Huang et al., 2004) and *in vitro* (Pinto and Ermentrout, 2001; Richardson et al., 2005). A paradigmatic continuous neural field model is the Wilson-Cowan equation for the voltage $u$ of a neuronal population at time $t$ and point $x \in \Omega \subseteq \mathbb{R}^d$:

$$\partial_t u(x,t) = -u(x,t) + \int_\Omega w(x,y) f(u(y,t)) dy + \xi(x,t) \quad [1]$$

where $w(x, y)$ is a spatial kernel approximating neural interactions between different cortical locations, typically declining roughly exponentially with distance (Markov et al., 2013; Ercsey-Ravasz et al., 2013), $f$ is a (usually nonlinear) gain function, and $\xi$ is a noise input term[8] (Wilson and Cowan, 1972; Cowan et al., 2016; Chow and Karimipanah, 2020). The problem is equipped with some initial and possibly some boundary condition.

Discretising field models such as Wilson-Cowan's involves considering the brain as a multi-body system of interacting units. Such multi-body systems can be described either at microscopic or at coarse-grained, phenomenological scales. The microscopic scale may be identified with neurons, or neuronal masses at various scales, and may contain more or less biological detail. Cortical columns are often treated as cortical systems's basic dynamical units, which are coupled through sparse long-range cortical connectivity. Thus, at system-level neocortical activity is often modeled as an array of weakly coupled dynamical units, whose behaviour is represented by dynamical attractors of various types (Breakspear and Terry, 2002). In its simplest form, the system's units are static, and are in essence similar to spins in the Ising model describing the main qualitative features of ferromagnetic phenomenology (Kadanoff, 2009). Each unit can also be thought of as a *dynamical system* (Golubitsky and Stewart, 2002a), e.g. a spiking neuron or a neural mass. A dynamical system is a pair $(\mathcal{M}, X)$ where $\mathcal{M}$ is some space, e.g. a manifold, and $X$ is a vector field on that space. Overall, the networked system, whose nodes are dynamical systems, is then a discrete space, continuous time dynamical system, i.e. a pair $(N, \mathcal{P})$, where $N$ is a graph and $\mathcal{P}$ is a function assigning to each node $v \in V$ a phase space, which can be thought of as a manifold $\mathcal{P}(v)$ (DeVille and Lerman, 2015). For instance, the field equation [1] can be discretised into a set of separate dynamical systems; the kernel $w(x, y)$ is replaced by the coupling weight matrix $w_{ij}$ describing the connectivity between any two nodes $i$ and $j$, and $I_i$ is an external input from distant nodes.

$$\dot{u}_i = -u_i + f_j\left(\sum_j w_{ij} u_i + I_i(t)\right) \quad [2]$$

In the simple case in which each node is identified with a spiking neuron, the equivalent expression represents the neuron's spiking activity, and equation [2] takes the form[9]

---

[7] Riemannian manifolds are spaces which are locally homeomorphic to $\mathbb{R}^n$, and which are endowed with rules for measuring distances and angles, subject to restrictions ensuring that these quantities behave analogously to their Euclidean counterparts. In particular, a Riemannian manifold is equipped with a smooth varying inner product $g(\cdot,\cdot)$ on the tangent space.

[8] In its simplest form, the noise term is thought of as additive Gaussian white noise. However, noise of a different type, e.g. multiplicative noise (Bressloff and Webber, 2012), and different statistics may also apply (Faisal et al., 2008).

[9] Spike trains are stochastic processes, subject to various noise sources, i.e. dendritic and synaptic activity (Tsuda, 2001; Faisal et al., 2008), but also from endogenous balancing of excitation and inhibition, which generates chaos-like dynamics (van Vreeswijk and Sompolinsky, 1996, 1998; Renart et al., 2010). They are often modelled as sums of delta

$$x_i(t) = x_i^0(t) + R_i(t)\left(I_i(t) + \sum_j W_{ij}(t)x_j(t)\right) \quad [3]$$

where $x_i^0(t)$ is neuron $i$'s baseline firing when considered in isolation; $R_i(t)$ is neuron $i$'s trial-averaged linear response to synaptic input perturbation, $I_i(t)$ is an external input to neuron $i$, and $W_{ij}$ the magnitude and time course of synaptic connections between node $i$ and any other node $j$ in the system. In the frequency domain, this expression would characterise the trial-averaged firing and linear response rate.

More generally, omitting for now the noise term, a typical firing-rate neural network model describes the time evolution of $N$ recurrently connected nodes $x_i$,

$$\dot{x}_i = \mathbf{F}\left(x_i, \frac{\sigma}{N}\sum_{j\neq i}^{N} A_{ij}\mathbf{H}(x_j)\right) \quad [4]$$

where $\dot{x}_i(t)$ denotes the state variation of the $i$th node at time $t$, $\mathbf{F}(x_i, 0)$ the dynamics of the isolated node, $\sigma$ is the overall coupling strength, $A_{ij}$ the coupling matrix, $k_i = \sum_{i=1}^{N} a_{ij}$ node $i$'s degree, and $\mathbf{H}(x_j)$ a coupling function representing the drive from other nodes on the $i$th node[10]. A common approach involves additive interactions:

$$\dot{x}_i = \mathbf{F}(x_i) + \frac{\sigma}{N}\sum_{j=1}^{N} A_{ij}\mathbf{H}(x_j) \quad [5]$$

which may be justified for systems with essentially pairwise interactions summed according to some linear weight given by connectivity. A force term can also be added to account for external forcing.

A similar network representation can also be associated with the phase space of the dynamics (Dorogovtsev et al., 2008; Baiesi et al., 2009; Papo et al., 2014c). In such a representation, the phase space is a network whose nodes represent local energy minima of the system and links transition pathways connecting two neighbouring minima (Baronchelli et al., 2009). The corresponding Hamiltonian is then:

$$\mathcal{H} = -\sum_{i<j} J_{ij} a_{ij} \sigma_i \sigma_j - \sum_i h_i \sigma_i \quad [6]$$

where $J_{ij}$ equals 1 if nodes $i$ and $j$ are linked and 0 otherwise, and $h_i$ is a local applied field. Thus, nodes identify mesoscopic states and links transitions between them (Schnakenberg, 1976). A graph may also express the conditional dependence structure between random variables of a probabilistic model, encoding a distribution over a multi-dimensional space (Koller and Friedman, 2009).

Unlike in the previous scenario, in this context, nodes and links both emerge from the dynamics[11]. Similarly, the trajectories of a discrete dynamical system form a directed network in phase space, wherein each node, representing a state, is the source of a link pointing to its dynamical successor (Shreim et al., 2007). Such networks may take various forms, e.g. that of a Lorentzian manifold[12], wherein causal relations between points define *accessibility*. Nodes can also correspond to the *causal states* of a process, i.e. to predictively equivalent sets (Shalizi and Crutchfield, 2001; Crutchfield et al., 2009). These spaces' topology and geometry are induced by the process from which they arise. For instance, the neighbourhood structure can be induced by the Markov blanket of a node, i.e. the set of states containing necessary and sufficient information required to predict the behaviour of that subspace and stemming from it (Pearl, 1988; Parr et al., 2020). In the Markov random field induced by the mean-field approximation, a node's Markov blanket is simply its adjacent nodes, but may otherwise be more complex.

*2.2.1 Elements of networkness*

Irrespective of the particular specification, networkness is incorporated into dynamical equations in two different variables, which turn out to be the relevant fields for systems of coupled oscillators qualitatively similar to those governed by Eq. [4] (Sornette, 2002; Osorio et al., 2010). The first is represented by the coupling factor $\sigma$, which can *prima facie* be thought of as some aspect of the anatomical network. For instance, for weak coupling of the Kuramoto phase oscillator (Kuramoto, 1975; Acebrón et al., 2005; Breakspear et al., 2010; Rodrigues et al., 2016), the oscillators run incoherently, whereas beyond a critical coupling threshold, collective synchronisation emerges spontaneously (Acebrón et al., 2005; Arenas et al., 2008). One limitation is of course its uniformity across the system, which is usually assumed. Morevoer, its meaning from an experimental viewpoint is not totally clear. The coupling term $\sigma$ constitutes a convenient control parameter in theoretical studies, but its identification and meaning in terms of neural structure are not entirely clear. In principle, it could be characterised in terms of the neuropil's composition and connectivity strength distribution (Ashwin et al., 2016; Pillai and Jirsa, 2017), but most often there would be no clear distinction between coupling term and connectivity matrix. The second variable encoding networkness is the matrix $A_{ij}$ which contains the system's topological structure. This term encodes the topology of direct interactions between linearisable dynamical systems. These linear operators generalise differential operators to arbitrary discrete interaction topologies, and their eigenvalues and eigenvectors capture the way topology affects collective dynamics and its stability at all scales. The coupling structure may be represented by different types of matrices, e.g. the *adjacency matrix* or the *graph Laplacian*, usually represented by a matrix $\mathcal{L}_{ij}$, or a transition matrix. $A_{ij}$ may come in different forms, e.g. as nearest-neighbour, hierarchical or random long-range coupling, or even as state-dependent interactions. $\mathcal{L}_{ij}$, which is a simple transformation of $A_{ij}$, allows extracting both topological and dynamical information contained in the graph spectrum, e.g. the number of connected components or cycles and the diffusion or vibration modes taking place on the graph (Monasson, 1999; Farkas et al., 2001; Goh et al., 2001; de Lange et al., 2014; Jost and Mulas, 2019). Solutions of the diffusion equation can be expressed as linear combinations of eigenvectors of the graph Laplacian $\mathcal{L}$. All these models are situated at a description level wherein nodes and coupling are unambiguously defined a priori. In practice, though, one key problem lies in the way single identifiable units are obtained from neuronal populations at lower scales (Fornito et al., 2010; Stanley et al., 2013; Papo et al., 2014b; Korhonen et al., 2021).

---

functions $x_i(t) = \sum_k \delta(t - t_i^k)$, where $t_i^k$ is the time of neuron $i$'s $k$th spike (Ocker et al., 2017a).

[10] Generic external inputs can be accounted for through the following equation: $\dot{x}_i = \mathbf{F}\left(x_i, \frac{\sigma}{N}\sum_{j\neq i}^{N} A_{ij}\mathbf{H}(x_j), \beta_i \mathbf{b}(t)\right)$, where $\beta_i$ quantifies the way a time-dependent input signal $\mathbf{b}(t)$ affects node $i$. (See also §3.4.3 Network structure and response to external fields).

[11] In a statistical mechanics approach, the system's degrees of freedom are identified with links, the maximum number of particles playing the role of volume in classical physical systems (Gabrielli et al., 2019).

[12] A *Lorentzian manifold* is a differentiable manifold with a metric tensor that is everywhere nondegenerate, but in which, contrary to a Riemannian manifold, the requirement of positive-definiteness is relaxed.

### 2.2.2 The ground level of networkness: from bare connectivity to complex networks

The first obvious logical step when addressing network structure relevance to brain dynamics and function is understanding what would constitute a ground level for *networkness*. This involves minimal assumptions on the ingredients through which networkness is introduced. When considering a connected space, the most important issue is perhaps whether it has some additional structure beyond bare connectivity.

A classical minimal assumption on structure is represented by the *mean-field representation* (Kadanoff, 2009; Le Bellac et al., 2004). In neuronal populations, perturbations may trigger reverberating loops where each neuron or subpopulation may influence and be influenced by other ones. Mean-field theory simplifies this scenario by imposing self-consistency between influencer and influenced and neglecting correlations and higher-order statistics. Concretely, the interaction between a node and its afferents is represented by a mean field generated by the latter. When there is a large enough number of coupled units and correlations between them are not too strong, the central limit theorem ensures that fluctuations vanish so that each neuronal population can be assumed to receive the mean input generated by all other populations and external currents. The general framework applies to situations in which fluctuations are irrelevant, and the input to one part of the system can be thought to result from the sum of a sufficiently high number of independent terms and the central limit theorem allows replacing the input current with a Gaussian variable. In this framework, it is often assumed that the synaptic weight distribution only depends on the distance between interacting populations, i.e. $w(x,y) = w(|x-y|)$, with $w(s)$ a decreasing function of separation $s$. (Ermentrout and Cowan, 1998; Bressloff et al., 2001; Henderson and Robinson, 2013). While steep connectivity decreases with distance have been reported (Kaiser et al., 2009), non-trivial structure in a wide range of scales of cortical anatomy has also been reported (Hagmann et al., 2008). (See also *§Network structure and pattern formation in brain dynamics*).

The general mean-field framework can be realised by two configurations with different order and symmetries. On the regular side of the spectrum, nodes may for instance sit upon each site $r$ of a lattice graph[13] embedded in $\mathbb{R}^3$, interacting with each other through the lattice edges. In the simplest case, whereby each node may be identified with a spin $\sigma_r$, and take one of two possible values, the corresponding Hamiltonian is that of equation [6], where $J_{ij}$ reduces to a dimensionless coupling strength $\kappa$. On the other hand, randomness is often thought of as a null condition in network analysis (e.g., van Vreeswijk and Sompolinsky, 1996; Brunel, 2000; Kozma and Freeman, 2016). For instance, in its original version, the Wilson-Cowan model models the activity of *randomly coupled* heterogeneous threshold neurons (Wilson and Cowan, 1972, 1973). In an *Erdős-Rényi graph*, each node is randomly and uniformly connected to a finite fraction of the other nodes, so that every node plays essentially the same role. Second, mean-field theories of networked systems are predicated upon assumptions on the absence of some property, e.g. dynamical correlations between nodes, local clustering, i.e. the states of neighbouring nodes are independent of each other, and modularity, i.e. all nodes of a given degree $k$ are described by the average over all nodes of degree $k$. For instance, in the *heterogeneous mean-field* prescription (Dorogovtsev et al., 2008), while fluctuations are still assumed to be negligible, nodes within degree classes are lumped together under the hypothesis that all nodes within a degree class have the same dynamical properties (Pastor-Satorras and Vespignani, 2001). As a result, mean-field networks have particular topological properties: they have no loops, and are tree-like or directed acyclic graphs (Chow and Karimipanah, 2020).

In spite of their initial popularity in brain modelling (van Vreeswijk and Sompolinsky 1996; Amit and Brunel, 1997; Brunel 2000; Renart et al., 2010), and of their success in describing neural activity (Cowan et al., 2016), uniform random networks cannot accommodate the connectivity patterns found in biological experiments (Yoshimura and Callaway 2005; Yoshimura et al. 2005; Tomm et al., 2014), and mean-field models' main assumptions appear to be violated at various levels of neural structure. For instance, correlations have been shown to play a major role in neural dynamics (Salinas and Sejnowski; 2001; Schneidman et al., 2006; Averbeck et al., 2006). Moreover, the spatial homogeneity assumption is in general violated in both anatomical and dynamical brain networks. As a consequence, models of brain dynamics going beyond the mean-field approach (Buice and Chow, 2013), with structure represented by *complex* rather than *random* networks (Yoshimura et al., 2005; Song et al., 2005; Bullmore and Sporns, 2009; Perin et al., 2011) seem in general more appropriate. However, under what conditions and to what extent mean-field predictions cease to constitute an accurate model of brain dynamics and function and complex network structure becomes a more accurate model are still poorly understood questions at both theoretical and experimental levels.

### 2.2.3 Disorder, symmetries, phases, and phase transitions

A natural way to conceive of networkness is as a form of (strong) disorder (Dorogovtsev et al., 2008). In a many-particle system, order and disorder are usually thought of in terms of the presence or absence of some symmetry or correlation. A rather general interpretation of disorder is given by Landau's approach (Landau and Lifshitz, 1958). Phase transitions are associated with symmetry changes and their order is determined by the index of the broken symmetry phase's subgroup $H \subset G$, where $G$ is the global symmetry group of the system's Hamiltonian. Phases are described in terms of an *order parameter*, whose value vanishes in the presence of symmetry and is different from zero in the broken-symmetry phase. Thus, an *ordered medium* is a space described by a function assigning an order parameter to every point in the space. The possible values of the order parameter constitute an *order parameter space*.

From these general ingredients, i.e. disorder, symmetry, phases and phase transitions, it is possible to understand various key aspects of networkness. First, disorder can be understood as *heterogeneity*. Heterogeneity may come in different forms, e.g. topological disorder or disorder in interaction parameters, with different underlying causes. For instance, in computational models, heterogeneity is often represented not only by local oscillation frequency dispersion, i.e. dispersion of the natural frequency probability distribution of **F** in equation [4] (Scafuti et al., 2015) or by oscillators' connectivity strength distribution, but also by the topology induced by the connectivity, including, at the simple level, through *sparsity*[14] (Osorio et al., 2010). Second, the presence of disorder affects important properties of a system. For instance, disorder can affect transport coefficients' value, e.g. velocity, or the system's governing laws and scaling relations (Klafter and Shlesinger, 1986; Bouchaud and Georges, 1990). (See *§Quenched network structure and brain criticality*). Third, insofar as both local transport coefficients and driving fields tend to be non-trivial, it is natural to consider a given instance of a heterogenous system as a statistical realisation of a given ensemble, in which local quantities are extracted from some probability distribution. In some sense then, networkness can be understood as a state of matter parametrisation, characterised by order and symmetries intermediate between those of a solid crystal and those of a liquid. Altogether, these properties could allow classifying networked systems in a way similar to that of materials (Papo, 2013a). (See also §3.4.3 Network structure and response to external fields). Disorder may not just be a property of the ambient space within which relevant processes take place and may itself be a functionally significant process[15]. (See §3.3 Annealed network structure and dynamics). Fourth, defining an appropriate order parameter describing the collective behaviour of a networked system is a non-trivial endeavour that first requires identifying a broken symmetry. (See §The role of network symmetries in brain dynamics). In turn, an order parameter induces an associated field describing the properties and *topological defects*[16] emerging in a given phase, which can be described in topological terms (Mermin, 1979). (See §Quenched network structure and brain criticality). Outstanding questions are how heterogeneity affects collective dynamics and how it can be reconciled with the emergent universality observed across many diverse real-world phenomena.

---

[13] A lattice is an array of regularly spaced points arranged in a space of given dimension where the local structure repeats periodically in all directions.

[14] For a definition of *sparsity*, see *Table 1*.

Finally, insofar as networkness is associated with disorder relevance, its assessment boils down to a *Ginzburg-Landau criterion*[17] evaluating when heterogeneities cause properties to deviate from those of homogeneous systems, and fluctuations become larger than the mean-field ones (Bradde et al., 2010). The *maximum entropy principle*[18] can help identifying the simplest possible model of neural activity (Jaynes, 1957), e.g. in terms of particular topological or geometric properties.

## 3. Signs of brain networkness: dynamic relevance of network structure

It is natural to suppose that a biological system such as the brain has a structure that enables it to fulfil some function. The structure-function relationship is often identified with the anatomy-dynamics one, the former playing the role of structure and the latter of function. However, this identification is incorrect for at least two main reasons. On the one hand, dynamics should not be equated to function, the latter being a non-trivial aspect of the former (Papo, 2019). On the other hand, insofar as both anatomy and dynamics can be equipped with topological and geometrical structure, structure-dynamics and structure-function relationships should apply to both. It is then appropriate to first deal with structure-dynamics relationships, bearing in mind that structure need not be referred to that of the anatomy, nor be defined in the real anatomical space.

Dynamic patterns observed in coupled dynamical systems, e.g. multistability or metastability, arise under different local dynamics and coupling configurations (Heitmann and Breakspear, 2018). Both in theoretical and in experimental studies, spatial and temporal complexity are often treated as separate dimensions[19] (Delvenne et al., 2015; Schiff et al., 2007; Barzel and Barabási, 2013). Topological properties can then be thought of as generated by a given global activity field or, conversely, dynamical properties as emerging from the network structure. Within this framework, understanding network structure's relevance to brain dynamics requires addressing two dual aspects: the dynamical consequences of network structure and the structural consequences of brain dynamics. What brain physics, and in particular what dynamics can we extrapolate from the brain network structure that we know? For a network of given connection topology, what dynamics can be expected? Conversely, what can we say about brain network structure given the brain physics that we already know? These questions can be mapped onto the following dual problems: the *direct problem* aims at predicting the dynamical properties of a network from its topological parameters (Donetti et al., 2005), and the *inverse problem* involves reconstructing the network topological features from dynamics (Timme, 2007; Coombes et al. 2014; Burioni et al., 2014; Tirabassi et al., 2015; Cocco et al., 2017; Rings et al., 2022). For instance, in the inverse Ising problem, this involves reconstructing the coupling strength matrix $J$ from spin dynamics (Nguyen et al., 2017). Solving these problems would allow constructing networks giving rise to given topological properties or dynamical fields (up to some precision). However, on the one hand, collective dynamics turns out to be hard to infer even for networks with known node dynamics, interaction type and topology (Timme, 2007). On the other hand, most existing approaches use mean-field theories and no general method has as yet been proposed to address the inverse problem of uncovering internode couplings together with their strengths, signs, and directions (Song et al., 2014). An interesting method involves inferring network connectivity from the system's stable response dynamics (Timme, 2007). For *strongly connected networks*[20] of otherwise arbitrary topology and to some extent for sparsely connected ones, network connectivity can be recovered from the network's stable response dynamics to constant driving (Timme, 2007).

Interestingly the topology-dynamics relationship may be topology-dependent. In directed networks of oscillators, a topologically induced transition from ordered, synchronised to disordered dynamics has been reported (Timme, 2006). While in the former case all nodes display identical dynamics, independent of the node's location in the network, in the presence of disorder, node dynamics strongly depends on the node's *topological identity*[21], particularly the fine-scale topology of the strongly connected component it belongs to, as well as on the initial condition. Thus, in the presence of disorder, node dynamics encodes information about its topology and thus characteristic for it (Timme, 2006). Conversely, partial information about network topology may be inferred from the disordered dynamics of its nodes (Timme, 2006).

---

[15] The disorder is said to be *quenched* if the time scale of some observed dynamical property is much shorter than the turnover time of the disorder. This frozen heterogeneity constitutes a background random potential for the fluctuating degrees of freedom. Quenched disorder is distinct from *annealed* disorder, where the random degrees of freedom are ergodic.

[16] *Topological defects* are irregularities in the order parameter field that cannot be fixed through local rearrangement (Sethna, 2006). Defect structures can be thought of as an emerging property of spontaneous symmetry breaking. While for a macroscopic system energy is minimised when symmetry is uniformly broken throughout the system, in some circumstances symmetry may be broken differently in different parts of the sample. Under such circumstances, defects appear, for instance, at the boundary between spatial regions characterised by different values of an *order parameter* (Nishimori and Ortiz, 2010), through which it is possible to classify topological defects and their stability, while ordered phases can be described in terms of the interactions of its defects (Sethna, 2021).

[17] The mean field approach is valid whenever the mean amplitude of thermal fluctuations is much smaller than the value of the order parameter $\phi$ near the critical point, i.e. $\langle(\delta\phi)^2\rangle \ll \langle\phi\rangle^2$.

[18] According to the *maximum entropy principle* the best estimate of a probability distribution is the one which leaves the largest remaining uncertainty consistent with available constraints, i.e. the one with largest information entropy (Jaynes, 1957). In network science, the maximum entropy principle is applied to the probability distribution $P(G)$ of observing a given graph $G$ of finite size $N$ in an ensemble of random graphs (Squartini et al., 2018; Cimini et al., 2019). Each constraint leads to a particular network model. For instance, when constraining network properties' expected values, the resulting $P(G)$ is a Gibbs-like distribution of exponential random graphs. While the maximum entropy principle *per se* cannot explain why given properties (e.g. heterogeneities) arise so often or predict the spatial distribution of a given set of nodes (Radicchi et al., 2020), it allows building a hierarchy of models with the minimal structure needed to reproduce the expectation values of a given distribution (Yeh et al., 2010), and identifying differences between models in terms of the statistical structure that they capture (Savin and Tkačik, 2017).

[19] In its simplest form, this bipartition into topology and dynamics can be represented as $Dx = Mx$, where $M$ is a real matrix encoding mutual influences in the network, and $D = F^{-1}$ is a generic operator acting on the trajectory of each node $x_i$.

[20] A directed network is *strongly connected* if there is a directed path between any two nodes. A maximal strongly connected subnetwork is called *strongly connected component* (Timme, 2006).

[21] The *topological identity* of a node $i$ corresponds to the part of the network directly or indirectly connected to $i$. It is defined in terms of $i$'s nearest connected neighbours, and its higher-order ones, and the connection weights between these sets of nodes (Timme, 2006).

Theoretical studies have extensively examined the effect of network topology, e.g. of degree heterogeneity or community structure, on dynamics, showing how generic dynamic properties may be produced by appropriate network topologies (Boccaletti et al., 2006; Dorogovtsev et al., 2008). One important question is that of the degree of *dynamical emergence*: to what extent and under what conditions is the overall system dynamics explained by local properties of the nodes? Likewise, in the neuroscience context, the general strategy consists in testing topological or dynamical properties capable of reproducing key empirical findings at various spatial and temporal scales, particularly at long time and global spatial scales, imposing as few as possible conditions on nodal properties. One recently explored strategy involves control methods (Kalman, 1963; Liu et al., 2011, 2016). This is often predicated upon assumptions both on the system to be controlled and on the technique employed to assess system controllability, which render such assessment strategy problematic in a neuroscientific context (Dehghani, 2018).

### 3.1 Structure of what?

Before addressing the question of network structure's relevance to brain dynamics and function, it is worth understanding what brain aspects can possibly be equipped with network structure. In principle, network structure can refer to both brain anatomy and dynamics[22]. At the system level, the former would typically consist of grey matter vertices connected by white matter edges, while in the latter the structure would be the one induced by brain dynamics equipped with some metric quantifying coupling between activity at different locations of the relevant space (typically the anatomical one). Straightforward as this dichotomy may seem, it conceals subtle conceptual issues, which are treated in a slightly different way in experimental and theoretical network neuroscience.

In network models, neural activity results from the interplay between a physical network structure and dynamics. The relationship between anatomy and dynamics is represented in a slightly different way in experimental network neuroscience. From an experimental viewpoint, anatomy and dynamics can both be endowed with network structure, and these two structures can be studied separately. On the one hand, considerable experimental effort has been devoted to the characterisation of the anatomical network structure (Bullmore and Sporns, 2009; van den Heuvel and Sporns, 2011), its development (Fan et al., 2011) and its possible functional role in the healthy brain (van den Heuvel et al., 2012; Betzel et al., 2017) and in pathology (van den Heuvel et al., 2010; Iturria-Medina, 2013). Local neocortical networks of pyramidal cells form densely interconnected communities through recurrent collaterals (Braitenberg and Schütz, 2013). Anatomical studies have shown that these networks are potentially fully connected, in the sense that axons of presynaptic pyramidal cells pass within a micrometer of dendrites of all neighbouring pyramidal cells (Kalisman et al., 2005; Stepanyants et al., 2002). On the other hand, brain dynamics can be endowed with a network structure of its own (Bullmore and Sporns, 2009; Papo et al., 2014a). Experimental estimates using electrophysiological techniques suggested a dynamical connectivity density of ~10% (Markram et al., 1997a; Holmgren et al., 2003). Bidirectionally connected pairs and specific higher-order network motifs turn out to be over-represented with respect to what may be expected of a random structure (Song et al., 2005; Perin et al., 2011; Ko et al., 2011; Brunel, 2016).

There are at least three ways in which the relationship between anatomical and dynamical structure can be understood. The first reflects the idea that the anatomical network constitutes the "true" network, which is mirrored in a somehow blurred way by the "virtual" one induced by dynamic couplings (Suárez et al., 2020). What is generally studied then is the set of conditions under which these two structures become isomorphic (Deco et al., 2013; Ponce-Alvarez et al., 2015; Diez et al., 2015; Fernandez-Iriondo et al., 2021). Typically, both structures are considered at experimental time scales. At these scales, the anatomical structure can be thought of as static, while the corresponding dynamical network is effectively multiscale. At slow time scales, the correlation structure of spontaneous resting fluctuations has been shown to be related to the underlying anatomical circuitry (Honey et al., 2007, 2009, 2010; Vincent et al., 2007; Hagmann et al., 2008; van den Heuvel et al., 2009; Pernice et al., 2011; Trousdale et al., 2012; Goñi et al., 2014; Zamora-López et al., 2016; Alexander-Bloch et al., 2018; Baum et al., 2020; Suárez et al., 2020; Matkovič et al., 2023), the underlying idea being that resting activity arises from neuronal noise correlations between brain areas coupled by the underlying anatomical connectivity. However, reports are mixed for such relationship at fast time scales (Honey et al., 2007, 2010; De Domenico, 2017; Sorrentino et al., 2021).

A second way to understand the relationship between anatomical and dynamical network structure reflects the idea that anatomy is the ambient structure within which dynamics takes place. On the one hand, anatomy may be thought of as the embedding space, without necessarily considering its network structure. Embedding brain dynamics into the anatomical metric space would force to treat the associated structure as a *spatial network* (Barthélemy, 2022). On the other hand, anatomy may act as a boundary condition for the dynamics (Gökçe et al., 2017), although under what conditions this constraint plays a role remains to be established. (See §*Boundary conditions*). In alternative, the anatomical network can provide the discrete structure on which dynamics takes place. It is then straightforward to think of neuronal populations as ensembles of oscillators interacting according to the coupling scheme mandated by the anatomical network (Deco et al., 2011; Pernice et al., 2011; Deco and Jirsa, 2012; Cabral et al., 2011; Haimovici et al., 2013; Atasoy et al., 2016; 2018; Sreenivasan et al., 2017). When nodes are identified with single neurons, each node can incorporate biophysically detailed conductance-based models of voltage-gated ion channels (Ashwin et al., 2016). At mesoscopic and macroscopic scales, one in principle considers systems wherein each node represents a neuronal population comprising both excitatory and inhibitory subpopulations of neurons (Sreenivasan et al., 2017). Nodal dynamics can either be the one observed experimentally, or some analytical mesoscopic model, e.g. the Wilson-Cowan model, while the spatial propagation term can be represented by the graph Laplacian of densely sampled MRI- or DTI-based static structural connectomes, which allows extracting the harmonic modes of macroscopic human cortical activity[23] (Atasoy et al., 2016; 2018). The graph Laplacian is identified with the fine-grained *effective connectivity* matrix and can be thought of as the partial correlation matrix of observed activity fluctuations, and its eigenfunctions are the functional connectivity's eigenmodes. It is important to note that, from an experimental viewpoint, at mesoscopic and macroscopic spatial scales, characterising nodes becomes increasingly non-trivial (Korhonen et al., 2021). Moreover, identifying not just the system's coupling but also its topology with that of the anatomical structure is only one possible proxy for the underlying physiology, which is still not well understood at the computational and algorithmic levels, and sometimes at the implementation level as well. The key point is determining the phenomenology that models with given ingredients allow or forbid, but also whether these ingredients are not only sufficient but also necessary conditions for this to occur.

---

[22] A network structure could represent an intrinsic structural property of other aspects of the neural system, e.g. its thermodynamics.

[23] Many physical systems' spatio-temporal patterns, e.g. the standing wave patterns associated with the sound emitted by string instruments' vibrations, can be characterised in terms of the Laplacian eigenfunctions.

In the third way, both anatomy and dynamics constitute (possibly interacting) dynamical systems. This could look like a mere generalisation of the former way where anatomy ceases to be a static system. However, an important difference lies in the fact that in the latter case the topological structure induced by the dynamics may also be of interest. On the other hand, while the first way essentially boils down to finding some structural similarity between two structures, the latter two entail an interaction between the networks associated with anatomy and dynamics, both of which have their own topology and dynamics.

*3.1.1 Structure-dynamics: the role of time scales*

Whether considering brain anatomy or activity, the brain can be thought of as a dynamical system, at some temporal scale. At fast time scales, the network structure induced by connectivity is approximately constant. For instance, at macroscopic spatial scales, brain anatomy may be thought of as essentially static at the typical timescales of a neuroimaging experimental session, while when considering local synaptic connectivity, the corresponding timescale may be of the order of milliseconds. However, such a representation is no longer tenable at much longer timescales, *viz.* at developmental or experimental timescales of the order of seconds and beyond, at which connectivity can undergo some dynamics (Papo, 2013b). Correspondingly, when equipping the brain with a network structure, both nodes and network topology can be regarded as dynamical systems (Gross and Blasius, 2008). Note that, at appropriately long scales, brain anatomy can be endowed with its own dynamics and, conversely, brain dynamics may be endowed with network structure.

The extent to which the system's behaviour depends on the network structure on which it unfolds hinges on how nodal and network dynamics interact with each other and this, in turn, on the relationship between the time scales at which they evolve. In this framework, the connectivity matrix is itself time-dependent and obeys some (stochastic) evolution equation, e.g. $\dot{A}(t) = -\alpha A(t) + \eta(t)$. Furthermore, some specific topological property of the connectivity, i.e. some particular function of the connectivity matrix, may be associated with its own dynamics and associated set of characteristic scales. (See also *§3.4.2 Topological phase transitions*). Overall, then, one treats the system as a genuine topological dynamical system, whose role depends on the relationship between the associated time scales and those of dynamics at various scales.

If node dynamics is much slower than the network's, what is studied is the evolution of *non-dynamic networks* with static nodes. If, on the other hand, the time scale over which the network evolves is much larger than that of dynamical fluctuations, overall, what is studied is the dynamics of static networks, with time invariant coupling $\sigma$ and adjacency matrix $A_{ij}$ (or, equivalently, Laplacian matrix $\mathcal{L}_{ij}$), and network structure heterogeneity acts as *quenched* disorder (Ódor, 2008; Maslennikov and Nekorkin, 2017). In the quasistatic limit, heterogeneity can cause memory effects, whose relevance needs to be evaluated. This representation of the system's dynamics is similar to that of a particle in a solid or a supercooled liquid, where the motion is split into a local rattling motion, superimposed on a much slower structural relaxation, with spatially correlated structural rearrangements[24] (Kurchan, 2005). When these two-time scales are not too different, network structure is considered to act as *annealed* disorder (Dorogovtsev et al., 2008). Contrary to the two previous scenarios characterised by scale separation, where the slow variable dynamics is only affected by the averaged state of the fast variables and the dynamical interplay between the time scales is relatively weak, in the latter case not only should both dynamics be considered, but these can potentially interact, a hallmark of *adaptive systems* (Gross and Blasius, 2008; Do and Gross, 2012; Maslennikov and Nekorkin, 2017; Berner et al., 2023).

The complex network approach allows representing phenomena as processes in terms of properties embedded in, but different from both the structure they take place in and the dynamics of its constituent parts. Topological disorder may affect not only networked systems' dynamics but also the dynamical *processes* taking place on them. Insofar as the processes unfolding on the network structure have their own time scales, the terms *quenched* and *annealed* can also refer to the relative scales of the network structure and of the process unfolding on the associated network. The time scales of these processes may not coincide with those of local dynamics or of network dynamics and may result in a non-trivial way from their mixing (Perra et al., 2012; Lambiotte et al., 2019; Papo et al., 2017). In the neuroscience literature, the first question is to do with the space to which a network structure is associated. Such a space is often the anatomical one (Deco et al., 2011; Cabral et al., 2011). Overall, in addition to the network structure's quasi-static timescale, the system has two, typically separated non-degenerate scales, *viz.* the slow one of the processes, and the fast one of nodes, each possibly being multiscale. The structure induced by brain activity can also be considered. Thus, in such context, local dynamics, network structure, and dynamical process all have characteristic times of their own, which can all in principle interact. A third, far less explored possibility is that nodes may themselves be non-stationary. This scenario can be understood by considering that nodes emerge from coarse-graining (typically, but not exclusively, of the anatomical space) according to some criterion (anatomical, dynamical, functional, etc.) (Korhonen et al., 2021). (See *§3.2.2 Dynamical processes on quenched networks*).

This representation should be considered a simplification for at least two partially interrelated reasons: first, in addition to local dynamics, one could also consider the dynamics of its control parameters; more generally, structure at all scales, from micro, to meso and macroscopic, can potentially be endowed with its own dynamics, giving rise to a wide range of time scales. Second, local dynamics may *prima facie* be broadband and fractal, although local fractality may emerge from spatial scales, and spatial and temporal scales may not be separable (Bianco et al., 2008).

It is important to understand how network structure at various levels may be relevant to brain dynamics and its organisation. Anatomical networks are often modelled as acting upon a dynamic activity field, of which they constitute the spatial structure, the topology and geometry induced by brain activity merely shadowing such anatomical structure. Dynamic networks, on the other hand, are often considered in relation to brain *function*, which is typically identified with subspaces of the (spatio-temporal) structure of dynamic connectivity. The extent to which these two structures may differ becomes clear when looking at the way synchronisation is integrated when considering anatomical or dynamical network structure. In the former case, synchronisation corresponds to a process on the network, where the process itself has in general no network structure of its own. On the other hand, when associated with dynamics, synchronisation induces a network structure from which function emerges.

---

[24] The essential difference between a regular crystalline solid and a complex network resides in that whereas in the former local quantities take positions in a lattice, in the latter the topology is non-trivial.

At experimental time scales, the anatomical network structure can be thought of as quenched, and the issue is to understand whether and how such a structure acts on dynamics. At longer (e.g. developmental or evolutionary) time scales, the anatomical network can no longer be considered to be static, and may act as annealed disorder under some conditions. But what is the role of the structure induced by brain dynamics? Does it act as annealed disorder with respect to local dynamics or is the dynamic network an epiphenomenon of anatomical topology and its interplay with local dynamics? And if so, when is disorder relevant?

### 3.2 Quenched disorder and dynamics

The effects of quenched network structure on dynamics have been the object of a large body of theoretical literature (Boccaletti et al., 2006; Dorogovtsev et al., 2008; Zhou et al., 2006, 2007; Kaiser et al., 2007, 2010; Arenas et al., 2008; Rubinov et al., 2011; Porter and Gleeson, 2016; Schöll, 2016). The typical scenario is that of networks wherein nodes are dynamical systems of some kind, e.g. oscillators, connected through quasi-static physical links. For linear or linearisable dynamical systems, coupling induces a topology which is encoded in linear operators generalising differential operators, whose eigenvalues and eigenvectors determine the effects of topology on collective dynamics. But how is this framework applied when studying brain dynamics, and what information can it provide? How is the quenched network structure related to neural systems' dynamics?

In a neuroscientific context, a quenched network structure is in general incorporated by assuming that the anatomical network constitutes the discrete structure on which dynamics takes place. Thus, while one could in principle consider the time-averaged structure induced by brain activity, what is typically considered is the anatomical structure, which is essentially static at experimental scales. Note that quenched disorder need not coincide with anatomical structure; on the other hand, at sufficiently long scales, anatomical structure may not be quenched. Suffice it to think about learning and its neural correlates, in which the structure itself is by definition changing over time. Overall, one studies the (spontaneous or driven) collective dynamics of systems whose nodes are coupled dynamical systems with various possible specifications but as few as possible ingredients. The basic ingredients are quenched network topology and nodal dynamics. In a pure model, both are specified theoretically: one takes dynamical systems with minimal though realistic specifications, couples them according to some given topology, and derives the emerging global dynamics, or characterises what topological properties give rise to particular dynamical patterns. On the other hand, in a purely empirical model (Deco et al., 2011; Cabral et al., 2011, 2017; 2022; Atasoy et al., 2016, 2018; Roberts et al., 2019), coupling topology and dynamic properties are both empirically determined, the former as connectomes (Van Essen et al., 2012, 2013; Larson-Prior et al., 2013; Alivisatos et al., 2012, 2013; Okano et al., 2015; Glasser et al., 2016; Amunts et al., 2016; Miller et al., 2016; Poo et al., 2016), the latter under the form of statistical or dynamical scaling properties or recorded brain activity at various scales. The bridge between these two approaches involves incorporating experimental information, particularly the spatial one, into the models, and deriving the model parameters that best reproduce the observed scaling properties of brain dynamics, or in a somehow similar way "stylised facts", such as the integration-segregation spatio-temporal organisation (Tononi et al., 1994), due to their dynamic and functional implications (Tognoli and Kelso, 2014; Breakspear, 2017). (See also *§3.2.4 Methodological issues in quenched network modelling*). Thus, this general strategy addresses the following general question: what aspects of the anatomical network structure and dynamics at individual nodes are associated with given macroscopic dynamic properties? For instance, what combinations of anatomical and dynamic properties underlie broadest fluctuations in synchronisation patterns? How do different types of bifurcations in node dynamics affect system structure?

Perhaps a deeper way to investigate the structure-dynamics interaction involves understanding the relation between the system's energy landscape and the topology of its network structure. For example, for a system of spins interacting via an energy function, the energy landscape's scaling properties were shown to have some degree of universality with respect to the link weight distribution (Burda et al., 2006). A connection between topology and dynamics can also be described by considering the dynamics as a random walk in a directed weighted complex network (Baronchelli et al., 2009). Glassy phenomenology such as ageing and non-Poissonian bursty dynamics emerges naturally when node attachment is proportional to some local topological property, e.g. the clustering coefficient, suggesting that complex network structure can be generated without imposing multiple dynamical mechanisms (Bagrow and Brokmann, 2013). These results are consistent with previously highlighted associations between dynamics or kinetics and phase space topology (Shlesinger et al., 1993; Zelenyi and Milovanov, 2004; Zaslavsky, 2002). However, the extent to which the emergence of complex scaling stems from network structure *per se* and the possible mechanisms through which this is accomplished in the context of brain dynamics are poorly understood. Models and simulations suggest that some generic properties of neural resting activity result from the interplay between local dynamics and large-scale brain structure (Cabral et al., 2011; Deco et al., 2011, Breakspear, 2017; Gollo et al., 2017; Vila-Vidal et al., 2021). In these models, large-scale dynamical patterns of spontaneous activity is shaped by local circuits' properties such as of excitatory and inhibitory synaptic connection strength (Deco et al., 2013), but the role of non-trivial network structure has not been systematically studied.

### 3.2.1 Brain dynamics: from connectivity to complex network structure

The role of bare connectivity has been examined at various levels. One important example, represented by the role of connectivity in the relationship between excitation, inhibition, and dynamical regimes, has been studied theoretically at the level of spiking neurons. Cortical neurons receive strong external and recurrent excitatory projections that in isolation would tend to saturate neuronal activity levels. Strong recurrent inhibition balances excitation, preventing network-wide synchrony, stabilising cortical activity (Denève and Machens, 2016). The activities of excitatory and inhibitory populations tightly track each other, generating asynchronous, uncorrelated spiking activity with Poisson-like variability, which cancels the effect of shared input (Renart et al., 2010). The role of various aspects of connectivity in the emergence of such dynamics in balanced systems has been examined. The most obvious one is connectivity density. Early studies of balanced networks considered sparse connectivity and negligible shared input between neurons (van Vreeswijk and Sompolinsky, 1996; Amit and Brunel, 1997), but homogeneous balanced networks have an asynchronous regime even with dense connectivity (Renart et al., 2010). This result, which suggests a deep and complex relationship between balance and asynchronous activity, also indicates that connectivity density *per se* may not be a decisive factor. Another potentially important factor is represented by the interplay between feedforward and recurrent connectivity. For instance, one crucial aspect associated with correlated spiking persistence in large networks with excitatory-inhibitory balance is represented by the spatial scales of these two types of connectivity (Ocker et al., 2017a).

Various studies investigated how recurrent connectivity and indirect interactions induce correlations and affect population dynamics and under what conditions network properties such as the scaling of connectivity with distance or disorder at various scales influence correlations and population activity in dynamic networks with linear pulse coupling (Pernice et al., 2011; Trousdale et al., 2012). For *random networks* with distance-dependent connectivity, network topology only influences higher-order properties of the correlation distribution, but not its average level, due to the homogeneity of the connectivity. Quenched network topology exerts its effect on dynamical correlations by affecting indirect connections leading to a broad distribution of activation levels with low average but highly variable correlations. On the other hand, in more disordered networks, average correlations depend on details of the connectivity pattern, and average correlations are strong (Pernice et al., 2011).

Over and above bulk connectivity density and the interplay between feedforward and recurrent wiring, topological features also play a role. For instance, it has been shown that local topological features can predict system-level network activity (Ocker et al., 2017a). At the lowest scale, node degree distribution has been shown to shape the dynamics of networks of sparsely connected spiking neurons, controlling the onset of oscillations (Roxin, 2011), while the synaptic weight distribution modulates neuronal population dynamics (Iyer et al., 2013). At mesoscopic scales, neural modes and their dimensionality are shaped by network properties such as degree distribution (Smith et al., 2018), clustering and motifs (Pernice et al., 2011; Litwin-Kumar and Doiron, 2012; Hu et al., 2013, 2014; Jovanović and Rotter, 2016; Ocker et al. 2017a; Recanatesi et al., 2019; Hu and Sompolinsky, 2022; Dahmen et al., 2020), and community structure (Aljadeff et al., 2015, 2016). For instance, the presence of even modest clustering motifs may substantially modify balanced state networks' activity, leading to the emergence of slow dynamics, during which clusters of neurons transiently modify their firing rate where the magnitude of spike train correlations depends on the network's motif structure (Litwin-Kumar and Doiron, 2012). Dynamic motifs tend to impose their intrinsic dynamics to the whole network, irrespective of its global topology, suggesting that motif statistics may be a crucial ingredient in the dynamics of networks of excitable units (Carvunis et al., 2006). Likewise, assuming an Erdős–Rényi directed neural network structure, degree fluctuations induce a dynamical regime characterised by intermittent quasi-synchronous events, where a large fraction of neurons fire within time intervals of few milliseconds, separated by intervals of uncorrelated firing activity one or two orders of magnitude longer (Tsodyks et al., 2000; di Volo et al., 2013). At macroscopic scales, though, various questions are still poorly understood. For instance, the extent to which the spontaneous generation of slow rate fluctuations requires a non-random structure is still unclear (Ostojic, 2014; Lajoie et al., 2014). The anatomical connectome's complexity has been suggested to constrain resting brain activity's repertoire (Weninger et al., 2022), which was proposed to reflect the largest complexity allowed by its anatomical connectome (Zamora-López et al., 2016).

Various important aspects of network activity result from the interaction between network structure and single-neuron nonlinearities. While single neurons' linear response only depends on mean-field rates and therefore not directly on the system's topology, the first nonlinear correction induces dependency on the connectivity (Ocker et al., 2017b). In addition to single-neuron transfer gain, the order of nonlinearity also determines how activity propagates through the network (Ocker et al., 2017b). Thus, to understand when connectivity starts being relevant boils down to determining the scale at which local nonlinearities start playing a role. At mesoscopic scales, the interaction between neural dynamics and structure in networks of spiking units can be assessed by studying stability and correlations of the dynamics linearised around mean firing rates where neurons fire asynchronously computed (Ginzburg and Sompolinsky, 1994; Brunel, 2000; Helias et al., 2014). In this context, mean-field stability of stationary states crucially depends on the connectivity structure, and can only be achieved in the presence of degree correlations, so that ultimately the stability of asynchronous balanced states is controlled by the motifs induced by degree correlations (Landau et al., 2016). However, this approach fails to capture individual neurons' response nonlinearity. In alternative, the impact of network connectivity on neural dynamics can be explored using the *path integral approach*[25] (Graham, 1977; Buice and Cowan, 2009; Chow and Buice, 2015; Ocker et al. 2017b; Crisanti and Sompolinsky, 2018) or its discrete equivalent, the *network propagator*[26] (Villegas et al., 2022, 2023). In this approach, each network results from an evolutionary path induced by a stochastic process from which the network properties emerge which constitutes one of the possible trajectories of the path integral characterising the system's state space (Bianconi et al., 2015; Bianconi and Rahmede, 2015).

The *inverse problem* of finding the network topology based on state variables' time series has also been addressed (Burioni et al., 2014; Pernice et al., 2013; Tlaie et al., 2019; van Meegen et al., 2021). Under sufficiently realistic assumptions, it is in principle possible to reconstruct the in-degree distribution of the underlying network from the average activity field (Burioni et al., 2014). The inverse problem can be addressed by dynamically perturbing the system and measuring the response to appropriate driving signals, under the assumption that the asymptotic response dynamics of a network to different exogenous perturbations is modulated by its topology. This strategy may help defining classes of network topologies consistent with perturbation–response pairs for sets of dynamical units of given functional form and coupling functions (Shandilya and Timme, 2011). Furthermore, for weak coupling, the network's degree distribution can be inferred from local dynamics, node degree turning out to be inversely proportional to node dynamics' complexity (Tlaie et al., 2019). Overall, though, how dynamics affects network topology and at what scales the former reflects properties of the latter are still poorly understood issues.

Finally, models of the effect of quenched anatomical disorder have been evaluated in terms of structural and dynamical robustness. Notably, while rich-club and hierarchical modal network organisation are associated with qualitatively different topological structure, both may confer a form of robustness, albeit of a different kind. The former, which does not preserve modularity, is vulnerable to targeted attacks, whereas the latter protects from node overload and overload spreading and alleviates the consequences of targeted attacks of network hubs (Song et al., 2006). Moreover, for networks with such a structure, robustness is maximised with fractal topology (Song et al., 2006). Thus, characterising the dynamical robustness of brain activity at long time scales seems to require a correct characterisation

---

[25] In Feynman's path integral formulation of quantum mechanics, the evolution of the universe can be thought of as the ensemble of all possible histories it can follow classically. In statistical physics, path integral methods can then be used to derive generating functionals for the moments and response functions of the stochastic differential equations used to model stochastic phenomena (Doi, 1976a,b; Peliti, 1985). In the former case, the expectation values of observables are dominated by a small subset of possible histories whose contributions are reinforced by constructive interference. Likewise, in the latter case a state space is dominated by maximum entropy contributions leading to thermodynamic behaviour.

[26] If the network undergoes diffusive dynamics, in analogy with the heat equation, its generic state at time $\tau$ is given by: $x(\tau) = x(0)e^{-\tau \mathcal{L}}$, where $x(0)$ is the network's initial state, and the exponential $\widehat{K} = e^{-t\mathcal{L}}$ is the network propagator (Masuda et al., 2017) which depends on the Laplacian matrix $\mathcal{L}$, and accounts for the sum of all possible paths connecting any two nodes within a time scale $\tau$ (Moretti and Zaiser, 2019). Diffusive dynamics samples paths in a statistical way and can therefore be thought of as the discrete counterpart of the path integral approach.

of the topological properties of the underlying quenched (anatomical) network.

*Network properties' self-averaging*

From the energy function [6] one can derive a probability distribution function as noise (i.e. inverse temperature) is varied, and in particular evaluate high-probability states, corresponding to low energy levels (Advani et al., 2013). In the simplest case with no external fields, the Hamiltonian in [6] reduces to $\mathcal{H} = -1/2 \sum_{i<j} J_{ij} \sigma_i \sigma_j$, and the probability density function is the Gibbs distribution $P_J(\sigma) = 1/Z(J) e^{-\beta \mathcal{H}(\sigma,J)}$, where $\beta$ is the *inverse temperature*[27], and $Z(J) = \sum_\sigma e^{-\beta \mathcal{H}(\sigma,J)}$ is the *partition function*. From the partition function one can derive aggregate thermodynamic variables of the system such as the free energy ($-\beta \mathcal{F}(J) = \ln Z(J)$), entropy, or specific heat. (See also §3.4.2 *Topological phase transitions*). At low temperatures, the probability is concentrated around discrete parts of the distribution, corresponding to minima of the free energy function, and associated with particular subsets of activity patterns.

An important question is the extent to which the system's statistical properties are *self-averaging*[28], i.e. are independent of the detailed realisation of the disorder. Renormalisation group and numerical studies have shown that self-averaging is lost if disorder is relevant (Aharony and Harris, 1996). The statistical properties of systems of many interacting degrees of freedom with quenched highly heterogeneous interactions can be studied with statistical physics methods viz. *cavity* and *replica* methods (Advani et al., 2013). In large random systems of binary neurons with quenched microscopic heterogeneity, quasi deterministic macroscopic order can arise in ways that do not depend on the details of the heterogeneity, and which are stable with respect to *thermal fluctuations*. However, the picture is much more complex and less studied for annealed disorder, and for non-trivial topology. Moreover, early models of neural structure contained various simplifying assumptions, e.g. binary neurons, symmetric connectivity, and considered an unperturbed system. Relaxing these assumptions can lead to qualitatively different results. For instance, the break down of connectivity symmetry has been shown to lead to complex dynamic phenomena, including ergodicity breaking and non-reciprocal phase transitions (Crisanti and Sompolinsky, 1998; Fruchart et al., 2020; Bowick et al., 2022). Over and above the universality of existing results with respect to network structure and disorder, the issues addressed by and variables found with such an approach acquire particular significance and are more straightforwardly interpreted in a functional rather than a purely dynamical context, where they may play a key role in the definition of classes of functional neural activity. Both aspects (universality and functional implications) will be discussed elsewhere.

### 3.2.2 Dynamical processes on quenched networks

The complex network approach allows representing processes on a given network structure, i.e. phenomena whose properties are embedded in but different from both the structure they take place in and the dynamics of constituent parts. An important issue relates to the type of process that can meaningfully be defined when describing brain dynamics. (See *§3.1.1 Structure-dynamics: the role of time scales*). Dynamical aspects of neural activity are relatively well understood at the single neuron level (Izhikevich, 2007; Ermentrout and Terman, 2010), but far less at the system level. Many processes are either essentially functional or require some hypothesis on function at microscopic scales and will be dealt with in a companion paper. Here, we address at least *prima facie* purely dynamical partially interrelated processes, viz. activity propagation, synchronisation, and criticality.

*The role of quenched network structure in activity spreading*

It seems sensible to assume that the way activity spreads within a networked system such as a neural population somehow depends on the network structure along which the activity takes place. At the most basic level, one may consider how dynamics spreads within the system, either following some external field or due to its endogenous dynamics. The main question involves determining how network structure affects dynamics propagation. Both theoretical and experimental studies of various networked systems have shown that propagation patterns in a networked system may somehow depend upon network structure (Boccaletti et al., 2006). While it is possible to think of some generic dynamics within a network, with no specific commitment to information spreading, we discuss elsewhere the various facets of spreading in the functional context of genuine *information transport*. Here, instead, we specifically address the topology-dynamics interplay associated with perturbation propagation (See also *§3.4.3 Network structure and response to external fields*).

A natural way in which propagation can be addressed is by perturbing the system locally and quantifying the spatio-temporal aspects of the response cascade that it causes (Barzel and Barabási, 2013; Mišić et al., 2015; Gollo et al., 2017; Hens et al., 2019). For instance, perturbing chosen connectome nodes results in global cascades, that early spreading is facilitated by hubs and main pathways, and that the connectome's shortest path structure enhances propagation, competing cascades ultimately converging in multimodal association areas (Mišić et al., 2015). In the linear response regime, associated with perturbations of the system's steady state regime, it is possible to quantify not only the response locally at each node, but also the impact on local dynamic stability, the spread of perturbation and the cascading effect of such a perturbation (Barzel and Barabási, 2013). (See also *§3.2.4 Methodological issues in quenched network modelling*). However, in the presence of nonlinearities, network topology does not uniquely determine propagation, which instead sensitively depends on the interplay between topology and the system's intrinsic dynamics, slight changes potentially leading to qualitative dynamical changes (Hens et al., 2019). For instance, the average inter-regional cross-correlation delay is not uniquely determined by neuronal conduction delays but also depends on local circuit parameters which ultimately determine the phase response properties of the coupled populations (Tiesinga and Sejnowski, 2010). Thus, it is important to understand how topology and dynamics may interact and in particular whether a general relationship between these two aspects can be found under sufficiently general conditions. In the limit of large weight degree distribution and for small perturbations of steady-state dynamics (i.e. in the linear response regime) node $i$'s response time $\tau_i$ scales with its weighted degree $S_i = \sum_{j=1}^N A_{ij}$ as:

$$\tau_i \sim S_i^\theta \qquad [7]$$

Under this approximation, $\theta$ determines how topological features translate into characteristic response times. Importantly, $\theta$ which is

---

[27] Large values of $\beta$ correspond to a low noise regime, conversely at high values, node connectivity is dominated by noise.

[28] A system described by a macroscopic variable $X$ is said to be *self-averaging* if the *dispersion coefficient* $R_X(L) \equiv \langle \Delta X \rangle^2 / \langle X \rangle^2 \to 0$ as $L \to \infty$, where $\langle X \rangle$ and $\langle \Delta X \rangle^2 = \langle X^2 \rangle - \langle X \rangle^2$ are respectively $X$'s mean and variance (Lifshitz et al., 1988; Aharony and Harris, 1996).

*prima facie* a local relation induces global dynamic classes of propagation patterns (Hens et al., 2019). In particular, for the perturbed steady state dynamics of a rate-based neural network model of neurons clusters with random interunit connections and strong local connectivity[29] (Stern et al., 2014), $\theta = 0$. Thus, $\tau_i$ is independent of $S_i$ (Hens et al., 2019).

While this result is of great generality, providing insight into the structure and characteristic temporal and spatial scales of perturbation propagation, it nevertheless hinges on some important assumptions limiting its scope. First, its starting point is a node's steady-state dynamics $x_i$, in the linear regime. Tracking the system's steady-state linear response provides a potentially limited picture of the system's dynamic range. Second, the scaling relation holds for random weighted networks albeit with arbitrary degree and weight distributions. Perhaps more importantly, although the model shows some robustness with respect to some structural properties such as clustering, other properties, e.g. degree-degree correlations, are poorly accounted for, as in random networks their contribution becomes negligible in the limit of sparse networks ($\langle k \rangle \ll N \to \infty$). (See *Table 1*). Finally, while brain signal propagation pathways are usually identified with the neural fibres, cortical wave modes' propagation in highly folded areas does not follow fibre directions (Galinsky and Franck, 2020a,b, 2021). The presence of inhomogeneous anisotropy can generate low amplitude but relatively long-lived (i.e. lasting longer than the spiking potential inverse decay constant) weakly damped wave-like modes propagating along a direction transverse to both the mean neural fibre direction and the cortical spatial gradient (Galinsky and Franck, 2020a,b, 2021).

*Synchronisation*

Coordination of neural activity over distant brain regions is essential to normal brain function (Fries, 2005, 2015). Thus, the general framework of coupled oscillators seems *prima facie* relevant to the study of brain dynamics and function, and the key question is to understand how such a system may achieve coordination and what role is played by network properties in such a process.

Synchronisation is perhaps the most studied process on networks of dynamical systems. Topological and spectral properties have been shown to play a key role in the synchronisation of systems of coupled oscillators (Boccaletti et al., 2006; Arenas et al., 2008). For instance, it has been shown theoretically that synchronisation can only be achieved if the network obeys specific topological conditions on subgraphs at all scales, highlighting the impact of mesoscale topological structure on collective dynamics (Do and Gross, 2012). Furthermore, non-trivial spectral properties such as the spectral radius are crucial for the stability of synchronisation processes (Barahona and Pecora, 2002; Donetti et al., 2005; Bunimovich and Webb, 2012). Similarly, synchronisation times vary for networks with different topologies and local dynamics (Almendral and Díaz-Guilera, 2007), and the *spectral gap* is used to determine dynamic properties on networks, including synchronisation thresholds and the rate of convergence to synchronisation (Almendral and Díaz-Guilera, 2007). More generally, one analytically tractable method to study how synchronisation may result from interactions between topology and dynamics of coupled dynamical systems is represented by the *master stability function* formalism[30] (Pecora and Carroll, 1998). However, even though this formalism has been generalised, relaxing various hypotheses (Sun et al., 2009; Huang et al., 2009; Zhang and Motter, 2018; Mulas et al., 2020; Gambuzza et al., 2021), it deals with the stability of the fully synchronised state in systems of diffusively coupled identical oscillators. On the one hand the synchronisation referred to in complex network theory is typically a steady-state process on a network of coupled oscillators. For instance, the stability of the fully synchronised state depends on the spectral gap of the graph Laplacian (Donetti et al., 2005), and is more generally optimal under specific topological conditions (Skardal et al., 2014b). On the other hand, in neuroscience, synchronisation typically refers to transient (usually bivariate) local coupling mechanisms between neuronal ensembles. Furthermore, the brain is characterised by anatomical and dynamical heterogeneity (Balasubramanian, 2015). Thus, neither the condition nor the system constitute ideal ingredients for a realistic description of global brain activity (Papo and Buldú, 2019).

An alternative method involves simulating systems of coupled oscillators and quantifying their collective behaviour in terms of a neurophysiologically plausible function. Phase synchronisation has been proposed as a plausible candidate mechanism behind the formation of dynamic links among brain regions and large-scale functional integration (Varela et al., 2001; Fries, 2005, 2015). It is therefore straightforward to consider the effects of network topology on the synchronisation properties of systems of coupled Kuramoto oscillators (Kuramoto, 1975; Acebrón et al., 2005; Rodrigues et al., 2016). This model describes the dynamics of *N* phase oscillators $\varphi_i$ with time-invariant natural frequency $\omega_i$ and sinusoidal coupling:

$$\dot{\varphi}_i = \omega_i + \frac{\kappa}{N} \sum_{i=1}^{N} A_{ij} \sin(\varphi_i(t) - \varphi_j(t)) \qquad [8]$$

where $\omega_i$ is extracted from some probability distribution function $g(\omega_i)$, e.g. a Gaussian distribution[31]. Various different forms for the coupling matrix in this particular model have been considered such as nearest-neighbour, hierarchical, random long-range, or state-dependent interactions (Acebrón et al., 2005). In the absence of coupling, each oscillator rotates at its natural frequency $\omega_i$, whereas the coupling term $\kappa$ tends to force it synchronise to all other oscillators. When the coupling is weak, the oscillators run incoherently, whereas beyond a certain coupling threshold collective synchronisation emerges spontaneously.

The collective behaviour of coupled systems is described in terms of an *order parameter*[32] (Sethna, 2021), as a function of some control parameter, e.g. the coupling term $\kappa$, at some time scale (typically in the long time limit), and for a given (finite or infinite) network size. For mean-field coupling among Kuramoto phase oscillators, the order parameter is represented by the amount of coherence in the system:

$$r(t)e^{i\psi(t)} = \langle e^{i\varphi_i(t)} \rangle \qquad [9]$$

---

[29] The network comprises *N* randomly coupled units, each evolving as $\dot{x}_i = -x_i + s \tanh(x_i) + x_i g \sum_{j \neq i}^{N} J_{ij} \tanh(x_j)$, where $s$ and $g$ respectively represent within-cluster and interunit coupling strengths, and the elements of the coupling matrix $J_{ij}$ are drawn from a Gaussian distribution $\mathcal{N}(0, 1/N)$ (Stern et al., 2014). Note that for vanishing self-coupling, the model reduces to the one proposed in (Sompolinsky et al., 1988).

[30] The invariant synchronisation manifold is defined by the $N - 1$ constraints $x_1(t) = x_2(t) = \cdots = x(t) = s(t)$, where $s(t)$ is a solution in $\mathbb{R}^d$ of the uncoupled system $ds/dt = F(s)$ of the type of Equation [5] (Boccaletti et al., 2006). The *Master Stability Function* is a parametric function of the Lyapunov exponent of the variational problem associated with the system, which tells how the dynamical system (through **F**) and the network topology (through the second term on the right side of equation [5]) concur in determining the *stability* of the synchronisation manifold.

[31] An additive noise term $\xi(x, t)$ can also be added (Acebrón et al., 2005).

[32] In Landau's theory of phase transitions, an *order parameter* is a quantity that characterises the state of a physical system during a phase transition. During a phase transition, the value of this quantity goes from zero in the disordered phase to a non-zero value in the ordered phase. The disordered phase is invariant under a transformation group $G$, while the ordered phase is only invariant under the action of a subgroup $G'$ of $G$. The order parameter is only an invariant quantity under the action of the subgroup $G'$ and it therefore vanishes in the disordered phase.

where $\psi(t)$ is the average phase of all oscillators and $r(t) \in [0,1]$ the overall phase coherence. A single measure for the phase ordering is given by the long-time average of the absolute value of the order parameter:

$$r^2 = \langle |r(t)e^{i\psi(t)}|^2 \rangle_t \quad [10]$$

In a similar way, it is possible to define a *local order parameter* averaging over the phase differences between directly connected nodes:

$$r_{loc} = \frac{1}{\sum_{i=1}^N k_i} \sum_{i,j=1}^N A_{ij} \left| \langle e^{i(\varphi_i - \varphi_j)} \rangle_t \right| \quad [11]$$

In fully connected and random Poissonian networks, Kuramoto model's dynamics undergoes a continuous phase transition (Hopf bifurcation) from an incoherent/asynchronous ($r \approx 0$) to a coherent/synchronous, oscillatory state ($r \approx 1$) at a critical coupling value $\kappa_c$. For intermediate couplings, part of the oscillators are phase-locked, the remaining ones rotating out of synchrony with the locked oscillators, and $0 < r < 1$. With frequency dispersion, a critical value separates the synchronous phase from the asynchronous one (Acebrón et al., 2005). The mean-field Kuramoto model for infinitely many oscillators can have different stable solutions or *phases* depending on $g(\omega_i)$, the values of the coupling strength $\kappa$, and input and noise properties (Acebrón et al., 2005). In real systems, however, when and how systems of oscillators synchronise also crucially depends on network topology. But what is the role of non-trivial network structure? Networks with isolated and independently synchronised moduli trivially exhibit an intermediate regime characterised by $r(t)$ oscillations. This regime, which can become chaotic when a small number of coherent moduli become coupled together (Popovych et al., 2005), can emerge in complex networks of zero-lag coupled modular structure with weakly interconnected moduli, and in general in networks with hierarchical modular structure (Moretti and Muñoz, 2013; Villegas et al., 2014). The interplay between structural and quenched intrinsic frequency heterogeneities at various scales gives rise to *frustrated synchronisation*, a regime characterised by well-separated synchronisation domains of different frequencies, which the dynamics can only resolve at well-separated coupling values (Villegas et al., 2014; Millán et al., 2018a). This regime is dominated by large complex spatio-temporal fluctuations (Moretti and Muñoz, 2013; Millán et al., 2018a), metastability, and chimera[33]-like states (Shanahan, 2010; Villegas et al., 2014; Hizanidis et al., 2016; Chouzouris et al., 2018). In the absence of node frequency dispersion, in terms of which heterogeneity is typically accounted for in this particular context, modular structures constitute topological scales emerging at different hierarchically ordered time scales associated with gaps in the Laplacian spectrum (Arenas et al., 2006). The presence of structural bottlenecks alters this hierarchically nested synchronisation, inducing anomalously slow dynamics at very large timescales (Villegas et al., 2014). Note that in this context frequency heterogeneity is in general assumed, not an emerging property. Interestingly, such a regime, wherein the order parameter oscillates in a frequency-dependent fashion with the coupling value, emerges naturally when considering Kuramoto oscillators on the human connectome, indicating that hierarchical modular networks may constitute a parsimonious model reproducing the complex synchronisation phenomenology induced by the human connectome (Moretti and Muñoz, 2013; Villegas et al., 2014).

Another important aspect in the structure-dynamics relation is the route through which the system achieves synchronisation. To address this issue, the standard Kuramoto model (equation [8]) can be augmented with an additional term to account for individual units' excitability:

$$\dot{\varphi}_i = \omega_i + \frac{\kappa}{N}\sum_{j=1}^N A_{ij} \sin(\varphi_i(t) - \varphi_j(t)) + b\sin(\sigma t - \varphi_i(t)) \quad [8.2]$$

where $b$ is the forcing strength and $\sigma$ is the forcing frequency (Lindner et al., 2004; Childs and Strogatz, 2008; Montbrió and Pazó, 2020; Buendía et al., 2021). The collective dynamics and the transition diagram between regimes of systems of coupled oscillators of this type results from the relative magnitude of three competing effects: frequency variance, which favours phase scattering and desynchronisation; coupling, which tends to align the oscillators to the same phase, without favouring any frequency for the collective oscillation, and forcing which is frequency-specific (Childs and Strogatz, 2008). Such a system's phase diagram presents two main qualitatively different types of synchronisation transitions, corresponding to different subregions of the $b$-$\kappa$ plane (Lindner et al., 2004). In between these two subregions lies a small *hybrid region* of bistability where low and high activity levels coexist (Childs and Strogatz, 2008). This region is characterised by complex and wide range spatio-temporal fluctuations which are not seen in the rest of the space (Buendía et al., 2021).

What topological and dynamical mechanisms are responsible for synchronisation level variability across time and network scales? *Prima facie* conflicting results on the role played by network properties such as randomness, degree distribution or betweenness centrality have been reported (Arenas et al., 2008). These inconsistencies may stem from the sensitive dependence of the synchronisation properties on the specific combination of nodes and links, with a nonmonotonic, periodic structure of the synchronisation landscape as a function of both nodes and links (Nishikawa and Motter, 2010).

Topology can have counterintuitive effects on synchronisation. For instance, while removing an edge from a globally connected network decreases its ability to synchronise, targeted removal of additional edges may have the opposite effect (Nishikawa and Motter, 2006). Coupling strength or connectivity distribution heterogeneity tend to prevent synchronisation (Nishikawa et al., 2003), but may compensate each other when simultaneously present (Motter et al., 2005). Similarly, while negative interactions tend not to support synchrony, but stabilise synchronous states (Nishikawa and Motter, 2016). For instance, in neuronal networks, inhibitory interactions may facilitate synchronous bursting (van Vreeswijk et al., 1994; Wang and Buzsáki, 1996; Belykh and Shilnikov, 2008). Moreover, cortical synapse density first increases and then decreases during early brain development (Huttenlocher, 1979; Bourgeois and Rakic, 1993).

While various network properties at all scales, e.g. connectivity weight distributions, link directionality, may play an important role, hierarchical-modular networks which also exhibit a *core-periphery* structure (Bassett et al., 2013) have been shown to be associated with a particularly broad dynamical repertoire compared to other network topologies, including a random hierarchically modular one (Buendía et al., 2022). In networks with *core-periphery* architecture, the presence of central connector hubs induces a scale-free degree distribution, in contrast with hierarchical modular networks, which lack genuine hubs and whose degree distribution is in general exponential (Zamora-López et al., 2016).

---

[33] *Chimera states* are spatiotemporal patterns formed by two or more domains of qualitatively different dynamics, some in which the oscillators are synchronised and others in which they evolve incoherently (Kuramoto and Battogtokh, 2002; Abrams and Strogatz, 2004). Contrary to situations in which the symmetric state loses stability, in this scenario, the fully synchronised symmetric and the chimera asymmetric states are simultaneously stable. Chimera states emerge naturally from the Kuramoto model's partially synchronised state via a symmetry-breaking bifurcation (Kotwal et al., 2017). While chimera states have initially been studied in networks of identical oscillators, such states can also arise in heterogeneous networks for which the natural frequency of oscillators is chosen from a distribution (Laing, 2009). (See also *§The role of network symmetries in brain dynamics*).

An important issue is how synchronisation is affected by the presence of noise. At a critical value of node coupling strength, noise enhances both local and global dynamical coupling, a regime termed *stochastic synchronisation* (Pang et al., 2021). On the one hand, moderate noise levels may promote integration within functional modules by forcing node phase alignment, decreasing the number of phase clusters while increasing their size; on the other hand, it may modify functional connectivity between brain subnetworks (Pang et al., 2021). A systematic evaluation of the contributions of network topology and node dynamics indicates that stochastic synchronisation is driven by a complex interaction between brain anatomy's topological heterogeneity (Hagmann et al., 2008) and the hierarchy of natural frequencies associated with region-specific abilities of cortical circuits to store information over time (Chen et al., 2015), via the promotion of *dynamically frustrated* states, which may disrupt synchronous dynamics, allowing coexisting zero-lag synchrony metastable configurations (Gollo and Breakspear, 2014a,b). While neurological and neuropsychiatric disorders have been associated with alterations of both anatomical network heterogeneity (Bassett et al., 2008; van den Heuvel et al., 2010; Gollo et al., 2017; Griffa and van den Heuvel 2022; Stam, 2014) and neural timescale hierarchy (Watanabe et al., 2019), whether and how the optimal working point changes across brain states, clinical conditions and time are still poorly understood issues.

The significance of these results should be gauged by considering some important caveats. Macroscopic brain dynamics is characterised in terms of a convenient order parameter for the resulting network model. This way, phenomenology at mesoscopic and macroscopic scales emerges from dynamical processes at microscopic scales not directly accessed by models or experiments. However, the initial step requires defining a dynamical connectivity metric. This typically involves assumptions at both computational and algorithmic levels, to which a local rule is associated, based on which the order parameter is specified. While various hypotheses on the way neural populations communicate at various scales (Varela et al., 2001; Fries, 2005, 2015; Luczak et al., 2015) and corresponding quantitative specifications (Pereda et al., 2005; Huber et al., 2021) have been proposed, the true mechanisms of brain information transfer and processing are in general unknown and may even be topographically- and context-specific (Malagarriga et al., 2017).

A somehow related issue is represented by limits to the generality and validity of the Kuramoto model in the description of brain dynamics should be acknowledged. On the one hand, local brain dynamics is in general not well described by a single natural frequency, particularly at mesoscopic and macroscopic scales. Note in addition that while natural frequencies' spatial distribution is generally assumed to be random, it may in fact bear some relation with network location and node connectivity (Pang et al., 2021). Moreover, phase-amplitude coupling gives rise to a rich phenomenology that more readily accounts for very high frequency (100-600 Hz) long-range coupling (in spatial ranges of the whole system order) in both healthy (Buzsáki et al., 1992; Jones et al., 2000; Engel and da Silva, 2012; Buzsáki and da Silva, 2012; Buzsáki, 2015; Arnulfo et al., 2020) and pathological (Bragin et al., 1999, 2002; Köhling and Staley, 2011; Jefferys et al., 2012; Frauscher et al., 2017) brain dynamics, than phase or amplitude coupling considered separately (Galinsky and Frank, 2020a,b, 2021).

*Quenched network structure and brain criticality*

Empirical evidence shows that at long time scales, brain activity is characterised by generic non-trivial scaling with non-Gaussian statistics of both local brain activity, synchronisation events between distant neuronal populations or neural avalanches of all sizes (Novikov et al., 1997; Linkenkaer-Hansen et al., 2001; Beggs and Plenz, 2003; Bianco et al., 2008; Freyer et al., 2009, 2011; Expert et al., 2010; He et al., 2010; Wink et al., 2012; Freeman et al., 2003; Stam and de Bruin, 2004; Ciuciu et al., 2012), i.e. spatial patterns of propagated synchrony (Plenz and Thiagarajan, 2007), intermittent switching between periods of low and high activity, resulting in fat-tailed intercommunication time distributions (Gong et al., 2003, 2007; Allegrini et al., 2010).

The presence of generic non-trivial scaling properties, which are typical of systems operating far from equilibrium (Täuber, 2017), has been proposed to mean that the brain operates in a state that is situated at or very near to a nonequilibrium phase transition between qualitatively different dynamical regimes or phases (Chialvo, 2010; Hesse and Gross, 2014; Zimmern, 2020). Typically, the system operates transitions between quiescent and active states (Beggs and Plenz, 2003; Kinouchi and Copelli, 2006), synchronuous and asynchronuous states (di Santo et al., 2018; Fontenele et al., 2019) or laminar and chaotic dynamical regimes (Langton, 1990; Wainrib and Toboul, 2013; Kanders et al., 2017a; Dahmen et al., 2019).

The specific role of the system's topology and geometry on its scaling behaviour, e.g. the values of the critical exponents or universal scaling functions (Sethna et al., 2001), has been the focus of numerous studies. At macroscopic scales, although uncorrelated spatial and temporal disorder are irrelevant for the universal critical behaviour (Ódor, 2004), correlated, diffusive spatiotemporal disorder has been found to be a relevant perturbation (Vojta and Dickman, 2016). Insofar as temporal complexity can be thought of as an emerging property of spatially non-local interactions (Lindner et al., 2005; Bianco et al., 2008) it may result from, or at least depend upon, network structure. Quenched disorder has profound effects on critical points around which systems undergo phase transitions (Vojta and Sknepnek, 2004; Vojta, 2006). Close to a non-equilibrium phase transition, rare domains with large disorder fluctuations whose internal state changes exponentially slowly as a function of their size (Ódor, 2014a) can give rise to strong power-law singularities in the free energy, smearing and even wiping away phase transitions (Harris, 1974; Vojta and Sknepnek, 2004; Vojta, 2006; Villa Martin et al., 2014, 2015). At a smeared transition, the system is fractionated into spatial regions, which undergo the transition at different control parameter values. Once a true static order has developed on some of the rare regions their order parameters can be aligned by an infinitesimally small interaction or external field. Therefore, global order develops very inhomogeneously over a range of control parameter values (Vojta and Sknepnek, 2004). These singularities, which occur in a large control parameter region, are known as Griffiths phases (Muñoz et al., 2010), a form of dynamical criticality, characterised by slow power-law dynamics and high sensitivity to external fields in an extended parameter space (Vojta, 2006).

In graph theoretical studies, a great deal of attention has been devoted to the effects of topological properties on the behaviour of equilibrium and non-equilibrium processes defined on networks (Dorogovtsev et al., 2008; Barrat et al., 2008; Castellano and Pastor-Satorras, 2010; Muñoz et al., 2010; Ódor, 2014a,b; Porter and Gleeson, 2016). Various studies showed that heterogeneous degree distributions and spectral properties can induce specific universality classes and phase transitions (Pastor-Satorras and Vespignani, 2001; Barahona and Pecora, 2002; Nishikawa et al., 2003; Bradde et al., 2010). For instance, the critical behaviour for the Ising model, percolation, and spreading processes on scale-free networks shows an explicit dependence on the degree distribution's power law exponent (Dorogovtsev et al., 2008; Barrat et al., 2008). Furthermore, the critical behaviour of the ferromagnetic Ising model on complex networks has been shown to be richer than the one predicted by mean-field theory, an effect that can be ascribed to the strong heterogeneity of networks, which is neglected in the simple mean-field theory (Dorogovtsev et al., 2002; Leone et al., 2002). Increasing network heterogeneity changes the system's critical behaviour, varying the control parameter value at which the transition occurs and making the ferromagnetic phase transition less sharp (Dorogovtsev et al., 2008). At small scales, dynamics and network topology concur in determining avalanche size and duration distributions (Radicchi et al., 2020).

Whether and how a network structure underlies the emergence of non-trivial brain activity fluctuations are important though to a large extent still open questions. At the most basic level, spatial interactions generate multiple timescales, each associated with fluctuations at a particular spatial frequency, making a hierarchical contribution to the correlations (Shi et al., 2023). In random networks with recurrent connectivity, heterogeneity in the distribution of recurrent couplings may induce a large range of time scales corresponding to neural cluster strengths (Stern et al., 2022). In particular, anatomical heterogeneity may affect the system's critical behaviour by perturbing the excitation-inhibition balance (Landau et al., 2016). Various scenarios considered the dynamics of randomly connected neural networks (Sompolinsky et al., 1988; Aljadeff et al., 2015; Crisanti and Sompolinsky, 2018). The general model considers the dynamics of a set of $N$ randomly connected nonlinear units with homogeneous weight connectivity (Sompolinsky et al., 1988). Each unit is described through the dynamics of the local field $h$ (which can for instance be thought of as a neuron's membrane potential):

$$\dot{h}_i = -h_i + \sum_{j=1}^{N} J_{ij} \phi(h_j) \qquad [12]$$

where $\phi(h_i(t))$ is a sigmoid-shaped time-varying gain function defining each node's input-output function (e.g. the way in which the membrane potential is related to the firing rate). In mean-field theory ($N, t \to \infty$), the total synaptic input current of each neuron, i.e. the second term in [12], is modeled as a time-dependent Gaussian random variable $\eta_i(t)$. At a critical value of the gain parameter, the resulting dynamics undergoes a transition between an ordered, stationary silent phase to a chaotic regime with global homogeneous delocalised fluctuations (Sompolinsky et al., 1988; Wainrib and Toboul, 2013; Aljadeff et al., 2015). The Gaussian assumption, which is consistent with systems having a characteristic scale (Wainrib and Toboul, 2013; Aljadeff et al., 2015), yields a discontinuous transition to chaos, something that would prevent the system from lying close to the edge of chaos. A continuous transition to chaos with scale-free avalanches requires instead a heavy-tailed synaptic weight distribution (Kuśmierz et al., 2020). For random neural networks with a heavy-tailed weight distribution, an extended critical regime of spatially multifractal fluctuations between the quiescent and active phases may emerge, characterised by complex properties such as long-range correlations, temporal multiscaleness and low spatial dimensionality relative to system size. Contrary to chaotic fluctuations, which tend to be maximally delocalised, heavy-tailed neural networks are dominated by multifractal chaotic fluctuations, a scenario characteristic of *Anderson transitions*[34] (Anderson, 1958). On the other hand, node heterogeneity may not be a necessary ingredient for avalanche generation (di Santo et al., 2018).

At a higher level, the type of topological disorder often thought to play a fundamental role in critical behaviour is represented by hierarchical modularity (Kaiser et al., 2007; Rubinov et al., 2011; Wang et al., 2011). In hierarchical modular networks, weakly coupled modules may act as effective rare regions, whose phase may greatly differ from that of the rest of the system and where activity remains mostly localised, lingering for a very long time before being exstinguished by fluctuations (Vojta, 2006). Hierarchical network organisation has often been associated with and suggested to be responsible for the emergence of complex fluctuations, including broad avalanche distributions (Friedman and Landsberg, 2013), subdiffusive dynamics (Kaiser et al., 2007), ergodicity breakdown (Tavani and Agliari, 2016; Agliari and Tavani, 2017), localisation phenomena (Ódor, 2014a, 2019), rounding of first-order phase transitions (Martín et al., 2015), and universality even when the underlying dynamical process has no critical points (Friedman and Landsberg, 2013). Perhaps most importantly, this particular form of disorder may help explaining how brain activity achieves criticality without an unrealistic need for tuning, which characterises criticality in homogeneous systems. In hierarchical modular networks, scale invariance is observed for a whole range of spreading rates, and the origin of this scaling behaviour may be structural (Moretti and Muñoz, 2013). Griffiths phases emerge irrespectively of the synchronisation route through which they are entered, although with qualitatively different bifurcation-specific properties (Buendía et al., 2022). Remarkably, this is the type of phenomenology that emerges when considering a system of oscillators coupled according to the drastically coarse-grained representation of the human connectome proposed by Hagmann and colleagues (2008) (Moretti and Muñoz, 2013; Villegas et al., 2014; Safari et al., 2017; Cota et al., 2018). Thus, quenched disorder of the anatomical network has been suggested to constitute the structural mechanism through which brain dynamics attains criticality without fine parameter tuning (Moretti and Muñoz, 2013). The mechanisms through which quenched disorder affects dynamics is reminiscent of the role played by impurities and *topological defects* (Mermin, 1979; Nelson, 2002) in changing the underlying system's global symmetries and as drivers of phase transitions in condensed matter physics (Egolf, 1998; Thiffeault, 2005; Ódor, 2008; Nishimori and Ortiz, 2010; Sethna, 2021; Bowick et al., 2022; Shankar et al., 2022).

Some important points need to be discussed. First, a recent study (Morrell et al., 2021) showed that uncoupled binary neurons with common time-correlated external inputs can produce non-Gaussian scaling similar to the one reported for neural networks (Morales et al., 2023), raising the possibility that rather than reflecting critical dynamics, neural scale invariance may emerge as an evoked response to shared external driving.

Second, hierarchical modular networks have been suggested to constitute a parsimonious model reproducing the human brain connectome's synchronisation phenomenology (Moretti and Muñoz, 2013; Villegas et al., 2014; Safari et al., 2017; Esfandiary et al., 2020). But is this specific type of disorder a necessary condition for this form of extended criticality? There are indications that this may not be the case, as Griffiths phases can also arise in finite non-modular systems (Cota et al., 2018), so that their presence in brain activity *per se* is no guarantee of hierarchical modularity. Griffiths phases can be observed in sparse networks (Ódor et al., 2015), possibly explaining why they may feature at scales beyond the transition from scale-free to small-world topology (Gallos et al., 2012). Furthermore, generic slow dynamics with fat-tailed intercommunication time distributions, resulting in intermittent switching between periods of low activity and high activity (Allegrini et al., 2010), may stem from small-world networks of nodes with non-Poissonian bursty dynamics (Ódor, 2014b). In alternative, bursty behaviour can constitute an emergent collective behaviour in quenched networks of Markovian variables close to criticality or in extended Griffith phase regions (Ódor, 2014b).

---

[34] The phenomenon known as *Anderson localisation* describes the transition from the metallic phase to the insulating phase, occurring in disordered electronic systems for certain values of the disorder strength (Anderson, 1958). These two phases are respectively associated with delocalised and exponentially localised eigenfunctions. Scaling theory of localisation has shown that the Anderson transition is related to conventional second-order phase transitions (Abrahams et al., 1979). Importantly, at the Anderson transition, wave functions exhibit strong multifractal amplitude fluctuations (Evers and Mirlin, 2008), and quantum dynamics can be formulated in terms of a *network model* (Shapiro, 1982).

Third, do brain anatomical networks have a hierarchical modular structure? Various authors have proposed that both anatomical and dynamical brain networks have a hierarchically modular organisation (Chatterjee and Sinha, 2007; Ferrarini et al., 2009; Kaiser and Hilgetag, 2010; Meunier et al., 2010; Bassett et al., 2010; van den Heuvel et al., 2016; Bardella et al., 2016; Smith and Escudero, 2017). Modular networks have an exponential degree distribution so that genuine high-degree hubs are absent, and the distance between highly connected nodes is in general higher than in scale-free networks (Song et al., 2006). However, hubs have been reported to characterise both anatomical (van den Heuvel and Sporns, 2013) and dynamical (Tomasi and Volkow, 2011) healthy brain networks and to be altered in various neurological and psychiatric conditions (Fornito and Bullmore, 2015; Navas et al., 2015; Jin et al., 2015; Rittman et al., 2016; Roger et al., 2020; Yu et al., 2021; Royer et al., 2022). A *rich-club* structure (Colizza et al., 2006) of the anatomical network has also been proposed (van den Heuvel and Sporns, 2011; Senden et al., 2014). In networks with such a structure, some nodes can be identified as hubs and tend to be more densely connected among themselves than nodes of a lower degree. Moreover, while extended criticality has been shown to be supported by hierarchical modular networks, human connectomes with broad link weight distribution may not show criticality (Ódor, 2016, 2019). Available connectome data are inherently coarse-grained, but the coarse-graining dependence of the resulting topology is still poorly understood (Zalesky et al., 2010, 2014; Papo et al., 2014b; Kujala et al., 2016; Korhonen et al., 2021). Ultimately, the anatomical networks' topological character hinges on the coarse-graining level at which the system is considered (Rozenfeld et al., 2010; Gallos et al., 2012; Wang and Kennedy, 2016; Millán et al., 2022a) as well as on the particular way in which modules are defined (Korhonen et al., 2021). Also worth noting is that in models such as Kuramoto's, oscillatory dynamics is explicitly assumed (Roxin et al., 2004; Kinouchi and Copelli, 2006; Laing, 2016). However, self-sustained oscillations may instead constitute a global property arising spontaneously in a way that depends on network structure, even when nodal dynamics is not of an explicitly excitable nature (Ruiz-García and Katifori, 2020).

Finally, an important though often overlooked point is that criticality may designate a heterogeneous set of singularities, with different underlying neurophysiological bases (Dahmen et al., 2019; Gross, 2021; O'Byrne and Jerbi, 2022). For instance, in what is perhaps the best studied scenario, criticality is associated with phenomena characterised by neuronal avalanches, i.e. activity bursts whose size distribution scales as a power law (Beggs and Plenz, 2003, 2004). In this model, criticality coincides with a regime where excitatory and inhibitory activity are approximately equal. Alongside this phenomenology, another type of criticality, without avalanches and with instead weak and fast correlations, has been proposed (Bertschinger and Natschläger, 2004). This form of criticality is associated with an inhibition-dominated state, which is driven to an edge-of-chaos type of criticality by network connection heterogeneity (Langton, 1990; Dahmen et al., 2019), where the system shows a rich repertoire of coexisting and topologically complex dynamics (Wainrib and Toboul, 2013). Avalanche criticality does not necessarily entrain dynamical edge-of-chaos criticality, suggesting that the different fingerprints may pertain to distinct phenomena (Kanders et al., 2017a,b). However, whether network properties play a role in these and other forms of singular behaviours and the extent to which such a role is specific to each of them remain open questions.

*Epidemic models of brain criticality*

Oscillatory activity and broad-band scaling fluctuations have long been thought of as incompatible models of brain activity. Recently, however, empirical (Poil et al., 2012) and theoretical evidence (Yang et al., 2017; di Santo et al., 2018; Dalla Porta and Copelli, 2019) has shown that these two regimes can coexist. These findings may in fact indicate that mesoscopic cortical dynamics may operate close to a *synchronisation phase transition* where oscillations coexist with scale-free avalanches, and activity is neither totally synchronous nor completely incoherent (Yang et al., 2017; di Santo et al., 2018; Dalla Porta and Copelli, 2019), allowing for transient and flexible connectivity patterns and a correspondingly rich dynamical repertoire (Breakspear, 2010; Markram et al., 2015). In the long time limit, this regime can be thought of as intermediate between one where activity wanes and another where it undergoes runaway excitation, and which is characterised by scale-invariant dynamical patterns.

It is straightforward to think of these regimes as steady state patterns resulting from activity propagation. In turn, propagation can be thought of as the dynamics of an ecological system, wherein active states correspond to the invasion of a given spatial domain by a population, and the subsequent waning of activity as the extinction of such population, the variables of interest being for instance the time to extinction (Bressloff, 2012). Heterogeneities play an important role in wave propagation (Xin, 2000). This is why in the context of continuum neural field models, much attention has been devoted to the role of inhomogeneity in determining spatio-temporal brain activity propagation (Mendez et al., 2003; Bressloff, 2001, 2012; Bressloff and Webber, 2012). In these models, heterogeneity may be incorporated by assuming a modulation of distance-dependent scaling $w(x,y) = w(|x - y|) M(x)$, where $M(x)$ is some spatial modulation of the connectivity weight distribution (Coombes and Laing, 2011). In the simplest case, an approximately periodic microstructure supports a heterogeneous periodic modulation of long-range connections (Bressloff, 2001). However, there is no clear evidence for such a heterogeneity in the cortex and at least locally, the cortex may be better modelled as a disordered medium with random spatial fluctuations in an underlying homogeneous medium. Time-dependent heterogeneity can also be introduced as a particular form of noise (Xu, 1998; Xin, 2000; van Saarloos, 2003; Panja, 2004). For instance, time-dependent heterogeneities in the form of extrinsic multiplicative noise may induce subdiffusive front wandering on short timescales and rare noise-driven transitions to an absorbing state of vanishing activity at long ones (Bressloff and Webber, 2012). More generally, heterogeneities may give rise to wave scattering (Goulet and Ermentrout, 2011) but also to complete extinction (Bressloff, 2012). The effects of connectivity structure of front propagation have also been examined. Bare connectivity properties, i.e. connection probability and connectivity strengths, have been shown to play a crucial role in determining the dominant mode of spiking activity propagation in feedforward networks (Kumar et al., 2008, 2010). Furthermore, in networks of interacting Poisson neurons (Hawkes, 1971; Saichev and Sornette, 2011), system-level correlation reflects connectivity paths and motifs (Pernice et al., 2011, 2012). However, only seldom have the effects of complex spatial disorder on wave propagation explicitly been considered (Buice and Cowan, 2009).

A perspective similar to Bressloff's ecological one (Bressloff, 2012) has been adopted to study activity propagation within networked systems. This involves using the susceptible-infected-susceptible (SIS) epidemic model (Pastor-Satorras and Vespignani, 2001; Pastor-Satorras et al., 2015; de Arruda et al., 2018). In the SIS model, individuals, represented by nodes, are either susceptible or infected. Susceptible individuals become infected by contact with infected individuals at a given rate $\kappa$ weighted by the number of infected contacts, whereas infected individuals spontaneously recover at some other rate $\mu$. The order parameter through which the collective state of the system is described is represented by the density of infected nodes $\rho$. The model has an absorbing state phase transition between a healthy and an endemic phase at a critical value of the infective rate (Pastor-Satorras and Vespignani, 2001). The SIS model exhibits an absorbing state phase transition between a disease-free *absorbing state*, i.e. a state that the dynamics cannot escape from once into it, and in which activity is absent, and an *active* stationary phase, where a fraction of the population is infected, in which the activity lasts forever in the thermodynamic limit (Pastor-Satorras and Vespignani, 2001; Henkel et al., 2008). These regions are separated by an epidemic threshold $\kappa_c$. The location and nature of the epidemic threshold in this kind of network have been a matter of debate (Castellano and Pastor-Satorras, 2010; Goltsev et al., 2012; Boguñá et al., 2013). Insofar as the epidemic threshold has been shown (both analytically and numerically) to asymptotically vanish for degree distributions decaying slower than exponentially (Boguñá et al., 2013), this model turns out to be particularly relevant in the case of highly heterogeneous networks[35].

Spreading dynamics in quenched networks is often described using the *quenched mean-field approximation* (Castellano and Pastor-Satorras, 2010; Boguñá et al., 2013). In this approximation, dynamical correlations are neglected[36]. The quenched mean-field approximation is essentially exact for networks with large Laplacian *spectral gap*[37], e.g. small-world networks (Ódor, 2013). A large spectral gap characterises networks with mean-field interactions, which lack a clear notion of *locality* and cannot easily be divided into separated moduli (Millán et al., 2021b), and can therefore be used to characterise the relevance of network inhomogeneities. (For the notion of *locality*, see also *§3.4.1 Network structure, scaling, and the emergence of locality*). For such networks, steady-state activity is dominated by the principal eigenvector and the critical activity spreading threshold $\kappa_c$, at which scale invariant dynamic patterns appear, is given by the inverse of the adjacency matrix's largest eigenvalue (which is unique for connected networks), something that does not hold for networks with small spectral gaps. On the other hand, the quenched mean-field approximation fails for networks with small or vanishing spectral gaps (Ódor, 2013; Moretti and Muñoz, 2013; Safari et al., 2017). For instance, it does not provide a correct prediction for the epidemic threshold in hierarchical modular networks, in which a finite number of unstable eigenmodes corresponding to active localised modules can become connected in the presence of transient fluctuations.

Several authors have proposed to use the SIS model to describe healthy (Kaiser et al., 2007; Moretti and Muñoz, 2013; Safari et al., 2017) and pathological (Peraza et al., 2019; Millán et al., 2022b) resting brain activity propagation. In a system-level description of brain dynamics, infected individuals are represented by active brain regions which may activate previously quiet regions, playing the role of susceptible individuals, and may be inactivated, for instance due to synaptic vesicle depletion.

In principle, epidemic spreading models allow addressing various important issues related to brain dynamics and the role of network structure in its determination. The most obvious one is represented by the way the topological properties of the underlying anatomical network affect epidemic spreading onset (Safari et al., 2017). Epidemic models suggest that the mechanisms of activity spreading onset are topology-specific: while in scale-free networks epidemic spreading may be promoted by a hub-mediated reactivation mechanism (Boguñá et al., 2013), in hierarchical modular ones, reactivation occurs at the inter-module level and across hierarchical levels (Safari et al., 2017). Once a true static order develops on some of the rare regions their order parameters can be aligned by infinitesimally small interactions or external fields (Vojta and Sknepnek, 2004). Spreading can be framed in terms of *localisation*[38], i.e. the extent to which activity in one region may spread to other parts of the system in the long time limit (Goltsev et al., 2012; Pastor-Satorras and Castellano, 2016, 2018). In steady state, this question may be understood in terms of network heterogeneities and structure of the connectivity matrix's principal eigenvector, which can be quantified in terms of the inverse *participation ratio*, i.e. the number of nodes in which the eigenvector is non null[39] (Goltsev et al., 2012; Ódor, 2014a). Thus, spreading methods allow understanding locality as an asymptotic emergent property of the interaction between structural disorder and intrinsic dynamics at criticality. (See also *§3.4.1 Network structure, scaling, and the emergence of locality*).

Epidemic spreading methods can also help determining the universality class of the underlying process. A seminal study of neuronal avalanches in local field potentials *in vitro* reported a power-law avalanche size distribution $P(s) \sim s^{-\tau}$ with $\tau \approx 2$ up to an upper cutoff (Beggs and Plenz, 2003)[40]. The exponent value, which was suggested to result from a critical branching process, was consistent with a vision of the brain hovering around a critical point belonging to the universality class of a particular absorbing phase transition, the mean-field-directed percolation (Muñoz et al., 1999), and is consistent with field theoretical studies of system-level brain activity (Buice and Cowan, 2009). But are absorbing state phase transitions a faithful representation of brain dynamics? Various experimental and theoretical findings are inconsistent with this conjecture. On the one hand, although directed percolation does not involve oscillations, signs compatible with criticality may appear in neuronal populations in ways that depend on the level of synchronisation (Ribeiro et al., 2010; Hahn et al., 2017). On the other hand, recent evidence addressing this issue and showing that a phase transition occurs under specific conditions of spiking variability intermediate between complete synchronisation and desynchronisation, reported critical exponents incompatible with the mean-field directed percolation universality class (Fontenele et al., 2019).

---

[35] In a finite system, due to dynamical fluctuations, the unique true stationary state is the absorbing state, even above the critical point.
[36] The original approach to the dynamics of the SIS model was based on the *heterogeneous mean-field* theory (Dorogovtsev et al., 2008), which neglects both dynamical and topological correlations. The quenched structure of the network, given by its adjacency matrix $A_{ij}$ is replaced by an annealed version, in which edges are rewired at a much faster rate than that of the process running on the network, while keeping the degree distribution $P(k)$ constant (Boguñá et al., 2013).
[37] When referred to a network structure, the spectral gap usually designates the second smallest eigenvalue $\lambda_2$ of the network's Laplacian matrix $\mathcal{L}$, or equivalently the gap between first and second largest eigenvalues of the adjacency matrix $A_{ij}$. The spectral gap can be used to characterise various important aspects of the dynamics on networks, e.g. the relaxation time of a random walk on the network or the coupling strength threshold for synchronisation (Motter, 2007; Watanabe and Masuda, 2010).

[38] In condensed matter physics, *localisation* designates the absence of wave diffusion in a disordered medium e.g. a semiconductor with impurities or defects. For instance, in the Anderson localisation, electron localisation is possible in a lattice potential, provided spatial disorder is sufficiently large.
[39] An eigenvector is *localised* on a subset $V$ of size $N_V$ if a finite fraction of the normalisation weight is concentrated on $V$ even though $N_V$ is not proportional to $N$. This includes localisation on a finite set of nodes, with $N_V$ independent of $N$, and on a mesoscopic subset of nodes, for which $N_V \sim N^\beta$ with $\beta < 1$. The eigenvector is *delocalised* if a finite fraction of the nodes $N_V \sim N$ contributes to the normalisation weight (Goltsev et al., 2012).
[40] Avalanches fulfill additional scaling relations. For instance, avalanche duration $T$ was found to scale as $P(T) \sim T^{-\alpha}$ with $\alpha \approx 2$ (Beggs and Plenz, 2003). Moreover, avalanche size and duration are related through a hyperscaling relation $\gamma = (\alpha - 1)/(\tau - 1)$ (Friedman et al., 2012).

Structural disorder may be reflected by non-degenerate eigenvalues at the lower edge of the quenched network's Laplacian spectrum (or at the higher edge of the adjacency matrix) which together with the corresponding eigenvectors are not considered in the quenched mean-field approach. Insofar as each eigenvalue is associated with a characteristic relaxion time $t_i \sim 1/\lambda_i$, the lower edge of the Laplacian spectrum accounts for disorder-induced slow-down (Ódor, 2014a). Disorder, e.g. hierarchical modularity, is necessary in order to transform this hierarchy of discrete levels into a continuous *Lifshitz tail* (Villegas et al., 2014). Eigenvalues in the Lifshitz tail have strongly *localised* eigenvectors, a condition that may be conceptualised as absence of diffusion in epidemic spreading (Goltsev et al., 2012; Ódor et al., 2015; Pastor-Satorras and Castellano, 2016, 2018; Esfandiary et al., 2020). It has been proposed that the human anatomical connectome is characterised by a tail of small non-degenerate eigenvalues, each corresponding to nested submodules, which may reflect the depth of the hierarchical modular organisation (Villegas et al., 2014).

The results of epidemic models in neuroscience need to be carefully discussed. First, the nodes are thought of as passive relays of a transport network and are not local processing units transforming incoming information via some complex dynamics (Zamora-López et al., 2016). Moreover, limitations in their predictive power have been reported, suggesting that they may not capture the underlying mechanisms of healthy and pathological brain dynamics (Millán et al., 2022b). It is unclear whether these stem from the excessive simplification of model specifications or from more fundamental inadequacy of the SIS and similar models in representing actual brain dynamics. In the specific context of epilepsy, one possible reason for the limited predictive power of SIS away from seizure onset may stem from the different mechanisms involved in the various phases of epileptiform activity (Pinto et al., 2005).

*3.2.3 The dynamical role of frustration*

Spontaneous brain activity is characterised by complex fluctuations, with properties including metastability, ageing, weak ergodicity breaking, which are typical of glassy materials (Papo, 2014a). As it is the case in glassy materials, complex fluctuations may not be explained by disorder *per se*. An ingredient that may be required to break ergodicity and induce multiple metastable states is *frustration*, a condition where not all constraints can simultaneously be satisfied so that local order, favoured by physical interactions, cannot propagate throughout the system (Bowick and Giomi, 2009). For example, metallic glasses are frustrated because the natural tetrahedral cluster cannot tile the space so that short-range order is incompatible with crystallinity. Frustration prevents the system from freezing into a single state, inducing complex energy landscapes with multiple local minima of the effective energy function, separated by regions of high strain in both real and phase space. In spin glasses, frustration arises when the interactions have different signs, preventing the dynamics from freezing into a single state, the resulting energy landscape (Toulouse, 1977; Mézard et al., 1987). But does frustration play a role in neural activity, and specifically, if this is indeed the case, how does it interact with network structure?

We saw that the interplay between quenched structural and intrinsic frequency heterogeneities at various scales can give rise to *frustrated synchronisation*, a regime characterised by frequency-specific synchronisation domains, which the dynamics can only resolve at well-separated coupling values (Villegas et al., 2014; Millán et al., 2018a). (See *§3.2 Quenched disorder and dynamics*). This is consistent with experimental evidence suggesting that anatomical (cytoarchitectonic, cytochemical, etc.) heterogeneity, and time scale hierarchy (Chen et al., 2015) may create dynamically frustrated states underlying stochastic synchronisation (Pang et al., 2021).

Frustration may arise in qualitatively different ways, at different scales, e.g. at micro- or mesoscopic scales (Schneidman et al., 2006; Gollo et al., 2014a) and through different mechanisms. One possible scenario arises when considering canonical three-node motifs (Milo et al., 2002; Sporns and Kötter, 2004). In such structures, strong zero-lag synchronisation in the weak coupling regime requires at least one pair of reciprocally connected nodes (Gollo et al., 2014a,b). Indirectly connected node pairs synchronise in-phase, while directly connected ones synchronise in anti-phase. Mutually connected nodes enhance synchronisation, an effect that can propagate along chains of connected nodes (Gollo et al., 2014b). For some configurations, e.g. when three nodes are all mutually connected, anti-phase synchronisation is frustrated. Frustrated closed-loop motifs fragment the synchronised landscape, allowing coexisting zero-lag synchrony metastable configurations (Gollo and Breakspear, 2014a). While anti-phase synchronisation has been suggested to play an important role in long-range relationships between cortical regions (Vicente et al., 2008; Canolty et al., 2010), possibly representing the dominant regime in the presence of substantial conduction delays (Li and Zhou, 2011), how general this type of frustration may be hinges on the presence and stability of this kind of synchronisation in neural activity.

Interestingly, phase frustration may interact with quenched network symmetry, as it may force directly connected oscillators to maintain a constant phase difference, pushing the dynamics towards a stationary state in which oscillators at two symmetric nodes have exactly the same phase, which differs from the phases of nodes with different symmetries, for a wide range of values of the frustration parameter (Nicosia et al., 2013). (See also *§Network symmetry and synchronisation*). However, whether such a role is actually played but the static anatomical network is still an insufficiently corroborated hypothesis.

*3.2.4 Methodological issues in quenched network modelling*

The interplay between quenched network structure and brain dynamics in various states (resting, awake, anesthetisised, etc.) is in general addressed by modelling sets of dynamical systems with various possible specifications but as few as possible ingredients, using experimentally defined structure as gauge. Theoretical modelling generally proceeds by assuming basic stylised, often experimentally derived, facts, and deriving the properties that are most compatible with these facts, via a general "*If…then…*" approach. Basic starting points can for instance be the existence of an asynchronous state, of an overall balance between excitation and inhibition, or of a critical state. This approach allows structure to emerge, conditional on ground truths. An important example is represented by models representing global brain activity as oscillators coupled according to static anatomic connectomes (Cabral et al., 2011, 2017, 2022; Deco et al., 2013; Luppi et al., 2022a). The best agreement between model-generated correlations and experimentally measured ones is obtained when the dynamics at individual mesoscopic nodes lies close to a Hopf bifurcation, so that, high levels of overall resting brain activity are best reproduced if each mesoscopic unit is at the edge of the oscillatory regime (Cabral et al., 2011, 2017; Deco et al., 2017). One obvious advantage of the quenched anatomical structure approach is that it avoids conjectures on the nature of coupling processes. Conversely, its most obvious limitation lies in its assuming that dynamic connectivity is a rather trivial function of the underlying anatomical structure and is, in some sense, time-invariant. In other cases, the emerging structure is the one associated with scale-free fluctuations, under a more or less explicit assumption of optimality of this regime. For hierarchical modular structure, fluctuations are scale-free for a range rather than for a single value of the coupling $\kappa$ (Moretti and Muñoz, 2013; Villa Martín et al., 2015). This general modelling framework also provides conditional obstruction rules. For instance, if connection probabilities scale with metric distance, stable balanced firing rates require that the spatial spread of external inputs be broader than that of recurrent excitation, which in turn must be at least as broad as that of recurrent inhibition. Thus, network models with broad recurrent inhibition are inconsistent with the balanced state (Rosenbaum and Doiron, 2014; Rosenbaum et al., 2017).

At least in their standard form, most of these approaches to brain modelling are predicated upon three interrelated assumptions: 1) the dynamical structure reflects the anatomical one and signal propagation along anatomically defined pathways is sufficient to deduce the dynamical characteristics of brain activity at different spatio-temporal scales (Honey et al., 2009; Betzel et al., 2013; Zamora-López et al., 2016; Sorrentino et al., 2021); however, dynamical structure may not mirror the anatomical one (Muller et al., 2016, 2018; Galinsky and Franck, 2020a,b); 2) dynamical dysfunction necessarily involves anatomical damage and, reciprocally, anatomical damage necessarily involves dynamical pathology; 3) a given observed property results from an optimisation process, possibly at evolutionary timescales. However, the fact that a given property may be optimised under certain structural conditions does not entail that this is indeed the one optimised by the neural system under consideration. Evolution may simply have optimised some other functional (perhaps to serve a different function), or the structure itself may simply not be optimised.

One fundamental issue relates to the degree of generality of the chosen model ingredients, and that of empirical findings. Empirical findings may reflect a specific cut into the system, e.g. a scale-specific view, into the underlying system's structure. More generally, the stylised facts that modelling strives to reproduce may not be independent of prior assumptions. For instance, not only is the precise dynamical characterisation and functional meaning of non-trivial fluctuations still a matter of debate (Beggs and Timme, 2012; Papo, 2014a; Muñoz, 2018; Priesemann and Shriki, 2018; Morrell et al., 2021), but the very existence of genuine scaling is also still disputed (Bédard et al., 2006; Ignaccolo et al., 2010a,b). Likewise, connectivity estimates have significantly been revised over the past few years (Kennedy et al., 2013; Wang and Kennedy, 2016), and it is unclear to what extent models predicated upon highly sparse connectivity (see e.g. van Vreeswijk and Sompolinsky, 1996) may be robust to such a change. A related point is that of the often assumed hierarchical structure of the quenched anatomical structure, which has been discussed above. (See §*Synchronisation*). While hierarchy has been proposed as a brain organisational feature since Hughlings Jackson's theory of neurological disorders (York and Steinberg, 2011), and its neurobiology explored at various levels of the central nervous system (Hubel and Wiesel, 1962; Rockland and Pandya, 1979; Felleman and Van Essen, 1991), it often designates different constructs in complex network terms (Hilgetag and Goulas, 2020).

A similar problem relates to the general form of quantitative models of brain activity. In the field of complex systems in general, and in brain sciences in particular, the interplay between topology and dynamics is often modelled using some variation of equation [5]. (See §*2.2 Modelling brain networks*). This ansatz can generate a wide range of dynamical behaviours, as a function of nonlinearity and connectivity pattern, suggesting its appropriateness as a model of neural activity. But how universal is this model? One way to assess its degree of universality involves perturbing the system's dynamics at some node and evaluating how this modifies the dynamics of all other nodes within the system (Barzel and Barabási, 2013). In the linear response regime, if the system's dynamics is such that the ratio between the term capturing the dynamical mechanism governing the pairwise interaction in equation [5] and the one representing dynamics of the isolated unit can be factorised as $\mathbf{H}(x_i, x_j)/\mathbf{F}(x_i) = f(x_i)\, g(x_j)$, where $f(x_i)$ represents the impact of $i$'s activity on itself and $g(x_j)$ that of $i$'s neighbours on $x_i$, the system's response to perturbation can be captured by the leading term of the Laurent expansions[41] of the system's dynamics around its steady state (Barzel and Barabási, 2013). Irrespective of the detailed structure of $\mathbf{F}$ and $\mathbf{H}$, the number of distinct dynamical patterns equation [5] can exhibit is finite, predicting the existence of a limited number of universality classes governing network dynamics. The corresponding scaling exponents can be derived by direct measurement of perturbation propagation, so that the leading terms of the dynamical functions $f(x_i)$ and $g(x_j)$ can in principle be estimated from empirical data. This would allow constructing an effective theory for systems of unknown dynamics. However, how this generalises to dynamical processes that cannot be modelled by equation [5] and in general to non-stationary phenomena is largely unknown. This self-consistent approach separates the contribution of network topology and dynamics, in essence replicating the assumption implicit in equation [5]. But to what extent does such a framework constitute a valid representation of brain activity? A clear time scale separation between local and network dynamics may not exist, so that topology and dynamics may not be separable aspects. so that there is no way, other than heuristic, to state whether dynamics is an emergent property of network structure or vice versa (Garlaschelli et al., 2007). From a phenomenological view-point, this would mean that the order parameter with which the dynamics is described retroacts on the control parameter (Sornette, 2006; Do and Gross, 2012; Dai et al., 2020a,b). Theoretical studies have shown that networks in which topological and dynamical scales mix are typically associated with three characteristic properties: formation of activity clusters; emergence of complex topologies, and presence of transient topological and dynamical regimes (Maslennikov and Nekorkin, 2017), properties typically reported in empirical studies of brain dynamics.

Furthermore, the meaning of observed structure depends on the underlying system's general characterisation (e.g. non-equilbrum steady state, critical, etc.), which may be instrumental in interpreting their significance. A given property may be generic in a space of given characteristics, e.g. spatial inhomogeneity is generic in systems with long-range interactions (Chavanis, 2008), long-range correlations are a generic property of stationary nonequilibrium states (Bertini et al., 2015), and ensembles of interacting heterogeneous threshold oscillators generically exhibit self-organised behaviour (Sornette and Osorio, 2010). Whether scaling laws are the result of short or long-range potentials determines the role of topology in producing singularities (Campa et al., 2009). Likewise, if the brain is in a non-equilibrium steady-state regime, under the action of an energy current flow eventually dissipated as heat, then strong long-range spatio-temporal correlations would simply arise from brain thermodynamics and would constitute a straightforward manifestation of steady-state conditions rather than of criticality (Livi, 2013).

---

[41] Laurent series expansions are a tool in complex analysis allowing to work around the singularities of a complex function $f$. While a Taylor series can only be used to describe the analytic part of a function, the Laurent expansion allows a series representation in both negative and positive powers in a region excluding points where the original function is not differentiable. If $f$ is differentiable in the entire region, then it is analytic and the Laurent series centered at a singular point $z0$ reduces to the Taylor series.

Finally, what information does asymptotic activity actually convey? Asymptotic methods renormalise fast degrees of freedom, ultimately yielding effective Markov processes induced by transitions between suitably defined states of the resulting slow dynamics (Bo and Celani, 2017). Considering steady state dynamics allows characterising brain activity as a unique equilibrium state, and corresponding generic topological structure with given properties. However, to what extent such representations constitute a good description of essentially transient brain dynamics (Friston, 2000; Tsuda, 2001) is not entirely clear.

### 3.3 Annealed network structure and dynamics

Experimental studies have shown that spontaneous activity fields exhibit sparse spatio-temporally correlated activity (Bressloff et al., 2016), and this is mirrored by reports of time-averaged non-trivial network structure (Bullmore and Sporns, 2009). But what is the meaning of such a structure? A genuine network structure is often exclusively ascribed to brain anatomy in the quasi-static limit, but annealed network structure, either induced by brain activity at experimental timescales or associated with anatomy itself but at developmental or evolutionary scales, is not necessarily a shadow, as it were, of the underlying static anatomical structure.

Most key issues associated with annealed brain network structure, ranging from the way the system is equipped with a network structure to the significance that may be ascribed to such a structure, are better treated in a genuinely functional rather than in a purely dynamical framework. This is because aspects such as the metric with which brain activity is mapped into edges and the meaning of the structure that these induce are inherently functional ones. However, annealed disorder is often considered in a purely dynamical context wherein function is not explicitly considered. To do so, it is thought of as the structure of the unperturbed (spontaneous) dynamics. The resulting network structure can then be considered in real space, wherein nodes are identified with particular regions of the anatomical space, or in the phase space of the dynamics.

While in quenched network models structure is in general not an emerging property and the focus is on activity propagation and on some dynamical or scaling property associated with brain activity, when considering annealed structure, the focus is on all of these aspects, i.e. on propagation and on the emergence of network structure and scaling properties.

An important question when dealing with activity-induced network structure relates to how it is constructed, particularly the way edges are defined (Korhonen et al., 2021). Although the problem is more acute in a genuinely functional context, where the functional role of dynamical connectivity is explicitly considered, even in a purely dynamical one the particular metric used to quantify connectivity can in principle affect the picture that models may end up showing. For, instance, a given connectivity metric may induce a space of qualitatively different properties from those associated with other *prima facie* equally plausible ones.

The most fundamental issue in this context relates to the very status of dynamic brain networks, viz. has this structure its own separate meaning independently of its relationship with the static anatomical structure? To what extent is annealed structure not a direct consequence of anatomical network structure? Does the former contain information not contained in the latter? If so, what could it be its meaning? In condensed matter physics, the standard case is one in which structure determines dynamical behaviour (Chaikin and Lubensky, 2000). For instance, the level of order determines a given system's ability to flow and relax. On the other hand, systems close to a glass transition, show massive changes in dynamics but no appreciable structural one. Correlations are revealed not by structural disorder but by dynamic heterogeneity (Kob, 1999; Berthier, 2011).

Annealed disorder relevance, independently or even in the absence of quenched anatomical one, would indicate that brain dynamics is effectively in a state intermediate between those of a liquid and a solid.

A further important point is the relationship between the activity-induced network structure and local dynamics (dynamics *on* the network). Typically, in system-level studies of brain activity, local dynamical units communicate with each other via the time-invariant anatomical network structure, but do not interact with the topology of the network induced by the dynamics. Even when it is considered, topological dynamics (dynamics *of* the network) is studied separately, and not much is known about the relationship between dynamics *on* and *of* the network.

A related question involves considering the relationship between the activity-induced network structure dynamics and the processes unfolding on such a structure. In particular, it is interesting to examine the possible role of annealed network topology in criticality. An interesting starting point is offered by reports of the co-presence of topological network properties and complex scaling. For instance, in a near-critical regime, hierarchical structure of the dynamical network becomes a robust feature (Safari et al., 2021). However, it is difficult to assess the exact relationship between the two aspects, and in particular whether and how they coevolve. While temporal hierarchical behaviour may emerge from critical behavior under certain conditions, e.g. complex fixed-points and maximal correlation between the system components (Pérez-Mercader, 2004), criticality may in principle be an emerging property of adaptive networks, i.e. networks in which local state dynamics (dynamics *on* the network) is topology-dependent and topological changes *of* the network occur in a state-dependent manner.

### 3.4 Network structure and emergence of spatio-temporal structure

Irrespective of whether dealing with quenched or annealed disorder, when analysing the scaling properties of neural systems, at least three main aspects should be considered: scaling of the (networked) system and of local activity, but also of the topology (topological phase transitions). Moreover, the role of network structure in producing the system's scaling properties should be examined both in the unperturbed system and as the system undergoes external fields.

*3.4.1 Network structure, scaling, and the emergence of locality*

The notion of heterogeneity is intimately related to that of *locality*. Though traditionally discussed in a functional context, locality also plays an important role in dynamics. In terms of dynamics, *localisation* essentially means that local fluxes can be described as a function of the local gradient of some intensive property (Paradisi et al., 2009). This assumption is appropriate for Gaussian statistics of the random fields and linear scaling of fluctuation's variance with time and can be related to short range (exponential) decay of space-time correlations. The dynamic origin of non-local flux-gradient relationships and the emergence of coherent metastable structure arises from the random occurrence of *crucial events*, i.e. abrupt dynamical transitions with random life-times, occurring independently with a power-law waiting time distribution (renewal point process) (Paradisi et al., 2009; Allegrini et al., 2009, 2010a,b), without no specific mention to an underlying network structure.

The notion of *locality* is implicit in the discretisation underlying network structure, particularly in the definition of the network's nodes. Locality means that nearby points in a given space are related by spatial, temporal, or contextual proximity (Robinson, 2013). The sets in which proximity is meaningful are precisely topological spaces. Locality may have a clear basis in the quenched anatomical structure, particularly at the microscopic network scales of single neurons. Dynamics may become localised as a result of the interaction between quenched disorder and the system's intrinsic dynamics at criticality (Villegas et al., 2014). Note that part of the disorder would be represented by anatomical connectivity and part by heterogeneity in nodal dynamics, typically in terms of frequency distribution and global dynamic disorder results from quenched anatomical and local dynamic disorder. (See §*Quenched network structure and brain criticality*). Such fragmentation would for instance seem consistent with reports that connectivity *within* regions characterises epilepsy duration and treatment outcome (Chen et al., 2021). But can dynamics be localised in the absence of quenched anatomical disorder?

Locality is also strongly associated with information routing, which is more appropriately defined in a genuinely functional context. Much as in a dynamical context, the essential question in a functional one is whether and the extent to which function may be thought of as local in real space, the extent to which this is associated to neural aspects unfolding at evolutionary or developmental and which may be thought of as static when considering experimental time scales or, in alternative whether localisation results from specific properties of brain activity (Palmigiano et al., 2017). (See also §*Network topology and symmetry breaking in dynamical pattern formation*).

### 3.4.2 Topological phase transitions

Most of the time, when considering the brain as a topological dynamical system, the network structure is thought of as an ingredient of a complex dynamics, which is then typically studied in the time domain or in phase space. For instance, when considering a quenched network structure, one studies the scaling properties of the system's dynamics, given some connectivity pattern. While both quenched and annealed connectivity dynamics may play an important role in brain dynamics, it is important to examine whether particular properties of the connectivity structure are also associated with non-trivial dynamics and, if so, to characterise such dynamics, and in particular discuss whether it undergoes phase transitions or *crossovers*[42].

At perhaps the most basic level, abrupt transitions observed in brain activity have been proposed to be induced and maintained by neural percolation processes (Kozma and Puljic, 2015). Neuropercolation considers neural populations as large-scale random networks undergoing sparse connectivity rewiring and transitions between transient dynamical regimes are controlled by the ratio of non-local connections as well as the global strength of inhibition and noise. Such a scenario can in essence be thought of as a connectivity phase transition (Kozma and Puljic, 2015).

In a statistical mechanical model of network dynamics, where energy is assigned to different network topologies, a variety of topological phase transitions associated with singularities in global network structure properties can emerge as temperature, quantifying the level of noise associated with rewiring, is varied. The order parameter through which the transition is monitored, the type of phase transition and the network topology emerging at the critical temperature value all depend on the chosen energy function (Derényi et al., 2004, 2005; Palla et al., 2004). But does brain network structure also undergo genuine *topological* phase transitions? Metastable-state transitions with topological changes in the minimal-spanning-tree of the network induced by cross-correlations in electrical brain activity have been reported (Bianco et al., 2007). However, whether brain dynamics is characterised by significant topological fluctuations and, more specifically whether network topology acts as a control parameter enforcing bifurcations in brain dynamics (either globally or in some well-defined subsystem) are poorly explored issues.

There is an additional, qualitatively different way in which the topology induced by dynamical connectivity may be at the root of singularities in brain activity. Phase transitions, which are in general identified with singularities of thermodynamic potentials, may be understood in purely topological terms (Caiani et al., 1997; Franzosi et al., 2000; Donato et al., 2016; Pettini et al., 2019), their occurrence being connected to non-trivial changes in the configuration space topology (Caiani et al., 1997). Under some circumstances, topological changes of the level subsets of some property of the configuration manifold e.g., for a Hamiltonian system, the surfaces of constant potential energy (Pettini, 2007), are related to microcanonical entropy's non-analytic points (Pettini, 2007; Kastner, 2008; Santos et al., 2014; Casetti et al., 2000). According to Morse theory[43] (Matsumoto, 2002), this topological change is signalled by the presence of singularities of the Euler characteristic $\chi$, an integer-valued topological invariant describing the structure associated with the system[44]. In short-range systems, the presence of topological changes in the configurational space constitutes a necessary (though not sufficient) condition for phase transitions to occur[45] (Franzosi and Pettini, 2004), some information on dynamics turning out to be necessary (Ribeiro Teixeira and Stariolo, 2004). For systems with long-range interaction potentials, on the other hand, topological properties of the potential energy may not change remarkably at a second-order phase transition (Campa et al., 2009). When considering dynamical brain networks, it is straightforward to ascribe the role of height function or energy levels to dynamic coupling levels emerging as a connectivity threshold (or some other similarity measure) is varied (Santos et al., 2019; de Amorim Filho et al., 2022). This may allow identifying topological phases and phase transitions induced by brain dynamics and understanding how brain pathology may affect them (Santos et al., 2019). Importantly, such properties are *intrinsic*, i.e. they do not depend on the embedding of the space induce by the dynamics. Since integers cannot change smoothly, spaces with different $\langle\chi\rangle$ are topologically distinct as they cannot be deformed into one another. Typically, though, topological changes may be numerous even away from a phase transition, so that a purely topological criterion alone may not allow identifying phase transitions, and a geometric criterion may turn out to be useful for such an identification (Kastner and Schnetz, 2008). A necessary criterion for the occurrence of a thermodynamic phase transition relates it to the curvature at the saddle points of the potential (Kastner and Schnetz, 2008). Note that while in the former scenario phase transitions are *diachronic*, in the latter they are *synchronic*. (See also §*Curvature*).

Alongside these genuine dynamic transitions, other types of topological transitions can be identified through renormalisation methods (Rozenfeld et al., 2010; DeDeo and Krakauer, 2012;

---

[42] *Crossovers* are associated with drastic changes in the phase of the system but differ from canonical phase transitions in that they are not associated with a change of symmetry, or a discontinuity in the free energy functional. Typically, crossovers occur in a phase space *region*, rather than at a singular point, and in the presence of more than one critical fixed-point, as the system depends on several *relevant* (in a renormalisation group sense) parametres (Nishimori and Ortiz, 2010).

[43] *Morse theory* investigates how functions defined on a manifold are related to the manifold's geometry and topology. In particular, Morse theory shows that it is possible to study the topology of a differentiable manifold by analysing its level sets. The first important result is Morse's lemma, which gives a relationship between the critical points of a sufficiently general function and changes in the topology of the manifold.

[44] $\chi$ can be expressed in terms of the *Betti numbers* $\beta_i$, each quantifying the number of holes of a given dimension characterising an object: $\chi = \sum_k (-1)^k \beta_k$. This reflects the fact that the Euler characteristic counts the zeroes of a generic vector field (Ghrist, 2014).

[45] This relation between topology and phase transitions constitutes an example of *contextual emergence*, a notion referring to systems in which the microscopic scale contains necessary but not sufficient conditions for the appearance of macroscopic properties (Atmanspacher and beim Graben, 2007; Atmanspacher, 2015; Fülöp et al., 2020). While demonstrating whether topology may indeed constitute a sufficient condition for brain transitions is rather arduous, recent results suggest that this may indeed be the case, under certain conditions (Di Cairano et al., 2021).

Villegas et al., 2023). For instance, at macroscopic scales, brain activity may show hierarchical organisation into modules with large-world self-similar properties, but the addition of only a few weak links can turn the network into a non-fractal small-world one (Gallos et al., 2012). The overall significance of such singularities can only be determined by considering the brain function they are associated with. Furthermore, phase transitions can be found using the Laplacian matrix. For instance, it was shown that Shannon entropy of a particular function of the network Laplacian[46] acts as an order parameter for structural phase transitions and the peaks of its derivative (acting as an effective *specific heat*) with respect to the time, identify characteristic network scales (Villegas et al., 2023). It is important to understand that these singularities are in essence those induced by the diffusion process used to explore the system's network structure, which should not be automatically identified with neural processes occurring within the system.

*3.4.3 Network structure and response to external fields*

Up until now, we considered an isolated system and the role of network structure on such a system's dynamics. However, a fundamental property of physical systems is the way they respond to external fields[47]. In fact, the response to external perturbations contains fundamental information on complex systems' microscopic dynamics (Timme, 2007), and observed dynamics can be explained in terms of classes of correlated dynamics (Barzel et al., 2015). In a neuroscience context, this issue is in general addressed in a functional context, particularly in sensory-motor system modelling, but some fundamental aspects can also be naturally introduced and discussed in a purely dynamic context. Two main questions need to be addressed: how does network structure affect the response to external fields and, conversely, how do external fields affect network structure?

On the other hand, how a networked system responds to perturbations and noise depends not only on the driving signal's properties but also on network connectivity (Timme, 2007; DeDeo and Krakauer, 2012). For instance, how a quenched network structure can affect a response to a generic external excitatory field, affecting several key dynamical aspects, i.e. the synchronisation transition, can be studied by examining the synchronisation transition of the Shinomoto-Kuramoto model (Shinomoto and Kuramoto, 1986), a modification of the standard Kuramoto model [8] incorporating a force term accounting for excitatory fields:

$$\dot{\varphi}_i(t) = \omega_i + \frac{\kappa}{N}\sum_{i=1}^{N} A_{ij} \sin(\varphi_i(t) - \varphi_j(t)) + F\sin(\varphi_i(t)) + \epsilon\eta_i(t)$$

[8.1]

The model incorporates a site-dependent periodic force term, proportional to some coupling $F$, describing excitation, and a Gaussian annealed noise term $\eta_i(t)$, with amplitude $\epsilon$ (Shinomoto and Kuramoto, 1986; Ódor et al., 2023). When run on top of the human connectome, it is possible to quantify the effect of both excitatory input and thermal noise on the system's scaling properties.

In particular, the presence of an excitatory force term changes the nature of the bifurcation from the Hopf transition of the standard Kuramoto model to an extended non-universal scaling tails and reduces the scaling exponent value (Ódor et al., 2023).

On the other hand, while the effect of external fields on the temporal scaling properties of networked systems has received much attention (West et al., 2008; Priesemann et al., 2018), and can for instance be thought of as a rejuvenation process (Papo, 2013a), much less is known about the effects on spatial properties and, not surprisingly, most of what is known is of an inherently functional character. In the presence of nonlinear spatial interactions, external input can modulate the correlation timescales in a way that depends on the dynamics at which the system operates (Shi et al., 2023).

*From dynamics to thermodynamics: the role of network structure*

The fluctuation-dissipation theorem (Kubo, 1966), a fundamental result of statistical physics, suggests a different approach to the characterisation of neural responses to external fields (Papo, 2013c). The fluctuation-dissipation theorem posits that in systems operating at or close to equilibrium, the autocorrelation of some observable's fluctuations in the unperturbed system is related to the response to small external perturbations through temperature (Kubo, 1966; Puglisi et al., 2017). In terms of brain activity, this establishes a substantial equivalence between stimulus-evoked and ongoing brain activity, where the former can be understood by quantifying the correlation of fluctuations of the latter. The statistical properties of observed spontaneous fluctuations reveal the particular fluctuation-dissipation relation that applies, which in turn contains important information on brain activity's dynamical regime. Equilibrium fluctuations have Gaussian statistics and exponentially vanishing memory. Thus, the non-trivial fluctuations characteristic of neural fluctuations (Papo, 2014a) suggest that the brain operates away from equilibrium and violates the standard fluctuation-dissipation theorem (Papo, 2013c, 2014b; Sarracino et al., 2020; Lindner, 2022; Nandi et al., 2023). For disordered systems, both close and far from equilibrium, alternative (Marconi et al., 2008; West et al., 2008; Baiesi et al., 2009; Aquino et al., 2011; Evans et al., 1993; Crooks, 2000; Evans and Searles, 2002), typically scale dependent (Cugliandolo et al., 1997; Egolf, 2000; Papo, 2014b; Battle et al., 2016) relations hold, reflecting an altogether different underlying thermodynamics.

A hallmark of systems operating out of equilibrium is *time-reversal symmetry breaking*[48] (Gnesotto et al., 2019). Not surprisingly, brain activity has been found to be characterised by significant irreversibility (Paluš, 1996; Zanin et al., 2022; de la Fuente et al., 2023; Gilson et al., 2023; Bernardi et al., 2023). The breakdown of time-reversal symmetry reflects the irreversible increase in entropy of the environment due to heat dissipation associated with non-equilibrium transformations: the greater the dissipation, the more pronounced the irreversibility. Remarkably, dissipation is proportional to the violation of fluctuation-dissipation relations (Harada and Sasa, 2005), and therefore quantifies how far the system lies from equilibrium (Fodor et al., 2016). Thus, dissipation is related to a system's response function.

---

[46] The *Shannon entropy* $S(K(\tau)) = -(1/\log N) \sum_{i=1}^{N} f(\lambda_i(\tau)) \log f(\lambda_i(\tau))$, where $K(\tau) = e^{-\tau \mathcal{L}}$ is the network propagator, whose elements are the sum of diffusion trajectories along all possible paths connecting any two nodes $i$ and $j$ at time $\tau$, and the argument is a function of the Laplacian matrix $\mathcal{L}$, is a measure of the residual information encoded in a network having undergone $\tau$ time steps (De Domenico and Biamonte, 2016). For simple graphs, $\mathcal{L}$ is Hermitian and for systems at equilibrium it can be thought of as a Hamiltonian operator. In turn, $dS/d(\log \tau)$ plays the role of the *specific heat*, and $\tau$ that of (inverse) temperature (Villegas et al., 2023). For a system at equilibrium, the former quantifies the number of accessible states per temperature unit, whereas the latter regulates the rate at which the system makes microstates available as a function of energy level fluctuations.

[47] Note that the term 'external' designates any type of perturbation, including activity originating from other parts of the brain, when the network structure under consideration is one particular region and not the whole system.

[48] *Time-reversal symmetry* or *irreversibility* is a measure of the extent to which it is possible to discern a preferred time direction of a stationary stochastic process (Pomeau, 1982).

While the fundamental relations between dynamics and thermodynamics (Seifert, 2012, 2019; Wright and Bourke, 2021) can be used to characterise brain phenomenology and its dysfunction (Papo et al., 2023), the possible role played by network topology in a system's thermodynamics, particularly that of the brain is still poorly understood. If topology has any effect on the type of fluctuation-dissipation (or fluctuation) relation, then it should also have an effect on the system's thermodynamics. But does network structure have a role in entropy production and dissipation? If so, what aspects of brain network structure can affect a neural system's thermodynamics?

In phase space, one such aspects is represented by connectivity asymmetry of the Schnakenberg network, whose nodes and links respectively represent mesoscopic states of the system and transitions between these states (Schnakenberg, 1976). This reflects the breakdown of *detailed balance* within network cycles resulting from the presence of thermodynamic forces. How such a fundamental macroscopic property can be shaped by properties at lower (meso- and microscopic) scales is still poorly understood. Nonequilibrium steady states of Markov processes give rise to nontrivial cyclic probability currents, and expectation values of observables can be expressed as cycle averages[49]. Recently, it was shown that, in the steady state, the cross-correlation asymmetry achievable by a nonequilibrium system modeled as a continuous-time Markov chain is bounded from above by the *cycle affinity* $\mathfrak{F}_c$, i.e. the sum of thermodynamic forces acting on the system's state cycles[50] (Ohga et al., 2023). The response function of such a class of relatively simple non-equilibrium systems is regulated not only by thermodynamic forces but also by topological constraints on fluctuations (Fernandes Martins and Horowitz, 2023). Constraints on the steady state equilibrium response can be expressed in terms of state space microscopic topology and the strength of thermodynamic driving and requires no kinetic information beyond the one encoded in the state space structure (Fernandes Martins and Horowitz, 2023).

Much less is known about the role of network structure in real space. The dominant contribution to the break down of detailed balance can be traced to node interactions (Lynn et al., 2022). Irreversibility is at least partly intrinsic to brain activity and does not stem from exogenous perturbations such as stimuli. Recently, using a simple model of brain networked dynamics incorporating the complex quenched connectome topology, it was shown that entropy production correlates with consciousness levels (Gilson et al., 2023). The whole brain was modelled as a multivariate Ornstein-Uhlenbeck dynamics, where the topology is incorporated in the friction term and plays a role in the process covariance. While this important result suggests that the connectivity matrix contains relevant information, the exact role played by topology is not totally clear.

## 3.5 Beyond topology: the role of dimension and symmetry

So far, we mainly considered the role of topological network structure (defined in various ways and at various scales) on several aspects of brain dynamics. Here, we examine two important properties of such structure which are often insufficiently explored, particularly when investigating brain dynamics in a network neuroscience context, i.e. symmetry and dimension.

### 3.5.1 Symmetry and topology in brain dynamics

Symmetries are the invariant properties of an object under a set of transformations (Weyl, 1952; Brading and Castellani, 2002). The symmetries of a given object form a group $\mathcal{G}$. Thus, symmetries are the properties that remain invariant under the *action* of a group of transformations[51]. While symmetries are often used to characterise and classify static objects and patterns, they also play a prominent role in dynamical laws and patterns (Gross, 1996; Golubitsky and Stewart, 2002b). For a classical dynamical system, a symmetry is a transformation, $\mathbf{x}(t) \to \mathcal{R}[\mathbf{x}(t)]$ of the dynamical variable $\mathbf{x}(t)$ that leaves the action $S[\mathbf{x}(t)]$[52] unchanged. Thus, if $\mathbf{x}(t)$ is an extremum of the action, so is $\mathcal{R}[\mathbf{x}(t)]$. Likewise, if $\mathcal{R}$ is some spatial transformation, e.g. a rotation, then if $\mathbf{x}(t)$ is a solution of the equations of motion, the spatially rotated $\mathbf{x}(t)$ is also a solution. Symmetries affect fundamental properties of dynamical systems, forcing particular types of interactions (Yang, 1980), changing their critical behaviour and universality class (Hinrichsen, 2000), and *observability* (Whalen et al., 2015). In particular, in systems whose dynamics can be represented by differential equations, the presence of symmetries affects the behaviour of their solutions, especially their symmetries (Golubitsky and Stewart, 2015). Not only can symmetries be used to derive new solutions, but they also play a key role in determining the types of solutions that systems can spontaneously generate (Ruelle, 1973; Sattinger, 1978; Golubitsky and Stewart, 2003). Furthermore, symmetric systems are typically characterised by more complicated bifurcations than those of systems lacking symmetry. For instance, critical eigenvalues are generically multiple in the presence of symmetry, and symmetry forces multiple bifurcations (Crawford and Knobloch, 1991).

But do brain networks possess symmetries? If so, on what space (e.g. real or phase space) and at what scales are brain networks asymmetric? How do symmetries emerge and how do they act? Can they be used to constrain models of brain dynamics? For instance, if brain networks have certain symmetries, how do they affect the brain's dynamical repertoire and bifurcations? Conversely, does network structure affect dynamic symmetries?

### The role of network symmetries in brain dynamics

At the most basic level, it is important to understand how connectivity structure symmetry may affect networked systems' dynamics, particularly their symmetries. This can for instance be framed in terms of *link reciprocity*, i.e. the tendency of node pairs to form mutual connections between each other (Garlaschelli and Loffredo, 2007). Symmetry-broken solutions may occur in symmetrically coupled networks (Abrams and Strogatz, 2004), e.g. in *dynamical solitary states* an entire extreme form of cluster synchronisation where the dynamics of a subnetwork differs from that of the symmetric network (Maistrenko et al., 2014; Berner et al., 2020; Schülen et al., 2021, 2022). While these states have been observed in symmetrically coupled systems with profoundly different properties, e.g. non-locally coupled phase, multilayer networks (Schülen et al., 2021), time-delayed systems (Schülen et al., 2019) oscillators, or adaptive networks (Berner et al., 2020), they tend to arise in heterogeneous asymmetric networks (Schülen et al., 2022). states. Local topology, viz. a minimum degree of solitary nodes' neighbours, turns out to be essential to the appearance of such states (Schülen et al., 2022). Moreover, the persistence of such states depends in a non-trivial way not only on dynamical properties but also on topological ones, i.e. the position in the network (Schülen et al., 2021). From a neuroscience viewpoint, while some effort has been made to quantify link reciprocity in brain activity (Zhou et al., 2007; de Vico Fallani et al., 2008), and more generally, connectivity asymmetry quantification has been the object of a huge number of studies (see Pereda et al., 2005; Friston, 2011; Vicente et al., 2011; Seth et al., 2015 for reviews on the topic), the effects of link symmetry on overall brain dynamics, are still poorly understood.

---

[49] For a system whose dynamics can be modelled as a Markov process making random transitions between states $i$ and $j$ at a rate $W_{ij}$, the probability for the system to be in a given state $i$ at time $t$ is given by the master equation $p_i(t) = \sum_{j=1}^{N} W_{ij} p_j(t)$. Non-equilibrium systems are characterised by the presence of non-zero steady-state probability currents $J_{ij} = W_{ij}\pi_j - W_{ji}\pi_i$, where $\pi_i$, is the steady state probability distribution of transitions from states $i$ to state $j$. These currents are driven by thermodynamic forces (Seifert, 2012), which can be identified by evaluating the balance along *cycles*, i.e. sequences of states starting and ending at a given node.

[50] The *cycle affinity* $\mathfrak{F}_c$ is the sum of all thermodynamic forces acting on the system along a given cycle $c$ (Seifert, 2012).
[51] The *action* of a group $\mathcal{G}$ on a set X is a homomorphism $\phi: \mathcal{G} \to S(X)$, where $S(X)$ is the group of all permutations of $X$.
[52] In classical systems, the *action* is a local scalar functional of $\mathbf{x}(t)$ describing how the system changes over time. The action can be written as the integral over time of the Lagrangian function $L$ of $\mathbf{x}(t)$ and its time derivative, i.e. $S[\mathbf{x}(t)] = \int_{t_1}^{t_2} dt\, L(\mathbf{x}(t), \dot{\mathbf{x}}(t))$.

In networked systems, symmetries are generally defined in terms of *network symmetry group*, i.e. the set of permutations of the nodes that preserves the coupling structure (MacArthur et al., 2008; MacArthur and Sánchez-García, 2009; Pecora et al., 2014; Golubitsky and Stewart, 2015; Sánchez-García, 2020). The set of symmetries of a graph $N$, known as *automorphisms*[53], i.e. the set node permutations preserving the coupling structure, can be represented by its automorphism group $Aut(N)$ (MacArthur et al., 2008; Garlaschelli et al., 2010). All elements of the group $\mathcal{G}$ represented as a permutation matrix $\mathcal{R}_\mathcal{G}$ in the space of network nodes commute with the matrix on which they act. If, for instance, the matrix is the adjacency matrix $A$, the invariance of $A$ under the action of $\mathcal{R}_\mathcal{G}$ implies that $\mathcal{R}_\mathcal{G} A = A \mathcal{R}_\mathcal{G}$[54]. Network symmetries can be used to define meaningful parcellations of the relevant space (MacArthur et al., 2008; Barrett et al., 2017). For instance, non-trivial automorphisms can be used to decompose any matrix $M \in C^{n \times n}$ appropriately associated with a graph, e.g. the adjacency matrix, or the Laplacian matrix, into strictly smaller matrices whose collective eigenvalues coincide with those of the whole matrix (Barrett et al., 2017). One reported topological condition for network symmetries is represented by similar degree nodes' tendency to share common neighbours (Xiao, 2008b), but alternative mechanisms are poorly explored. (See also §*Network symmetry and synchronisation*).

Network symmetries of other kinds can be defined (Holme, 2006; Garlaschelli et al., 2010). For instance, symmetry may be quantified in terms of the *duplication coefficient*[55], which quantifies the extent to which network nodes look like their most similar peer (de Lange et al., 2016). The presence of such nodal symmetry may be associated with specific spectral properties, i.e. mid-spectrum peak magnitude, (Banerjee and Jost, 2008). Furthermore, symmetries can feature as hidden scale-dependent properties (Smith and Webb, 2019). The existence of global network symmetries is often too strong a constraint, and weaker notions may be more appropriate, e.g. the *symmetry groupoid*[56] and *interior symmetry*[57] (Stewart et al., 2003; Golubitsky et al., 2004; Golubitsky and Stewart, 2006). Moreover, in real networks, symmetries are in general only approximate and of a stochastic nature and should be understood as transformations between networks within a given statistical ensemble (Garlaschelli et al., 2010). Finally, an interesting though unexplored possibility is that of hidden symmetries, i.e. symmetries not of the system itself but of its group of symmetries and of the associated *Cayley graph*[58].

While the symmetry group can be identified either in the brain space or in the system output (e.g. perceptual reports, motor behaviour), the most important information is related to the space on which symmetries act (Golubitsky and Stewart, 2002b; Golubitsky, 2006). Importantly, graph automorphisms may act on the system's phase space, inducing bijections on the set of attractors (Morrison and Curto, 2019). Thus, the presence of a graph automorphism can help predicting unobserved network attractors (Morrison et al., 2019).

*Network symmetry and synchronisation*

Network symmetries have a strong influence on dynamics (Pecora et al., 2014; Nijholt et al., 2016; Motter and Timme, 2018). In *coupled cell systems*, i.e. medium-sized directed networks, where nodes represent dynamical systems and edges are grouped into types (Golubitsky and Stewart, 2002a), the network's symmetry groupoid determines the space of *admissible vector fields* i.e. the dynamics consistent with the network structure (Stewart et al., 2003; Golubitsky and Stewart, 2015), and the robust synchrony patterns induced by the topology (Dias and Stewart, 2004). In large scale complex networks, while synchronisation patterns are often induced by network symmetries, network symmetry may control synchronisation patterns in a somehow counterintuitive way. In particular, for a wide class of diffusively coupled networks, stable synchronisation requires global structural symmetry breaking the network's (Hart et al., 2019). The network symmetry group *orbits*[59] can define synchronised clusters of nodes permuting under symmetry operations, in a way that is largely independent of local dynamic specifications (Pecora et al., 2014; Sorrentino et al., 2016; Abrams et al., 2016; Motter and Timme, 2018). Networked dynamical systems are in essence discrete-space continuous time-varying systems. The invariance under the action of $\mathcal{R}_\mathcal{G}$ implies flow invariance of the equation of motion. For instance, for equation [5]:

$$\frac{d}{dt}(\mathcal{R}_\mathcal{G} x_i) = \mathbf{F}(\mathcal{R}_\mathcal{G} x_i) + \frac{\sigma}{N}\sum_{j=1}^{N} A_{ij}\mathbf{H}(\mathcal{R}_\mathcal{G} x_j) \qquad [14]$$

so that all clusters of synchronised nodes remain synchronised under permutation. Equation [14] allows nodes within a cluster to be exactly synchronised. Remarkably, even random networks can exhibit a large number of non-trivial symmetry clusters with nodes often not directly connected (MacArthur et al., 2008).

The possible relationships between clusters can be understood in terms of symmetry group decomposition. A finite group $\mathcal{G}$ can be written as a direct product of $n$ subgroups: $\mathcal{G} = \mathcal{H}_1 \times ... \times \mathcal{H}_n$ where all the elements of a given subgroup commute with all the elements in any other subgroup (MacArthur et al., 2008, 2009). Thus, the set of nodes permuted by each $\mathcal{H}_i$ is disjoint from the one permuted by any other subgroup $\mathcal{H}_j$, although a given subgroup can permute more than one cluster. This decomposition ensures that each cluster is unaffected by the behaviour of clusters associated with different subgroups. This enables the clusters to have the same synchronised dynamics even when some other cluster desynchronises, explaining the appearance of a generic symmetry-breaking bifurcation termed *isolated desynchronisation*, wherein some clusters lose synchrony without necessarily desynchronising the rest of the system (Pecora et al., 2014)[60].

---

[53] For a graph $N$ with vertex set $V(N)$, a bijection $\alpha: V(N) \rightarrow V(N)$ is a *graph automorphism* if it preserves the edges of $N$, i.e. $i \rightarrow j$ in $N$ if and only if $\alpha(i) \rightarrow \alpha(j)$. *Graph automorphisms* characterise adjacency invariance to transformation operations on the node set. An automorphism acting on the node set is a permutation of the graph nodes which preserves the adjacency of the nodes. The set of automorphisms under permutation forms a group. In general, a network is deemed *asymmetric* if its graph only contains an identity permutation, and symmetric otherwise (Xiao et al., 2008a).
[54] Adjacency matrix $A$ invariance implies that of the graph Laplacian matrix $\mathcal{L}$ and vice versa (Pecora et al., 2014).
[55] The *node duplication coefficient* of a is the maximum similarity to any other node in the network. The *network* duplication coefficient is the average duplication coefficient over all nodes (de Lange et al., 2016).
[56] The *symmetry groupoid* consists of structure-preserving bijections between certain subsets of the cell network, the input sets (Golubitsky et al., 2010). A *groupoid* is a generalisation of a group, formalising *local network symmetries* in which products of elements are not always defined (Higgins, 1971).
[57] *Interior symmetry* is a construct intermediate between global group and local groupoid symmetries. It is a symmetry of some subsets of nodes that may not extend to symmetries of the full network, which fixes all nodes receiving input from those outside that subset (Golubitsky et al., 2004).
[58] A *Cayley graph* is a graphic representation of the structure of an abstract group $\mathcal{G}$, such that each element of the group can be written as a product of elements of its set of generators $S \subseteq \mathcal{G}$ (or their inverses) such that the identity element $I \notin S$. The Cayley graph associated with $(\mathcal{G}, S)$ is the directed graph having one vertex associated with each group element and directed edges $(g, h)$ whenever $gh^{-1} \in S$. A Cayley graph is connected if and only if $S$ generates $\mathcal{G}$. Intuitively, a Cayley graph shows how the generating set $S$ acts on $\mathcal{G}$.
[59] The *orbit* of an element $x$ of a set $X$ is the set of elements in $X$ to which $x$ can be moved by the elements of a group $\mathcal{G}$ acting on it. Moreover, the set of orbits of points in $X$ under the action of $\mathcal{G}$ forms a partition of $X$.

Subgroup decomposition can also help explaining a class of network symmetry-induced phenomena termed *remote synchronisation*, whereby distant oscillators can stably synchronise even when they are connected through asynchronous oscillators (Nicosia et al., 2013; Zhang et al., 2017; Motter et al., 2018). For instance, for some values of the phase lag parameter $\alpha$, networks of Kuramoto oscillators described by

$$\dot{\varphi}_i = \omega_i + \frac{\kappa}{N}\sum_{i=1}^{N} A_{ij} \sin(\varphi_i(t) - \varphi_j - \alpha) \quad [15]$$

undergo frustration, wherein directly connected oscillators cannot synchronise, while distant oscillators in the same symmetry cluster do. This results from equation [14]'s invariance under the action of the network symmetry group; the system admits a synchronous solution among those nodes, which is stable even when they are not directly connected.

Several variants of remote synchronisation have been propped, including *relay synchronisation*, whereby two delay-coupled oscillators can synchronise identically when connected through a third lagging oscillator (Fischer et al., 2006), and *incoherence-mediated remote synchronisation*, a scenario presenting properties of both remote synchronisation and *chimera* states (Abrams and Strogatz, 2004; Laing, 2009), chaotic oscillators connected via an intermediate cluster of oscillators incoherent both with the outer nodes and among themselves can stably synchronise (Zhang et al., 2017).

Symmetries of the quenched brain anatomical structure may induce correlated dynamics, where pairs of nodes with the same symmetry fully synchronise in spite of their distance (Nicosia et al., 2013). Likewise, network structure's mirror symmetry is the mechanism underlying *incoherence-mediated remote synchronisation*, whereby two non-contiguous parts of the network are identically synchronised while the dynamics of the intermediate part is statistically and information-theoretically incoherent (Zhang et al., 2017).

*Graph fibrations and hidden symmetries*

It has long been known that in the same way that dynamical systems with symmetry have symmetric solutions, network topology can force dynamical systems to synchronous or partially synchronous solutions, a phenomenon known as *robust network synchrony* (Stewart et al., 2003; Golubitsky et al., 2004, 2005; Golubitsky and Stewart, 2006; Golubistky et al., 2010). It was in particular shown that robustly synchronous network dynamics is determined by its *quotient networks*[61] (Golubitsky et al., 2005; Golubitsky and Stewart, 2006).

Quotient networks result from the identification of the nodes of the original network that evolve synchronously. While typically much smaller than the parent network, quotient networks can preserve geometric and topological properties such as geodesic distances, node heterogeneity and hubs (Xiao et al., 2008a). The quotienting of a network, i.e. a map $\phi: N \to N/\mathcal{G}$ from a network graph to its quotient, is an example of *graph fibration* (DeVille and Lerman, 2015). For networks of dynamical systems, *graph fibrations* $\phi: N_1 \to N_2$ (Boldi and Vigna, 2002) are a class of network morphisms where node clusters with isomorphic input trees, called *fibres*[62], are collapsed into a single representation, called *base*[63]. Thus, fibrations preserve the information flow within the network (Morone et al., 2020). Graph fibrations define *conjugacy*[64] between dynamical systems (Nijholt et al., 2016). *Topologically conjugate* functions have in some sense equivalent dynamics on their respective space. They have for instance the same number of fixed points. Graph fibrations give rise to maps between dynamical systems on networks (DeVille and Lerman, 2015). *Surjective* maps, e.g. arising from quotienting a graph by an appropriate equivalence relation, induce *embeddings* of dynamical systems, which can be used to characterise a networked dynamical system's modularity. *Injective* maps induce surjective maps of dynamical systems and *submersions* of the corresponding phase spaces and can depict the way a large dynamical system may be driven by one of its subsystems (DeVille and Lerman, 2015).

Graph fibrations induce a geometric network invariant with notable dynamical implications. Specifically, every networked dynamical system is conjugate to a network that is equivariant under the action of a semigroup (or semigroupoid). This generally holds true for *self-fibrations*, i.e. graph fibrations from a graph to itself $\phi: N \to N$[65]. The symmetry of the conjugate network thus acts as a *hidden symmetry* for the original network (Rink and Sanders, 2014). Thus, every networked dynamical system can be thought of as a dynamical system with a hidden symmetry acting on a *lift* rather than on it (see[64]). Importantly, contrary to network structure properties, hidden symmetries are invariant under coordinate changes and are therefore intrinsic properties of dynamical systems (Nijholt et al., 2016). Hidden symmetries constitute a geometric constraint inducing non-trivial network phenomenology such as *synchrony breaking bifurcations* which cannot be explained by the existence of robust synchrony (Rink and Sanders, 2014).

*Network topology and symmetry breaking in dynamical pattern formation*

An important, somehow complementary, issue is to do with the way network topology may interact with the system's dynamic symmetries. Networked dynamical systems can display complex bifurcations, e.g. anomalous steady states and Hopf bifurcations (Golubitsky et al., 2004; Aguiar et al., 2009; Rink and Sanders, 2013). These *synchrony breaking bifurcations* may be governed by spectral degeneracies similar to the symmetry breaking bifurcations in symmetric dynamical systems. Symmetry breaking leads to spatial and spatio-temporal *pattern formation*, e.g. synchrony and phase relations (Golubitsky and Stewart, 2002a,b). For instance, physical systems with $\mathbb{O}(2)$[66], as in a spherical space wherein connectivity

---

[60] Symmetry-induced cluster synchronisation and isolated desynchronisation have originally be illustrated for unweighted networks and identical systems, but the same general framework holds for directional coupling, heterogeneous coupling weights and non-identical nodes (Pecora et al., 2014).
[61] For a generic manifold $M$ with a symmetry group $\mathcal{G}$ acting on $M$, the *quotient space* $M/\mathcal{G}$ is the space in which two points in $M$ are identified if they can be obtained by the action of $\mathcal{G}$. If $M/\mathcal{G}$ itself turns out to be a manifold, the $M \mapsto M/\mathcal{G}$ map will then be a *local fibration*, and $M$ is built up of appropriately glued neighbourhoods in $M/\mathcal{G} \times \mathcal{G}$ (Dundas, 2018).
[62] The *fibre* over a node $x$ of $N_2$ is the set of nodes of $N_1$ that are mapped to $x$.

[63] A morphism of directed graphs $\phi: N_1 \to N_2$ is a *graph fibration* when each edge of $N$ can be uniquely lifted at every node in the fibre of its target. $N_2$ is called a *quotient* of $N_1$, while $N_1$ is a *lift* of $N_2$. A *self-fibration* is a graph fibration from a set to itself.
[64] Given two continuous functions on topological spaces $f: X \to X$ and $g: Y \to Y$, $f$ and $g$ are *topologically conjugate* if there exists a homeomorphism $h: X \to Y$ such that $h \circ f = g \circ h$. The conjugacy $h$ maps orbits in one space to orbits in the other.
[65] While not all networks admit nontrivial self-fibrations, every network is a quotient of a network with self-fibrations.
[66] The *orthogonal group in dimension n*, denoted $\mathbb{O}(n)$, is the group of geometric transformations of the Euclidean space of dimension $n$ preserving distances and a fixed

only depends on the relative distance between nodes, symmetry undergoing a Hopf bifurcation have been associated with standing and rotating waves (Fiedler, 1988; Golubitsky and Stewart, 2015).

A universal mechanism for the spontaneous generation of spatially organised patterns is represented by *Turing instabilities* characteristic of reaction–diffusion processes (Turing, 1952). The dynamics can be thought of as the evolution of spatially distributed species driven by microscopic reactions and freely diffusing in a medium. In its original formulation, one of the species acts as a self-catalyst, opposed by the competing species effectively stabilising the underlying dynamics. Introducing the notion of activator and inhibitor populations, Turing pattern formation can be conceptualised as the interaction between short-range activation (slow diffusion), long-range inhibition (fast diffusion) (Gierer and Meynhardt, 1972). In this framework, diffusion perturbs the homogeneous mean-field state, through an activator-inhibitor mechanism (Gierer and Meynhardt, 1972), promoting the emergence of patchy spatially inhomogeneous patterns. The system as a whole takes the form:

$$\frac{\partial}{\partial t}u(\pmb{x},t) - f(u,v,t) = D_{act}\nabla^2 u(\pmb{x},t)$$
$$\frac{\partial}{\partial t}v(\pmb{x},t) - g(u,v,t) = D_{inh}\nabla^2 v(\pmb{x},t)$$
[16]

where the first left-hand side term represents the unknown vector function, the second term is a non-homogeneous term accounting for all local reactions, while the right-hand side term is a diffusion term, where $D_{act}$ and $D_{inh}$ are diagonal matrices of diffusion coefficients[67]. The emergence of spatial order requires a marked difference of the diffusion constants associated with the interacting species, which destabilise the uniform state, leading to the emergence of periodic spatial patterns with alternating activator-rich and activator-poor domains from the uniform background (Turing, 1952; Gierer and Meynhardt, 1972)[68].

While in its original form the system typically lives on a continuous space, a regular lattice, or a small network (Horsthemke et al., 2004; Moore and Horsthemke, 2005), and the emphasis is on the form of the nonlinear interactions underlying the instability, it is important to consider how the dynamics may be modified when the ambient space is a discrete topology which is not trivially embedded in continuous domains (Othmer and Scriven, 1971). For networked systems, [16] is modified by expressing the diffusion part in terms of the graph Laplacian $L$ (Nakao and Mikhailov, 2010):

$$\frac{d}{dt}u_i(t) - f(u_i,v_i) = D_{act}\sum_{J=1}^{N} L_{ij} u_j(t)$$
$$\frac{d}{dt}v_i(t) - g(u_i,v_i) = D_{inh}\sum_{J=1}^{N} L_{ij} v_j(t)$$
[17]

While in continuous media, non-uniform perturbations can be decomposed into a set of spatial Fourier modes representing plane waves with different wavenumbers, in networked systems, the same role is played respectively by the graph Laplacian's eigenvectors and corresponding eigenvalues (Othmer and Scriven, 1971).

Important aspects of pattern-formation in networked systems stem from non-trivial network topology (Nakao and Mikhailov, 2010; Asllani et al., 2012, 2014, 2020). In large random networks of activator-inhibitor systems with broad degree distribution and undirected (symmetric) connectivity, the Turing instability emerges under the same conditions of those associated with a continuous support, but the spreading directions are determined by network topology (Nakao and Mikhailov, 2010). The critical Turing mode is localised on a subset of nodes with similar degree, which depends on the species' mobility. For stationary patterns, *multistability*, i.e. different coexisting stationary patterns for the same parameter values, and *hysteresis* are observed (Nakao and Mikhailov, 2010). On the other hand, topology can drive instabilities for systems with directional connectivity even when a regular lattice coupling structure would not be associated with such instability (Asllani et al., 2014). Moreover, the generated patterns are topology-specific. While the instability condition is model-dependent, each model is associated with a class of graphs with prescribed spectral properties for which the instability can occur (Asllani et al., 2014). However, the presence of directed connectivity extends the conventional Turing scenario, driving instabilities even under conditions for which a regular lattice coupling structure would not be associated with such instability and promoting the emergence of a generalised class of topology driven patterns which have no counterpart in systems with continuous domains (Asllani et al., 2012, 2014). Transport asymmetry across a sufficiently large discrete compartment allows robust patterns formation over an extended region of parameter values, which fractionate the embedding space in spatially extended individual units (Asllani et al., 2020).

Finally, while the instability and the corresponding emerging patterns may be intimately related to the degree of heterogeneity and, thus, be purely topological in nature (Nakao and Mikhailov, 2010), explicitly geometric patterns in simple network models have been reported in subsequent works (Cencetti et al., 2020; Hütt et al., 2022; van der Kolk et al., 2023). Turing instability triggers the emergence of purely geometric patterns that become evident in the latent space of real complex networks (van der Kolk et al., 2023).

*Network structure and pattern formation in brain dynamics*
Does network topology play any role in symmetry breaking and pattern formation in the brain? Insofar as macroscopic spatial patterns of cortical activation can be understood as arising from the self-organisation of interacting excitatory and inhibitory neuronal populations (Wyller et al., 2007), brain activity could in principle be thought in terms of pattern formation and Turing instabilities (Wilson and Cowan, 1973; Bressloff, 1996; Ermentrout, 1998; Jirsa and Kelso, 2000; Hutt et al., 2003; Verisokin et al., 2017). For example, in continuous attractor models, strong local inhibition drives pattern formation generating stable activity bumps organised in a triangular lattice on the cortical sheet, the cortical pattern periodicity being set by the spread of the lateral interactions (Khona et al., 2022). Phenomenology consistent with such dynamics has been reported at various neural levels. For instance, in a mean-field model of cortical dynamics, slow spatiotemporal oscillations arise spontaneously through a Turing spatial symmetry-breaking bifurcation modulated by a Hopf temporal instability (Steyn-Ross et al., 2013). In this model, interneuronal gap-junction synapses give rise to the diffusion term, and above a coupling strength threshold the dynamics undergoes a Turing bifurcation with stationary labyrinthine patterns (Steyn-Ross

---

point, with composition as its operation. Equivalently, it is the group of $n \times n$ orthogonal matrices, with matrix multiplication as group operation.
[67] Note that when the reaction term vanishes, the system reduces to the standard heat equation i.e. to a diffusion process.

[68] The system has a uniform stationary state $(\bar{u}; \bar{v})$, where $f(\bar{u}; \bar{v}) = 0$ and $g(\bar{u}; \bar{v}) = 0$, which can become unstable due to a Turing instability. The instability arises as the $D_{inh}/D_{act}$ ratio crosses a threshold. The Turing instability is revealed through linear stability analysis of the uniform stationary state with respect to non-uniform perturbations.

et al., 2007). Dynamic patterns may also emerge in pathological conditions. A notable example of coherent pathological global pattern formation is represented by epilepsy. Another important example of purely dynamical cortical pattern formation is represented by spreading depression[69] (Leão, 1944) and similar extreme cortical phenomena, which give rise to spatio-temporal patterns such as, spiral waves (Santos et al., 2014) or retracting waves (Dahlem et al., 2010). Interestingly, during spreading depression, stationary Turing-like patterns emerge from the interaction between vascular-mediated spatial coupling and local regulatory (Verisokin et al., 2017).

As in classical pattern formation models, spontaneous neuronal pattern formation in large scale neural activity has initially been defined on a continuous support or on a regular lattice (Cowan, 1982; Bressloff and Cowan, 2002; Bressloff et al., 2001, 2016). This may in some cases constitute an acceptable approximation, particularly at macroscopic scales. For example, global anatomical connectivity appears homogeneous (translationally invariant) and isotropic (rotationally invariant) across the cortex (Schüz et al., 2006; Braitenberg and Schüz, 2013; Kaiser et al., 2009). It was shown in particular that homogeneous networks in the primary visual cortex (V1) support travelling waves with propagating fronts separating regions of high and low activity (Bressloff, 2001). However, traveling wavefronts are not structurally stable solutions, and traveling wave solutions are sensitive to the degree of homogeneity of the connectivity pattern. In fact, weak heterogeneity is sufficient to dramatically affect wave propagation in excitable media (Bressloff, 2001). The main questions are then: to what extent can cortical dynamical properties be ascribed to homogeneity and isotropy? What is the role of inhomogeneous connectivity and anisotropy? In a spatially continuous homogeneously connected neural system, the introduction of heterogeneous two-point connectivity desynchronises connected areas, destabilising the associated spatial mode, and guiding the system through a series of phase transitions (Jirsa and Kelso, 2000). However, anisotropic coupling, characterising e.g. lateral connections in V1[70] (Hubel and Wiesel, 1974; Bressloff et al., 2001) is not a necessary condition for pattern formation, which may for instance be observed in spatially forced isotropic models. Overall, the role of non-trivial topology of cortical structure on pattern formation at mesoscopic and macroscopic scales is still poorly understood. Finally, it is worth noting that the role of homogeneity and anisotropy is scale-dependent. This in particular implies that isotropic homogeneous connectivity and modularity are not incompatible, the bulk within modules possibly being isotropic, the subgroup of connections between modules breaking this symmetry (Henderson and Robinson, 2013).

### 3.5.2 The role of network dimension in brain dynamics

For short-range equilibrium systems, second-order phase transitions can be cast into a relatively small number of universality classes, implying that simple representations provide quantitatively correct representations of seemingly rather different phenomena if the essential features responsible for the ordering are appropriately accounted for. Non-equilibrium critical phenomena too can be grouped into universality classes (Henkel et al., 2008). A straightforward consequence is that the dynamical phenomenology of a system may strongly depend on its dimension (Hinrichsen, 2000; Vojta, 2006; Ghorbanchian et al., 2021). Each universality class is fully specified by the system's global symmetry of the interactions and its dimensionality, all the detailed properties of microscopic interactions and other system properties being irrelevant in a renormalisation group sense (Kadanoff, 1971). For instance, the fluctuations of the order parameter about its average value tend to increase as dimension decreases and for a dimension below a lower critical one, they become strong enough to wipe away the ordered phase. This is why Landau's theory of phase transitions, where fluctuations are neglected, does not correctly describe the critical behaviour for systems of low dimension.

For regular networked spaces such as a lattice, it is straightforward to define a Euclidean dimension. In regular lattices, space dimensionality governs equilibrium and non-equilibrium systems' critical behaviour. Furthermore, the universal properties exhibited by dynamical processes such as diffusion, vibrational excitations, and scaling of fluctuations near a second-order phase transitions crucially depend on the *Euclidean dimension D* (Vojta, 2006). However, for more disordered spaces the concept of dimension needs to be generalised.

Several generalisations of the concept of dimension have been proposed for complex networks (Burioni and Cassi, 1996; Shanker, 2007, 2010; Smyth et al., 2010; Daqing et al., 2011; Lacasa and Gómez-Gardeñes, 2013; Skums and Bunimovich, 2020; Esfandiary et al., 2020). Dimension is usually defined as the scaling exponent of some property against some other one (e.g. volume within a given distance) in an appropriate limit. In the case of networks, it is straightforward to take the number of nodes as the volume.

In regular lattices, synchronisation is not observed for Euclidean dimension $D \leq 2$; for $2 < D \leq 4$ only entrained frequency synchronisation, but not phase synchronisation, is observed, while global synchronisation requires that $D > 4$ (Millán et al., 2018a). It has been shown both theoretically (Noest, 1986), using the Harris criterion[71] (Harris, 1974), and via simulations (Vojta et al., 2009) that spatially quenched disorder changes directed percolation's critical behaviour for $D < 4$ (Ódor, 2014a). For systems of the Kuramoto oscillators embedded in a discrete *D*-dimension small-world and modular manifold, frustrated synchronisation can arise for both $D = 2$ and $D = 3$, the latter condition is associated with more sustained synchronisation (Millán et al., 2018a), consistent with experimental evidence showing that neuronal cultures grown on 2*D* slices are less likely to maintain synchrony and to have simpler phenomenology than those grown on 3*D* scaffolds (Severino et al., 2016).

Geometrical investigations of network structure have considered *fractal dimensions,* which characterise the scaling of the number *N* of neighbours of a node as a function of some distance[72] (Rozenfeld et al., 2010). One such measure is the *topological dimension* $d_H$. In networks with finite $d_H$, disorder may turn out to be relevant (Muñoz et al., 2010). Interestingly, hierarchical modular networks have finite $d_H$ and intrinsically *large* diameters (Gallos et al., 2012). In hierarchical modular networks, $d_H$ tunes activity patterns, turning out to be the structural determinant of activity spreading in networks of oscillators (Safari et al., 2017). In fact, although the epidemic threshold in hierarchically modular networks depends on a large number of eigenvalues, it turns out to be inversely proportional to the topological dimension $d_H$. Insofar as hierarchical modular systems have finite $d_H$, a direct dependence of the epidemic threshold on $d_H$ indicates that the activity epidemic threshold never vanishes, i.e. information propagation is bounded, preventing runaway excitation and epileptiform activity. Importantly, this dependence suggests that the system may choose contextually appropriate dynamic regimes by tuning a single parameter (Safari et al., 2017). Furthermore, finite $d_H$ in hierarchical-modular networks is associated with non-trivial dynamical fluctuations (Tavani and Agliari, 2016; Agliari and Tavani, 2017; Moretti and Muñoz, 2013), and frustrated synchronisation (Donetti et al., 2006; Villegas et al., 2014; Millán et al., 2018).

---

[69] Cortical spreading depression is a wave of activity that propagates slowly across the brain cortex (Leão, 1944) which is associated with excessively high neuronal activity, leading to massive ion redistribution and corresponding increase of metabolic demands.
[70] In the primary visual cortex, *anisotropy* means that the neurons of neighbouring hypercolumns are connected only if they are tuned to the same orientation and oriented along the direction of their cells' preference (Golubitsky et al., 2003). Note that the bulk of the evidence for hypercolumns is not histological but stimulus-induced functional activity. In the developing cerebellar cortex, asymmetric Purkinje cell connectivity mediates traveling waves (Watt et al., 2009).

[71] The Harris criterion provides the conditions under which a system's critical exponents at a phase transition are modified by the presence of locally random impurities.
[72] The fractal dimension is metric-dependent (Boguñá et al., 2021). For instance, the topological or Hausdorff dimension of a network of *N* nodes is defined as the number $d_H$ such that the number of nodes in the neighbourhood a given node scales with the *topological distance r* as $N_r \propto r^{d_H}$. Low $d_H$ values are associated with long distances between nodes, while $d_H$ diverges for small-world networks.

Frustrated synchronisation was initially thought not to emerge in networks with stronger connectivity patterns, e.g. small-world or high-degree random networks with diverging $d_H$ (Villegas et al., 2014), which is thought to imply enhanced signal propagation and synchronisability (Esfandiary et al., 2020). Recent evidence shows that these networks can also display frustrated synchronisation, provided they have finite *spectral dimension* $d_S$ (Millán et al., 2018a, 2019). The spectral dimension $d_S$ characterises the Laplacian eigenvalue probability function's power-law scaling[73]. $d_S$ is thought of as an important factor in determining the resulting synchronisation properties and may constitute the relevant control parameter for the generic universal behaviour on inhomogeneous structures. Insofar as $d_S$ captures the dynamical signature of localisation and slowing down it has been used to characterise the effects of the quenched anatomical structure on resting brain activity in the long-time limit (Millán et al., 2021a). Interestingly, for hierarchical-modular brain networks, which have diverging $d_S$, the link between localisation of neural activity and spectral properties of the anatomical network may be related to a property of lower spectral edge of the Laplacian matrix, i.e. the existence of an anomalous *Lifshitz dimension*[74] (Esfandiary et al., 2020).

It is often stressed that while $d_H$ is a purely structural quantity, $d_S$ quantifies a network's propensity to diffuse information. Intuitively, it can be thought of as the network dimension that would be perceived by a diffusion process (Burioni and Cassi, 1996, 2005; Bianconi and Dorogovtsev, 2020; Torres and Bianconi, 2020). The meaning of this statement should be carefully parsed. On the one hand, the diffusion process is intended in a random walk sense. In turn, the random walk is sometimes intended as a proxy of information transport (Ghavasieh et al., 2020; Benigni et al., 2021; Villegas et al., 2022, 2023). However, there is no indication that such a simple model for dispersion phenomena, particularly in the simplest form of unbiased random walk with Gaussian steps often used in graph theoretical contexts, constitutes an appropriate model of actual brain activity and information spreading. Instead, it should perhaps better be conceptualised as a means to explore and quantify network topological properties, particularly community structure (Donetti et al., 2006).

Dimension has a qualitatively different nature and meaning for dynamical network structure. Particularly when explicitly considering dynamics, the very definition of dimensionality becomes less straightforward (Bialek, 2020, 2022). In a purely dynamical context, network dimensionality may be defined in terms of the number of modes into which it is organised (Dahmen et al., 2020). At macroscopic scales and in the strongly coupled regime, the dynamic network dimension may be controlled by connectivity strength (Dahmen et al., 2020). Connectivity, in turn, may be regulated by local synaptic features. Thus, while quenched dimensionality may be relevant to steady state spreading properties (Safari et al., 2017), annealed structure may be regulated by local network features at much shorter, experimental time scales. However, it is within a genuinely functional context, wherein its definition and the appropriate space which it quantifies change, that dynamic network dimensionality acquires its deepest significance.

### 3.6 Spatial embedding

Both anatomical and dynamical brain network models at various spatial and temporal scales are often embedded in the anatomical space, taking the spatial domain into explicit account (Mehring et al., 2003; Kaiser and Hilgetag, 2004a,b; Kumar et al., 2008; Bassett et al., 2010; Vértes et al., 2012; Ercsey-Ravasz et al., 2013; Stiso and Bassett, 2018). Spatial embedding means that the nodes and links live in a metric embedding space $(S, d)$ with metric $d: S \times S \to \mathbb{R}^+$. $S$ may for instance be a (bounded) Riemannian manifold, and $d$ the Riemannian metric. One important aspect of this space is the connectivity scaling function $\gamma: \mathbb{R}^+ \to [0,1]$, where $\gamma(s)$ gives the probability of an edge linking a pair of nodes at a distance $s$. Connectivity can be expressed in terms of a connection probability function: $\gamma(d(x,y)) = c(x,y)$ (Barnett et al., 2007), reflecting the intuition that spatial distance between nodes should be related to connection probability. Save for a few studies (Barnett et al., 2007; Bradde et al., 2010), the effects of spatial embedding on non-random network structure are still poorly understood. But what does spatial embedding mean? Is it justified or even necessary? Has spatial embedding any role in network structure emergence? Does it modify networked systems' dynamics?

### 3.6.1 The role of network geometry

Graph theoretical modelling in neuroscience is usually primarily concerned with the system's topology: vertices are thought of as an arbitrary set of points, and their location and metric distance within the anatomical 3D Euclidean space are not considered, distances between vertices being defined as the minimal number of edges in a path connecting them. However, brain networks' structural characteristics may not be uniquely determined by the adjacency matrix, and the network's layout and metric aspects may affect the system's structure, dynamics, and function (Wang and Kennedy, 2016). Embedding in anatomical space involves considering aspects such as distance-dependent couplings between nodes at the appropriate scale (Mehring et al., 2003; Kumar et al., 2008; Voges et al., 2010), or conduction delays (Voges and Perrinet, 2010).

A geometry can arise in various ways (Boguñá et al., 2021). Alternative ways in which geometry may become relevant are also possible. For instance, complex networks have an underlying continuous hidden metric structure, in which node distance reflects node similarity (Boguñá et al., 2008; Boguñá and Krioukov, 2009; Krioukov et al., 2010; Bianconi and Rahmede, 2017). Likewise, each process induces its own geometry with a particular (*kinematic*) distance, which defines an effective geometry of a system's function, which cannot be obtained by purely topological latent geometry (Boguñá et al., 2021). Dynamical processes induce a network structure. Furthermore, geometry may represent an emergent property of the underlying network structure (Bianconi and Rahmede, 2015, 2017).

---

[73] For lattices, $\rho(\lambda) \sim \lambda^{d_S/2 - 1}$, for $\lambda \ll 1$ (Rammal, 1984). $d_S$ can be measured by simple random walk analysis. The spectral dimension can also be quantified by random walk analysis. Given the average return probability $R(t)$ that a random walker starting at time $t_0$ at node $i$ ends up at node $j$ in $t$ time steps, the spectral dimension is defined as $R(t) \sim t^{-d_S/2}$. For finite $d$-dimensional lattices, $d_S = d_H$. However, in general, $d_S$ coincides neither with $d_H$ nor with the Euclidean dimension $D$ of the space it tessellates (Burioni and Cassi, 1996).

[74] The *Lifshitz dimension* is the scaling exponent $d_L$ of the integrated density of the Laplacian matrix, when this is dominated by the tail of the spectrum, i.e. the occurrence of eigenvalues above the lower Laplacian spectral edge is a large-deviation event. (Lifshitz, 1964). In this case, the density scales as $\rho(\lambda) \sim exp(\lambda^{d_L/2})$ (Esfandiary et al., 2020).

What brain aspects can meaningfully be endowed with a geometry? The most obvious geometry would be the one induced by anatomical wiring. At macroscopic scales, the anatomical space cannot be considered Euclidean, but can to some approximation be treated as a manifold $\mathcal{M}$, with some metric $g$ (Lenglet et al., 2006; Pennec et al., 2006; Krajsek et al., 2016; Pervaiz et al., 2020). (How this structure may emerge and the constraints it faces are discussed in §*Geometric constraints on wiring* below).

Is geometry playing any role in the annealed network structure? If so, in what space does network geometry play a role? A manifold representation may also be associated with the space into which dynamics and function live (Papo, 2019). For instance, thinking of brain dynamics as a probability distribution space (Papo, 2014a) allows representing brain activity as a manifold of probability distributions, where $g$ is the Fisher-Rao information metric, recovering a continuous space even when the underlying state space is discrete (Amari and Nagaoka, 2007) Furthermore, the presence of dynamic fluctuations with non-trivial scaling properties equips the system with a geometry (Papo, 2014a). Thes scaling laws, which can be seen as the statistical properties prescribed by the symmetries of a *semigroup* on the time-scale space equip the space with a geometry that is fractal rather than Euclidean (Lesne, 2008).

*Geometric constraints on wiring*

At the fine scales of a spatially embedded cortical network, physically constrained descriptions of connectivity may prove to be important for understanding cortical dynamics and function (Knoblauch et al., 2016). Network models usually assume dimensionless nodes and edges. However, the physical size of nodes and edges affects network geometry. In particular, the path chosen by edges, e.g. neural fibres, may be characterised by tortuosity as a function of node and edge size and density (Dehmamy et al., 2018). For small edge thickness, edge crossings can be avoided via local rearrangements, without altering the overall geometry. Above a thickness threshold, geometric quantities such as the total edge length and the edge curvature scale with edge thickness. As node and edge density increase, the non-crossing condition induces a cross-over to a strongly interacting regime where the overall network geometry is altered. The cross-over is associated with changes in the network's mechanical properties (Dehmamy et al., 2018). Networks in the weakly interacting regime display a solid-like response to stress, whereas in the strongly interacting regime they behave in a gel-like fashion (Dehmamy et al., 2018). Thus, in networks with a great number of nodes and links such as the brain, crossings cannot be ignored. Notably, the transition is topology-independent, suggesting that a complete structural description should integrate both topology and geometry. A quantitative estimate of link and node volumes' scaling properties may ultimately help determining the respective role of topology and geometry in brain structure.

The fact that physical wires cannot cross imposes limitations on the system's structure (Bernal and Mason, 1960; Song et al., 2008; Cohen and Havlin, 2010; Dehmamy et al., 2018; Liu et al., 2021), which is ultimately determined not only by operators associated with the connectivity matrix, but also by the network's 3D layout (Cohen and Havlin, 2010). For instance, in 2D, only *planar networks* are possible. On the other hand, all networks can be embedded without link overlap in 3D (Whitney, 1992), but a network can have an infinite number of configurations for a given adjacency matrix, differing in node positions and wiring geometry. Some of these configurations may be *isotopic*, meaning that they can be mapped into one another through continuous bending one-to-one at every step, and without link crossings or cutting. The layouts for which such a transformation is impossible are non-isotopic embeddings, each of which defines a distinct *isotopy class*. Whether two network embeddings belong to the same isotopy class can be determined using the *link number*, a quantity capturing the entangledness of a layout[75]. However, whether isotopic networks are necessarily dynamically and functionally equivalent and, conversely, non-isotopic networks are necessarily dynamically and functionally inequivalent is still not known. More generally, the following fundamental questions have yet to be addressed in earnest: what's the relationship between topology and geometry in anatomical networks? How do the geometric constraints on wiring affect structure and dynamics, but also development, evolution, and functional efficiency and robustness to various sources of damage?

### 3.6.2 Boundary conditions

One way in which spatial embedding may play a role is by imposing boundary conditions on the network structure. The most obvious boundaries are anatomical ones. For instance, at system level, the periodic boundary conditions associated with neocortical surface's closed shell geometry may exert an influence on brain dynamics, due to the interference of synaptic fields (Nunez, 2000). But boundary conditions need not be anatomical. For instance, in neural field models of brain dynamics, boundary conditions are in general not explicitly associated with anatomical landmarks, but are rather taken to represent dynamical and, as a result functional landmarks. While in these models the domain of definition is often infinite and essentially isomorphic to $\mathbb{R}^q$ (typically $\mathbb{R}$ or $\mathbb{R}^2$), in some studies (see for instance Faugeras et al., 2009; Gökçe et al., 2017), domains are limited to bounded sets $\Omega$ of $\mathbb{R}^q$. The corresponding boundary conditions specify the values that a solution of the system of integro-differential equations describing the neural population's time behaviour must have. The linear stability of localised states provides indications as to the way boundary conditions influence the creation and development of spatially extended patterns. For instance, the Dirichlet boundary condition, which specifies the values that the solution must take at the boundary of a domain, has been shown to limit the growth of highly structured stationary patterns (Gökçe et al., 2017). Likewise, different boundary conditions may affect spatiotemporal properties and dynamic stability. However, in these models, boundaries do not emerge from the dynamics. Instead, they would often be imposed *a priori* to evaluate their dynamical effects, e.g. on pattern formation.

An important question is whether and to what extent boundary conditions may affect brain scaling properties and criticality. In Landau's approach, the order parameter through which phases are described is unaffected by boundary conditions, for sufficiently large systems, as boundary conditions' contribution is typically sub-extensive. In this sense, phases are well-defined only in the thermodynamic limit. On the other hand, boundary conditions can indirectly modify the bifurcation behaviour of finite dynamical processes, inducing *nongeneric* bifurcations not expected for systems of given *codimension*[76] and symmetries. This could be done in various ways, e.g. by acting on symmetries (Crawford et al., 1991). For instance, in reaction-diffusion processes, boundary conditions may concur in extending the system to domains with larger symmetry groups (Crawford et al., 1991), and the extra symmetries may in turn be the source of the nongenericity (Field et al., 1991). However, in the context of brain dynamics, many aspects are still relatively unexplored, and many questions need to be addressed. Is there any empirical evidence for dynamic boundary conditions? If so, what type of boundary condition can be found? More relevant to the context being treated here, how do boundary conditions affect the network structure and the associated system's dynamics?

---

[75] The linking number $\mathcal{L}(\mathcal{E})$ is a knot invariant which measures the number of times two closed curves (cycles) wind around each other (Alexander, 1928; Kauffman, 2006). For a network with a given embedding $\mathcal{E}$, the graph linking number $\mathcal{G}(\mathcal{E})$ represents the sum of the linking numbers $\mathcal{L}(\mathcal{E})$ of all pairs of cycles in the graph (Liu et al., 2021). Although embeddings with the same *graph linking number* $\mathcal{G}(\mathcal{E})$ are not necessarily isotopic, embeddings with different $\mathcal{G}(\mathcal{E})$ belong to different isotopy classes. This allows distinguishing physical networks with identical wiring but different geometrical layouts. Importantly, different $\mathcal{G}(\mathcal{E})$s reflect differences in the associated system's function (Liu et al., 2021).

*Geometry- vs. topology-based brain eigenmodes*

One important question is the extent to which boundary conditions imposed by intrinsic brain geometry shape brain dynamics. (See *§3.1 Structure of what?*).

It has been proposed that brain activity is predicted by the harmonic patterns induced by the anatomy of the human cerebral cortex, the human connectome (Atasoy et al. 2016; Preti et al., 2019; Rué-Queralt et al., 2021). The underlying idea is that when excited with one of its natural frequencies, the system resonates, giving rise to stable dynamical states, called *eigenmodes*, within which standing waves are formed and sustained by the oscillation of each spatial location with the same natural frequency and synchronised throughout the system (Atasoy et al. 2016; Preti et al., 2019; Rué-Queralt et al., 2021).

The excitation spectrum of a given geometry stemming from the interaction between natural frequency and corresponding standing wave pattern is expressed by the eigenvalue–eigenvector pairs of the associated Laplace operator (eigenmodes): the eigenvalues relate to the natural frequencies, the allowed frequencies of standing waves emerging on that particular geometry, whereas the eigenvectors yield the associated wave patterns. Specifically, the connectome spatial harmonics $\psi_k$ are calculated by solving the eigenvalue problem

$$\Delta_{\mathcal{G}} \psi_k(v_i) = \lambda_k \psi_k \qquad [18]$$

where $\psi_k$ are the eigenvectors of the Laplace operator $\Delta_{\mathcal{G}}$ of a network $\mathcal{G}$ whose nodes $v_i$ uniformly sample the cortical surface space, and whose local and long-range edges are respectively associated with the connections of the vertices on the cortical surface and with cortico-cortical and thalamo-cortical white-matter fibres (Atasoy et al. 2016). The eigenvectors $\psi$ of the Laplace operator determine the shape of the vibrational (standing wave) pattern for the geometry of the underlying domain, and the corresponding eigenvalues λ relate to the activated natural frequency. The eigenvectors of the connectome Laplacian were found to be associated with frequency-specific spatial patterns across distributed cortical regions which overlapped with spontaneous brain activity patterns, suggesting that connectome-based harmonics emerge from the interaction between the brain's oscillatory activity and its connectivity structure (Atasoy et al., 2016).

Recently, an alternative mechanism for brain harmonic mode generation has been proposed, wherein rather than from connectome-induced modes with no geometric boundary condition global brain patterns would result from excitations of brain geometry's resonant modes (Pang et al., 2023). Specifically, it was argued that a mean-field solution with purely geometric (and ideally continuous) boundary constraints may better capture important properties of the system's spontaneous activity as well as its response to exogenous perturbations with respect to a connectivity-based equivalent model disregarding neural surface's geometry (Pang et al., 2023). The eigenmodes induced by continuous cortical geometry are obtained by solving the corresponding eigenvalue problem of the Laplace–Beltrami operator of the underlying manifold (Lévy, 2006).

The significance and implications of this suggestion must be evaluated. A possible implication is that cortical anatomical geometry may contain important information and should therefore be included in models of brain dynamics. But is brain dynamics defined by brain shape alone sufficient or is connectivity also necessary? Such a question constitutes a rephrasing of the celebrated inverse problem "Can one hear the shape of a drum?" (Kac, 1966)[77]. However, there are reasons to suspect that geometric modes alone may not be the unique determinant of brain modes. Mean-field approaches have been shown to constitute an insufficient description of brain activity in general (Buice and Chow, 2013), so that it would in general seem unlikely that the resonant modes induced by cortical resonant modes would uniquely be determined by the shape of the neocortical surface alone, i.e. not all disorder in the bulk can be retrieved by the geometry of the boundary. Fine details of the connectivity structure in the bulk have also been shown to be relevant, as neural modes may be shaped by network properties such as degree distribution (Smith et al., 2018), motifs (Recanatesi et al., 2019; Hu and Sompolinsky, 2022; Dahmen et al., 2020) and community structure (Aljadeff et al., 2015, 2016). Furthermore, brain waves may not be determined by neuronal geometry and connectivity, but by tissue inhomogeneity and anisotropy: waves can traverse regions between neurons providing a mechanism for ephaptic coupling and spiking synchronisation phenomenology (Galinsky and Frank, 2020a). For instance, cortical activity organised into circular wavelike patterns on the persistent temporally precise cortical surface was found to span areas not directly related to any structurally aligned pathways (Muller et al., 2016).

On the other hand, it is important to understand the relationship between connectome-based and geometry-based representations. This is naturally understood by considering that in both cases the eigenmodes are in practice reconstructed from data clouds approximating the true manifold structure of the system. One possible strategy for manifold reconstruction consists in equipping the point cloud with a network structure approximating the geometric structure of the manifold (Belkin and Niyogy, 2008). As the number of uniformly sampled data points taken from the underlying manifold increases, the discrete graph appropriately samples the corresponding manifold so that the graph Laplacian together with its corresponding eigenvalues and eigenvectors converges to the continuous Laplace–Beltrami operator of the underlying manifold, revealing intrinsic information about the smooth manifold. Likewise, the random walk defined by assigning transition probabilities between vertices according to the edge weights, thus a graph the continuous limit of which is a diffusion process over the underlying manifold (Belkin and Niyogy, 2007, 2008). In this sense, a network structure may be thought of as a discrete Riemannian manifold and the difference between these two representations may not be qualitative but rather a scale-dependent one, and in practice, the choice between discrete and continuous models may be dictated by the scales of description, observation, variations, and correlations (Lesne, 2007).

---

[76] The *codimension* of a bifurcation is the number of parameters which must be varied for the bifurcation to occur generically.
[77] The frequencies at which a drumhead can vibrate depend on its shape. A central question is whether the shape can be predicted if the frequencies are known. For a compact Riemannian manifold $(\mathcal{M}, g)$ the eigenvalues of the Laplace-Beltrami operator $(\Delta f = \nabla \cdot \nabla f)$ acting on the smooth functions on $\mathcal{M}$ are essentially the frequencies produced by a drumhead shaped like $\mathcal{M}$. But does knowing all $\lambda_i$s determine the underlying manifold (up to isometry)? In other words, if two domains in the Euclidean plane are isospectral, are they also isometric? The answer was found to be negative, and numerous examples are known of isospectral but nonisometric Riemannian manifolds (Gordon et al., 1992).

*Curvature*

A manifold structure endows the underlying space with intrinsic properties, i.e. properties that do not depend on the way the network structure is parametrised (Lee, 2006). Notably, a network structure can be endowed with *curvature* (Ollivier, 2007, 2009; Knill, 2012b; Weber et al., 2017). While many definitions of graph curvature have been proposed, most do not converge to any curvature in the continuous limit and are therefore not genuine curvatures (van der Hoorn et al., 2021). However, one particular curvature, the Ollivier curvature converges to the Ricci one under specific conditions, i.e. for weighted graphs with positive real weights (van der Hoorn et al., 2021).

The ability to define a network curvature has a number of potentially interesting consequences. First, curvature may in principle help characterising the system's latent geometry (Lubold et al., 2022). It may also indirectly define network topology, through the Gauss-Bonnet theorem and its generalisations to higher dimensions, which relate a surface's curvature to its Euler characteristic (Lee, 2006; Knill, 2012a)[78]. Curvature can affect dynamics by modulating frustration (Nelson, 1983, 2002; Sethna, 1983). (See *§3.2.3 The dynamical role of frustration*). Finally, network curvature is also related to the network's structural robustness (Demetrius, 2013).

How can neural systems be endowed with curvature and what meaning could it take? It is straightforward to equip the anatomical structure with a curvature (Farooq et al., 2019; Simhal et al., 2020). Curvature can affect important dynamic properties, such as wave propagation in 2D and though to a lesser degree, 3D surfaces (Kneer et al., 2014), and could therefore underly observed hemispheric differences in wave propagation (Kroos et al., 2016). Furthermore, the static anatomical network structure's Ollivier-Ricci curvature can in principle be thought of as a proxy of structural network robustness[79], when the latter is thought of as the rate function at which a network returns to its original state after a perturbation as a large curvature is associated with high rate of convergence to the equilibrium distribution (Sandhu et al., 2015)[80].

Various important questions have not yet been addressed: has anatomical network curvature any relation with anatomical mechanical robustness? Does it have an impact on brain dynamics and function? In particular, could it be thought of as a control parameter for brain dynamics, as somehow suggested by results linking treatment-induced changes in region-specific anatomic network curvature in autism (Simhal et al., 2020)? Could it play a role in brain dynamics' bifurcation structure? Does it reflect the presence of frustration? A curvature can also in principle be associated with dynamical networks, both in real space (Tadić et al., 2019) and in the parameter space (Janke et al., 2004). For instance, one may associate curvature to the manifold of isofiring rate neurons. But could annealed network curvature reflect relevant aspects of brain activity? Conversely, could curvature affect brain dynamics? Could it have a role in dynamical or functional robustness? Intuitively, the role of curvature can be grasped by thinking of network structure heterogeneity as topological defects, and considering how these interact with the curvature of the underlying space (Nelson, 1983; Vitelli and Turner, 2004; Turner et al., 2010)[81].

### 3.6.3 Critical dynamics

Spatial embedding may also affect a networked system's behaviour around a phase transition, producing critical fluctuations not captured by heterogeneous mean-field models (Bradde et al., 2010). The underlying dynamical system's critical behaviour is controlled by the spectral properties of the connectivity matrix, a Ginzburg-like criterion determining the conditions under which critical fluctuations become larger than the mean-field one (Bradde et al., 2010). Specifically, while for non-vanishing spectral gaps the fluctuations are always mean field, when the spectral gap vanishes, the critical behaviour depends on the scaling properties of the upper edge of the adjacency matrix spectral density. Likewise, in complex networks embedded in a low dimensional space the appearance of short-ranged connectivity changes the system's critical behaviour leading to a breakdown of the validity of (heterogeneous) mean-field arguments (Bradde et al., 2010).

### 3.6.4 Is the brain a spatial network?

Spatial embedding relevance would imply that the brain is a *spatial network* (Barthélemy, 2022). Spatial networks are graphs whose nodes have well-defined positions in space (Comin and da Fontoura Costa, 2018). In a statistical mechanics understanding of complex networks, the identity of single parts of the system is lost *prima facie*, as network properties are statistical in nature. On the other hand, spatial embedding is implicitly justified by the need to ascribe given properties to specific regions of the anatomical space. In this approach, an important network symmetry is lost, as nodes are no longer exchangeable.

Spatial embedding can in principle affect various network properties including degree and link distribution, centrality, clustering, community structure and modularity (Barthélemy, 2022). In a spatially embedded network, nodes are more likely to connect to their spatial neighbours, as spatial constraints and costs associated bound link length. The formation of cliques between spatially close nodes increases the clustering coefficient. Furthermore, spatial constraints favour the formation of regional hubs and locally reinforce preferential attachment, leading to a larger strength for a given degree than the one observed without spatial constraints. Long-range links tend to preferentially connect to hubs. Spatial distance selection also induces strong correlations between topological quantities, e.g. degree distribution, and non-topological ones, e.g. link weights (Barthélemy, 2022). Finally, spatial embedding also induces strong non-linear topology-traffic correlations.

But to what extent does the brain present the properties of a genuine spatial network? One way in which this can be evaluated is by constructing reference graphs preserving the scaling properties of connectivity, but lacking additional topological properties (Roberts et al., 2016). The connectome's structure cannot entirely be accounted for in terms of topological wiring rules and appears to partly stem from spatial embedding. Geometry has a strong influence on many topological properties of the anatomical network (Henderson and Robinson, 2011, 2013, 2014; Roberts et al., 2016). The connectome's embedding into $3D$ space determines some aspects of network topology compared to a completely random network (Henderson and Robinson, 2013). Cranial volume, folding patterns, and fibre packing add additional topological complexity (Henderson and Robinson, 2014).

---

[78] For a compact smooth oriented surface $M$ in $\mathbb{R}^3$, the Gaussian curvature $\kappa$ with respect to area $A$ on $M$ is related to the Euler characteristic $\chi$ as $\int_M \kappa \, dA = 2\pi\chi(M)$. Remarkably, curvature, a geometric quantity, which is invariant under translations and rotation, but not under stretching and deformation of the underlying surface, is related to the topological quantity $\chi$. In the discrete case relevant to graphs, Gaussian curvature on a graph $N$ with vertices $V$ and corresponding Euler characteristic $\chi$ are related as $\sum_{x \in V} \kappa = \chi(N)$ (Knill, 2011).
[79] Curvature and robustness are related to each other via entropy. Entropy and curvature are positively correlated, i.e. $\Delta S \times \Delta Ric \geq 0$, where $\Delta S$ and $\Delta Ric$ are respectively the change in entropy and Ricci curvature (Sandhu et al., 2015).

[80] A sytematic discussion of robustness/vulnerability and related constructs in neural systems is beyond the scope of this study and is more appropriately discussed in a genuinely functional context.
[81] In liquid crystal films coating frozen surfaces, each defect feels a geometric potential whose functional form is determined by the shape of the surface (Vitelli and Turner, 2004). In active matter, topographical changes in Gaussian curvature can regulate specific defect structure and direct flows (Bowick et al., 2022).

## 4. Concluding remarks: from dynamics to function

Such is the brain's complexity as to render the following question an open one (Beggs, 2015): can there be a physics of the brain? A system's physics is only possible if some structure can meaningfully be defined at some level. Thus, if a network structure was indeed intrinsic to the way the brain works, this would represent a fundamental step towards finding a positive answer to this question. Understanding the brain as a networked system and, as a result, equipping it with a network structure may seem both natural and simple and is indeed becoming standard in neuroscience. Here, we started discussing whether networkness can indeed be thought of as a genuine brain property and if so, how much network properties can tell us about the way the brain operates.

We have argued that there are multiple ways in which the brain can be thought of as a networked structure and, for various reasons, understanding the role of networkness in brain dynamics is far from straightforward. We first discussed the meaning of networkness and reviewed the possible role of combinatorial, topological, and geometric network properties in various aspects of brain dynamics. We illustrated how such roles hinge not only on the space equipped with a network structure and the scales at which that is done, but also on how such a network is defined. In particular, while we addressed a purely ontological (as opposed to methodological) question, we highlighted some important (though not always explicitly considered) implications of equipping a system with a network structure, generally stemming from the unavoidable fact that a network representation is in essence a kinetic equation, where nodes involve coarse-graining, and links some inferred metric.

Throughout, we addressed purely dynamical aspects, treating the system as a topological dynamical system, but avoiding as much as possible dealing with or incorporating details of brain function. This approach, which is rather common in neuroscience, wherein dynamics and function are used interchangeably, and the expression *functional brain activity* is employed without specific reference to function, stems from two underlying causes.

On the one hand, this approach is natural if one considers the statistical mechanics foundation of complex networks (Albert and Barabási, 2002; Dorogovtsev et al., 2008). Indeed, perhaps because it was not developed to describe living systems, statistical mechanics' standard formulation does not deal with function. Biological systems cannot be understood in full without a deep understanding of their function and bare dynamics only provides a partial picture on the brain can be appreciated at multiple levels. For instance, while neural populations' response function can be studied in terms of bare dynamics, it is far more natural to let it emerge from the functional properties of neurons' receptive fields. Likewise, low-dimensional manifold structure induced by neural activity typically has an inherently functional meaning and the same holds for pattern formation. At longer timescales, the relevant trade-offs underlying the formation of anatomical neural structure can only be characterised in terms of the function dictating it. Furthermore, while equivalence (topological, geometric, ...) classes of the dynamics may have important properties, their meaning must ultimately be gauged in terms of the system's ability to carry out the functions (transport, computation) it is assigned. Similarly, function induces spatial and temporal scales which cannot be trivially deduced from those of bare dynamics. Finally, and perhaps even more fundamentally, at microscopic scales, function is incorporated in the construction of dynamic coupling, even in purely dynamic models (Korhonen et al., 2021). Thus, a complete examination of brain networkness requires incorporating and accounting for function, i.e. explicity considering that neural structures in a wide range of scales are designed to carry out tasks, notably information transport and computation. A functional perspective in particular involves providing an adequate definition of genuinely functional brain activity, and essentially revisiting all aspects reviewed in dynamics, in the much richer functional domain.

On the other hand, a purely dynamical framework reflects network neuroscience's general tendency to borrow network constructs developed to study systems profoundly different from them and, at a deeper level, to generalise results and interpretations to a substrate for many reasons profoundly different from them (Papo et al., 2014c). The justification for such a generalisation lies precisely in the statistical mechanics approach, which is predicated upon the fact that at large scales, complex systems can be described in terms of only a small number of relevant features, microscopic details turning out to be irrelevant. But to what extent and at what scales brain dynamics and function depend on the specific details of their nodes and edges and, on the other hand, to what extent simple network structure retains sufficient information to account for brain structure, dynamics, and function are not yet totally clear questions. Thus, aspects of neural activity that should be incorporated into neural network modelling and somehow dually, network models that may help representing brain structure, dynamics, and function will have to be examined in detail.

Altogether, the characterisation of the role of network structure provided by purely dynamical approaches may profoundly differ from the one arising from a genuinely functional approach incorporating the necessary and sufficient neural and structural ingredients and should therefore be treated as partial and preliminary.